\documentclass[reprint,aps,prd,nobibnotes,nofootinbib,onecolumn]{revtex4-2} 

\usepackage{amsmath}
\usepackage{amssymb}
\usepackage{graphicx}
\usepackage{fullpage}
\usepackage{hyperref}

\usepackage{color}
\definecolor{darkblue}{rgb}{0,0,0.5}

\begin{document}

\title{More QCD Masterclass Lectures on Jet Physics and Machine Learning}%

\author{Andrew J.~Larkoski}%
\email{larkoski@aps.org}
\affiliation{American Physical Society, Hauppauge, New York 11788, USA}


\date{\today}%

\begin{abstract}
\noindent These lectures were presented at the 2026 QCD Masterclass in Saint-Jacut-de-la-mer, France.\footnote{\url{https://indico.cern.ch/event/1574368}}  They supplement the published notes from the 2024 school \cite{Larkoski:2024uoc} with presentation of a few new topics that have become popular in recent years.  These include event metrics, anomaly detection, scaling laws, the limits of classification performance, and regression that respects spacetime symmetries.  Many of the presented results are known, but some new results are derived, especially related to anomaly detection.  As was the case in those original notes, while all these problems are motivated from machine learning applied to QCD, no discussion of details of machine learning is presented in these notes.  All results follow from application of fundamental principles of QCD.  I begin these notes with an apology for humanist physics.
\end{abstract}

\maketitle

\vspace{-0.04cm}

\tableofcontents

\section{An Apology for Physics}

With the promise of lecturing on ``Jet Physics and Machine Learning'', I feel that an initial apology is needed.  Unfortunately, one meaning of ``apology'' is as a statement of regret in which one might mean to apologize that the content of the lecture is not delivering the promise of the title.  This common modern day use of the word is rather far from its Greek roots in which {\it apologia} means ``a speech in defense''.  It is in this, much more active and Platonic, sense that I mean apology.  I mean to defend writing lecture notes on jet physics and machine learning in which no machines are used.

In the time between the presentation of the lecture notes of Ref.~\cite{Larkoski:2024uoc} and these, a mere two years, the way that theoretical particle physics, and in particular jet physics, is done has dramatically changed.  Artificial intelligence (AI) agents, the Claudes, ChatGPTs, Geminis, and such, have become more and more capable, which is to say that they have been trained on a larger corpus and fine-tuned more accurately.  The past year has seen multiple papers written almost exclusively by an AI agent (with physicist oversight), e.g., Refs.~\cite{Guevara:2026qzd,Schwartz:2026ekw,Shih:2026lmy,Moreno:2026mqk}, many of which are on or related to the subject of jet physics.\footnote{This list is assuredly woefully incomplete because there is not yet a universal culture of acknowledging significant use of AI tools in publications.}  There are now numerous claims that AI agents will ``solve physics'' within a handful of years, see, e.g., the recent prognostications in talks or blogposts of Refs.~\cite{sohlant,brownsand,schwartz_anthro,openairesearch}.  Depending on your sensibilities, this future of ``physics without physicists'' may be depressing or exciting.  However, I feel that both of these emotional extremes, as well as claims that ``physics will be solved in five years'' misses the entire point.  I want to argue through these lecture notes (and throughout my career as a physicist) that ``solving physics'' is precisely not the point of why we do it, and not why {\it students} go to graduate school and attend summer schools like this.

Physicists get interested in physics because they looked at the starry sky as a child and contemplated infinity, or were gifted a microscope on their birthday which opened up the microverse, or they had a particularly inspiring math teacher in grade school who kept feeding them richer and richer problems that tickled a fundamental curiosity.  These are deep, intrinsically {\it human}, connections to our universe, our place in it, and is something that every human must confront and attempt to answer for themselves.  There are no right answers, there are an infinity of possible questions, or how to approach them.  Physics, as a human discipline, is an ordered, systematic way to pose those unanswerable questions, in the same way that Shakespearean scholars read between the lines of {\it Hamlet} to gain insight into the issues of identity and place.  Even assuming the perfect, idealistic future in which AI tools or agents can solve every possible physics question, develop a theory of quantum gravity, or whatever quixotic dream you might have, that in no way helps me as a human struggle with what it means to be human.  It is the struggle that makes us human, and physics, like art history, music, literature, or film criticism, is a discipline that provides us with a window, a perspective, through which to view what it means to be a human.

In this way, I think of myself as a ``humanist physicist''.  Merely producing results is not why I do physics, and I am certain not why most physicists do physics.  If we, as a field, reduce physics to assembly-line production of results, we will assuredly, and voluntarily, make ourselves redundant, as no human can compete with an AI agent trained using billions of GPUs in a data farm that exploits a major regional river for its energy needs.\footnote{On the penultimate night of the school, minutes before we were to be served the {\it fruits de mer} banquet dinner, OpenAI announced they had solved the Navier-Stokes problem, demonstrating that a finite-time singularity can develop from an initial smooth configuration \cite{openainavier}.  However, there are claims that their effort was influenced or even aided by others without their consent \cite{buckmaster_state}.   The oysters, drizzled with a light vinegar-shallot dressing, were excellent.}  We must remember we are {\it natural philosophers}, learning more about what it means to be human by plumbing the depths of Nature.\footnote{We must also defend liberal arts education otherwise we may, from the purely utilitarian, ``results'' based physics, have a future that is already the present of the more traditional humanities subjects.  In this way I note that the Leiden Declaration of the international mathematics community \cite{alper_2026_20302944} is both bold and brave, and is something that is sorely needed in physics.  For a couple different perspectives on the Leiden Declaration, see thoughts by Timothy Gowers \cite{gowersleiden} and Terry Tao \cite{taomathinai}.}

\clearpage

\section{Introduction}

In the lecture notes of Ref.~\cite{Larkoski:2024uoc}, significant background and review of fundamental material was provided, both of the fundamental theorems of machine learning, as well as foundational ideas of quantum chromodynamics (QCD) and the phenomenon of jets.  These supplemental lecture notes build on those earlier notes, and assume all of the background set up there.  These notes, then, will dive immediately into both new topics, and provide much more detail of few things that were just briefly mentioned in Ref.~\cite{Larkoski:2024uoc}.

Jet physics, the study of the high-energy, collimated streams of particles ubiquitous in QCD, has become and will remain a piece of the core research program of colliders like the Large Hadron Collider (LHC).  More and more, theoretical studies of jets have evolved to development of machine learning architectures for classification or regression tasks, though there are notable exceptions.  The lecture notes here, as was the case with Ref.~\cite{Larkoski:2024uoc}, will focus on development of the intuition and understanding of structures in QCD, the  way that they are imprinted on physical observables, and methods for their calculation and prediction.  Machine learning, through a strict definition of a trained network computing architecture, will essentially never appear.  The exception is that several of the topics we will discuss were motivated from machine learning studies, but here we will work to provide a first-principles description and understanding of them.

An outline of these lectures is as follows.  In Sec.~\ref{sec:smear}, I introduce a robust formulation of the sense of producing a continuous distribution from a discrete, finite sample of that distribution.  This is precisely what a machine does in a regression task, say, to estimate a matrix element from a finite dataset of collider events.  However, we can formulate the problem in generality and identify properties that we would want of such a prescription for ``smearing'' a discrete distribution into a continuous one.  This theme will appear again and again in these notes, so this section is sort of an overview of what is to come.

In Sec.~\ref{sec:antiqcd}, I approach the topic of anomaly tagging from a general perspective, and attempt to ground it in the theory of QCD.  There has been significant recent effort to construct machine-learned anomaly taggers in which ``optimal'' observables for isolation of deviations from a reference distribution are defined.  Typically, the ultimate goal of these taggers is to find signs of new physics, and so the reference distribution is (simulated) events that were generated through QCD processes.  Of course, in the binary discrimination problem, the true optimal observable is the likelihood (by Neyman-Pearson \cite{Neyman:1933wgr}), but away from that case, there is no universal optimum, because there is no unique definition of ``success'' for such a problem.  The approach advocated and presented here will use the smeared distribution to be able to define a local excess on a collider event's extremely high dimensional phase space.

In Sec.~\ref{sec:scaling}, I review scaling laws.  In 2020, the paper of Ref.~\cite{kaplan2020scaling} was basically individually responsible for the creation of the AI company Anthropic (compare the author list to the Anthropic C-suite).  The observation of scaling laws in large language models in which performance could be predictably improved by increasing compute resources meant that these corporations could (very literally) just throw more money at the problem.  These scaling laws are still only empirical observations, as there is no underlying theory for the structure of language or the space of images.  However, in particle physics we do have a fundamental theory and so we can directly predict and understand the relationship between performance (as defined by a classification task, for example) and compute resources (as defined by the size of a simulated data set, for example).

In Sec.~\ref{sec:toptag}, I present a general analysis of identification of a boosted top quark jet from jets in QCD initiated by light partons.  This problem is essentially as old as jet substructure itself \cite{Seymour:1993mx}, but has seen a renewed focus because of claims that there is a ``fundamental limit'' of this problem \cite{Geuskens:2024tfo,Pang:2025lbs}.  The study presented here will be essentially just (na\"ive) dimensional analysis, with only the simplest assumptions for what a ``top jet'' or ``QCD jet'' are.

In Sec.~\ref{sec:rambo}, I discuss regression tasks in jet physics and QCD.  Collider events of course are restricted by many symmetries and conservation laws, momentum conservation, Lorentz invariance, etc., and there has been significant effort to construct architectures that respect these symmetries (or at least have desirable transformation properties).  One can of course include violations of the conservation laws in the loss function of one's architecture, but this does not guarantee {\it exact} conservation, which can be significantly problematic practically, as well as for physics interpretation.  I present a simple solution that exploits the symmetries of massless phase space and the principle of maximum entropy, and which was written down 40 years ago in the RAMBO algorithm.

In Sec.~\ref{sec:concs} I briefly conclude, and present (more) thoughts on where jet physics and machine learning may go in the future.

\section{Smearing a Discrete Distribution}\label{sec:smear}

Given a large (simulated) data ensemble, our goal with machine learning is to ``learn'' the distribution from which that data are sampled.  This is the core of a discrimination task, in which we want to learn the differences between, say, a signal and background distribution to identify the optimal observable to discriminate them.  Or, this is the heart of a regression task, in which we would like to learn the probability distribution and then sample from that learned distribution in a controlled way in which we understand its systematic uncertainties, for example.  Machine learning is just a way to fit a distribution with an enormous number of parameters, but in a way in which minimal assumptions are used.  No constraint on the functional form of the distribution is assumed with machine learning; the machine just regresses to a best fit form, according to the architecture one uses and the specific loss function one employs.

However one does it, the machine outputs a distribution that can be sampled at any point on the event space manifold; not only those points that exist in the training sample.  However, the training sample does of course inform the output, according to the relationship between the training data and the particular point on the manifold in which one is interested.  This enables us to express the form of the output of such a regression task very generally.  Let's define $p(\vec x')$ as the distribution of the training data on the space $\vec x'\in {\cal M}_\text{data}$, where ${\cal M}_\text{data}$ is the data manifold. As the training data is discrete, this distribution is just a sum of $\delta$-functions:
\begin{align}\label{eq:discdist}
p(\vec x') = {\cal N}\sum_{i\in \text{data}} \delta(\vec x'- \vec x_i)\,,
\end{align}
where $\vec x_i$ is an individual instance or event in the training data and ${\cal N}$ is a normalization factor.  In what follows, we will consider completely general $p(\vec x')$, but it is good to keep this discrete form in the back of one's mind.

Next, we would like to establish the value of the distribution at a point $\vec x$ not in the training set, $p(\vec x)$.  The value at this point is informed by all points in the training set.  So, we can in general write that the value of the distribution can be expressed as a sort of convolution,
\begin{align}\label{eq:smeareddist}
p(\vec x|\epsilon) \equiv \frac{\int d\vec x'\, p(\vec x')\,\Theta\left(
\epsilon-d(\vec x,\vec x')
\right)}{\int d\vec x'\, d\vec x''\, p(\vec x'')\,\Theta\left(
\epsilon-d(\vec x',\vec x'')
\right)}\,.
\end{align}
Here, $\epsilon$ sets a sort of effective window of influence, and $d(\vec x,\vec x')$ is just some function that relates two points on the data manifold.  We have used a Heaviside step function, but this can be relaxed to a continuous or smooth window kernel.  This has been long considered in the statistics literature as kernel density estimation \cite{rosenblatt1956remarks,parzen1962estimation}, but we won't worry about possible forms for this kernel here.  Finally, the denominator ensures that the distribution remains unit normalized.

With this form of a ``smeared'', continuous distribution on the manifold ${\cal M}_\text{data}$, what are natural properties of the comparison function $d(\vec x,\vec x')$?  Honestly, very, very little constrains this in general, but with physics motivation, we can make progress.  First, we want $\epsilon$ to have some notion of ``distance'', so $d(\vec x,\vec x')$ must be non-negative, $d(\vec x,\vec x')\geq 0$.  Second, we would like to ensure that in the $\epsilon\to 0$ limit, the initial, discrete, distribution of Eq.~\ref{eq:discdist} is returned.  For this to be true, we must require that $d(\vec x,\vec x')\to 0$, as $\vec x'\to \vec x$.  Third, as written, Eq.~\ref{eq:smeareddist} is asymmetric between $\vec x$ (the point of interest) and $\vec x'$ (the influencing point).  However, it could have just as easily been that point $\vec x$ was in the training data, and $\vec x'$ was not, so we would like $d(\vec x,\vec x')$ to be symmetric in its arguments, $d(\vec x,\vec x')=d(\vec x',\vec x)$.

These three properties are referred to as non-negativity, identity of indiscernibles, and symmetry, respectively.  If we really want to interpret $\epsilon$ as a radius of a ``sphere of influence'', that is, as a true distance within which training data affects the point of interest, we need one more property.  That property is the triangle inequality which ensures that we can say that some points are closer to one another than other points.  The triangle inequality is
\begin{align}
d(\vec x,\vec x'')\leq d(\vec x,\vec x')+d(\vec x',\vec x'')\,,
\end{align}
for three points $\vec x,\vec x',\vec x''$ on the manifold ${\cal M}_\text{data}$.  These four properties of $d(\vec x,\vec x')$ define it to be a {\bf metric} and endows the manifold ${\cal M}_\text{data}$ with it.  As such, ${\cal M}_\text{data}$ is a metric space, on which distances can be meaningfully measured and compared.

Within physics, you are likely most familiar with a metric from a course on general relativity, in which the spacetime metric defines the local geometry of the spacetime you are studying.  Additionally, the spacetime manifold is Lorentzian; there is a {\it very} important relative minus sign between the temporal and spatial components of the metric.  For the dataspace manifolds we study here in collider physics, there are no dynamics, so there is no ``time''.  The manifolds we will study are truly Riemannian, which locally look Euclidean about every point.

The metric property of $d(\vec x,\vec x')$ is still not sufficient for robust physics interpretation of the smeared distribution of Eq.~\ref{eq:smeareddist}.  We additionally need to consider what we can, and cannot, measure in principle in a particle collision experiment.  At a particle collider, of course, we measure particle momenta, their energy and direction.  Additionally, we interpret measurements in terms of energies; collisions occur at a given center-of-mass energy, jets have a particular energy, pairs of photons have a measured invariant mass, etc.  Then, for the metric $d(\vec x,\vec x')$ to enable a similar physics interpretation, its units should be energy.  As such, we can interpret, that is, physically understand, the radius $\epsilon$ to be the energy scale range of influence of points on the manifold ${\cal M}_\text{data}$.  That is, points on ${\cal M}_\text{data}$ whose energy distribution (appropriately defined) differs by less than $\epsilon$ influence one another.

In that final sentence, the parenthetical ``(appropriately defined)'' is doing some rather heavy lifting.  This is also related to the identity of indiscernibles in the properties of a metric.  What does it mean for two points on ${\cal M}_\text{data}$ to be ``indiscernible''?  There should be no experimental way, in principle, for us to distinguish them.  If we are only able to measure particle energies and directions, then two events are indiscernible if their flow of energy on the celestial sphere (where measurements are performed) is identical.  With that identification, the flow of energy of an event is unchanged if exactly 0 energy particles are added (as no energy is manipulated).  Additionally, the flow of energy is unchanged if particles undergo exactly 0 angle splittings (because every detector has some finite angular resolution).  This observation means that for physical interpretation, the metric $d(\vec x,\vec x')$ must additionally be infrared and collinear (IRC) safe.  That is, if events $\vec x$ and $\vec x'$ differ only by exactly 0 energy particles or exactly 0 angle splittings, then $d(\vec x,\vec x') = 0$.  As discussed in Ref.~\cite{Larkoski:2024uoc}, IRC safety means that calculations in fixed-order perturbation theory are possible.

\subsection{IRC safe collider event metrics}

Also familiar from general relativity, given a Riemannian manifold, there is no unique metric that can be placed on it.  Different metrics emphasize different geometric features of the manifold; the Schwarzschild metric describes the region well outside a spherical mass, while the metric in Rindler coordinates describes the region near the horizon of a black hole, for example.  Additionally, the constraints of units of energy and IRC safety of a collider event metric are still rather weak, so there is no unique metric on collider events.  As such, many collider event metrics have been proposed and studied for various applications \cite{Komiske:2019fks,Mullin:2019mmh,Komiske:2020qhg,CrispimRomao:2020ejk,Cai:2020vzx,Larkoski:2020thc,Tsan:2021brw,Cai:2021hnn,Andersen:2021mvw,Kitouni:2022qyr,Alipour-Fard:2023prj,Larkoski:2023qnv,Davis:2023lxq,Ba:2023hix,Craig:2024rlv,Cai:2024xnt,Gambhir:2024ndc,Cai:2025fyw,DAgnolo:2025qqr,Andersen:2026ppe}.

\begin{figure}[t!]
\begin{center}
\includegraphics[width=0.95\textwidth]{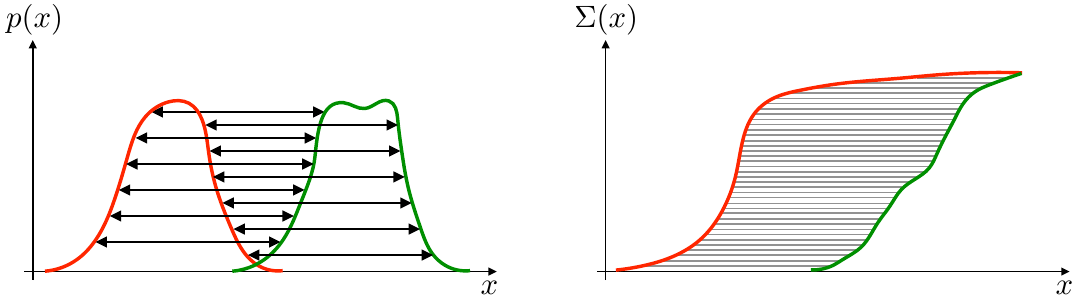}
\caption{\label{fig:wasserstein} Illustration of Wasserstein distance between two one-dimensional probability distributions.  Left: Optimal transport plan to transform one distribution into the other.  The arrows indicate how the ``mass'' of one distribution should be moved to create the other distribution. Right: Illustration of total Wasserstein distance between the distributions.  The total Wasserstein distance is the integral between their cumulative distributions, evaluated by summing over each grey horizontal line raised to the power $p\geq 1$.}
\end{center}
\end{figure}

The first modern event metric is referred to as the energy mover's distance (EMD) \cite{Komiske:2019fks,Komiske:2020qhg}, and its central innovation for defining a relevant metric is to quantify the energy cost of rearranging the energy distribution of one event ${\cal E}_A$ into that of another event, ${\cal E}_B$.  This idea actually dates back over 250 years, to French engineer Gaspard Monge who wanted to quantify the minimal cost, or minimal work, to move one pile of dirt into another pile \cite{monge1781memoire}.\footnote{Monge was also the founder of \'Ecole Polytechnique and is memorialized by a street in the Latin Quarter of Paris.}  This original metric that quantifies the distance between two distributions of dirt is the {\it earth mover's distance}, and much later was formalized mathematically into what is now known as a Wasserstein distance \cite{kantorovich1939mathematical,vaserstein1969markov,dobrushin1970prescribing}.  For two distributions $p_A(x)$ and $p_B(x)$, each a function in one variable $x$, their $p$-Wasserstein distance is defined as the integral of the difference of their inverse cumulative distributions:
\begin{align}\label{eq:wassdist}
d_{p\text{-Wass}}(p_a,p_b) = \left[
\int_0^1 ds\, \left|
\Sigma^{-1}_A(s)-\Sigma^{-1}_B(s)
\right|^p
\right]^{1/p}\,,
\end{align}
where the cumulative distribution is
\begin{align}
\Sigma(x) = \int_{-\infty}^x dx'\, p(x')\,.
\end{align}
The number $p\geq 1$ can be chosen to emphasize different scales on your manifold.  An illustration of this Wasserstein metric between two distributions is illustrated in Fig.~\ref{fig:wasserstein}.

Wasserstein metrics can be defined between distributions in any dimension.  However, unlike in one dimension, no closed-form expression can be written down; explicit minimization is required in general.  This is due to the fact that the real number line is totally-ordered, while points in $\mathbb{R}^n$ for $n>1$ have no intrinsic notion of an ordering.  The general theory of metrics between distributions in arbitrary dimensions is referred to as {\bf optimal transport}, and remains an active area of mathematics research.\footnote{His work on optimal transport (among other things) won C\'edric Villani the 2010 Fields Medal.  See Refs.~\cite{lott2009ricci,villani2009optimal} for more foundational mathematics research.}  The feature that points in $\mathbb{R}^n$ with $n>1$ lacks total-ordering has significant consequences for collider event metrics.  A particle collider detector lives on the two-dimensional celestial sphere, and this is the space on which event energy flows live.  The energy mover's distance quantifies the Wasserstein optimal transport cost to rearrange one flow of energy on the celestial sphere ${\cal E}_A$ into another ${\cal E}_B$ (with a given fixed total energy $E_\text{tot}$).  The energy flow of an event ${\cal E}(\hat n)$ is defined as the energy that flows in the direction of the unit vector $\hat n$, where
\begin{align}
{\cal E}(\hat n) = \sum_{i\in {\cal E}} E_i\, \delta^{(2)}(\hat n - \hat n_i)\,,
\end{align}
where $E_i$ is the energy of particle $i$, $\hat n_i$ is the unit vector in the direction of that particle's momentum, and the sum runs over all particles in the event.  The energy flow is normalized to the total event energy $Q$,
\begin{align}
\int d^2\hat n\, {\cal E}(\hat n) = Q\,.
\end{align}

Then, the EMD as the optimal transport metric between the energy flows of two events $A$ and $B$ can be expressed as
\begin{align}
\text{EMD}_\beta({\cal E}_A,{\cal E}_B) &=\min_{\{f_{ij}\}}\sum_{i\in {\cal E}_A,j\in{\cal E}_B} f_{ij}\,\theta_{ij}^\beta\,,
\end{align}
where the function $f_{ij}\geq 0$ is constrained as
\begin{align}
\sum_{i\in {\cal E}_A} f_{ij}\leq E_j\,,\qquad \sum_{j\in {\cal E}_B} f_{ij}\leq E_i\,,\qquad \sum_{i\in {\cal E}_A,j\in{\cal E}_B} f_{ij}=Q\,.
\end{align}
$\theta_{ij}$ is an angular measure, appropriate for the particular collider environment, and $\beta > 0$ ensures that the EMD is IRC safe.  Because the celestial sphere is two dimensional, one must perform an explicit minimization over the function $f_{ij}$ on a case-by-case basis.  In practice, this isn't necessarily a problem because there efficient algorithms, like the Hungarian algorithm \cite{kuhn1955hungarian,ollivier2009reduction}, and useful numerical implementations, like {\tt Python Optimal Transport} \cite{flamary2021pot,flamary2026pot}.  However, lack of a closed-form for the EMD metric is a rather significant problem for analytic calculations, as the form of the metric, or any observable derived from it, cannot even be expressed as an explicit function of the momenta of particles in the events.

\begin{figure}[t!]
\begin{center}
\includegraphics[width=0.45\textwidth]{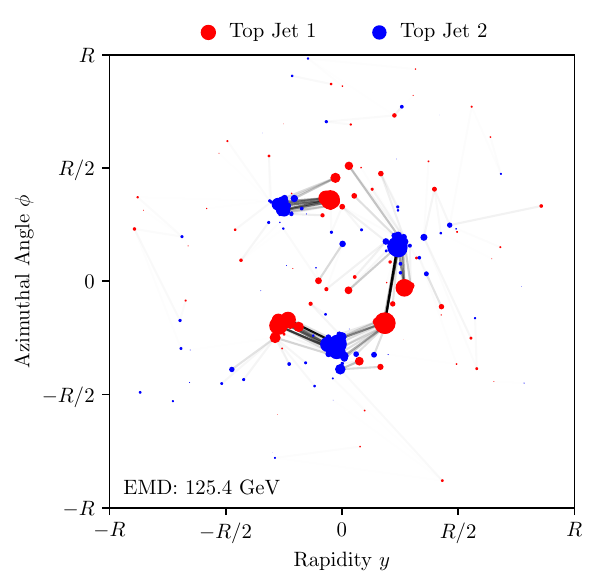}
\caption{\label{fig:emdex} Illustration of the optimal transport plan according to the EMD between two simulated jets (blue and red) from hadronic top decay, in the $(y,\phi)$ plane.  Particle transverse momenta are proportional to the area of the dots, and the transportation of transverse momentum is illustrated by the black lines, with the amount proportional to the width of the lines.  Figure from Ref.~\cite{Komiske:2019fks}.}
\end{center}
\end{figure}

An illustration of the optimal transport plan of the EMD between two simulated jets that were initiated by hadronic top quark decay is shown in Fig.~\ref{fig:emdex}, reproduced from Ref.~\cite{Komiske:2019fks}.  Here, the particles in each jet are illustrated by red or blue dots, respectively, on the rapidity-azimuth $(y,\phi)$ plane.  The jets have been translated so that their axes lie at the point $(0,0)$ and the area of the dots is proportional to their transverse momentum from the colliding beams.  The black lines that connect red-blue dots represent the transportation of transverse momentum between jets, with the width of the lines proportional to the amount of transverse momentum, according to the EMD.  The EMD between the jets is then the total ``work'' to move one jet into the other; just like work from in introductory physics, the EMD between the two jets is the product of the transverse momentum (the ``weight'') and the net angle it was carried (a ``distance'') from one jet to the other.

If we are to have a compact, closed-form for an event metric, requiring no numerical minimization, we will need some representation of our events onto a one-dimensional distribution.  Then, the one-dimensional distributions could be appropriately compared and their Wasserstein distance calculated through Eq.~\ref{eq:wassdist}.  To establish an appropriate one-dimensional distribution, we will seek inspiration almost 50 years in the past, from the energy-energy correlator \cite{Basham:1978bw}.  The energy-energy correlator (EEC) is an IRC safe observable defined as the pairwise energy-weighted, angular distribution of radiation in events:
\begin{align}
\text{EEC}(\chi) = \int d\sigma\, \sum_{i,j\in {\cal E}}\frac{E_iE_j}{Q^2} \, \delta\left(
\chi - \frac{1-\cos\theta_{ij}}{2}
\right)\,.
\end{align}
Here, $d\sigma$ is the differential cross section, $E_i$ is the energy of particle $i$, and $\theta_{ij}$ is the angle between particles $i$ and $j$.\footnote{Note that the energy-angle coordinates used here are appropriate for colliders with spherical symmetry, like a lepton collider.  Coordinates appropriate for other colliders can be substituted in.}  The EEC has recently seen significant interest, see, e.g., Refs.~\cite{Hofman:2008ar,Dixon:2018qgp,Chen:2020vvp,Moult:2025nhu}, but for about 40 years, was not an especially exciting collider observable.\footnote{As of summer 2026, about half of all citations of Ref.~\cite{Basham:1978bw} have been published since 2018.}  Unlike so-called ``event shape'' observables that return a single number for each event, the EEC is only defined on an ensemble of events, as the average of the {\bf spectral function} of individual events
\begin{align}
\mathfrak{s}_{\cal E}(\chi)\equiv \sum_{i,j\in{\cal E}}E_iE_j \, \delta\left(
\chi - \frac{1-\cos\theta_{ij}}{2}
\right) = \sum_{i\in {\cal E}}E_i^2\,\delta(\chi) + \sum_{i<j\in {\cal E}}2E_iE_j\,\delta\left(
\chi - \frac{1-\cos\theta_{ij}}{2}
\right)\,.
\end{align}
Studies of this spectral function in its own right has seen some interest recently, as well \cite{Jankowiak:2011qa,Lim:2018toa,Chakraborty:2019imr,GarciaCaffaro:2025gkm,Altakach:2026gdg}.  While this particular form is directly motivated from the EEC, all that is required for IRC safety is that the appropriate angular distance is monotonic in the pairwise angle, $\theta_{ij}$.  It will be beneficial to consider the more general form of the spectral function
\begin{align}
\mathfrak{s}_{\cal E}(\omega)\equiv \sum_{i,j\in{\cal E}}E_iE_j \, \delta\left(
\omega - \omega_{ij}
\right) = \sum_{i\in {\cal E}}E_i^2\,\delta(\omega) + \sum_{i<j\in {\cal E}}2E_iE_j\,\delta\left(
\omega-\omega_{ij}
\right)\,,
\end{align}
where $\omega_{ij} = \omega_{ij}(\theta_{ij})$ is monotonically increasing in $\theta_{ij}$ and $\omega_{ij}(0) = 0$.  Note that this form is a simple change of variables of $\chi$ to $\omega$ in the EEC.

This spectral function seems to be a rather interesting object, and the way that it encodes relative angles and particle energies in an event is intriguing.  Further, it is a one-dimensional distribution with a fixed normalization, 
\begin{align}
\int_0^{\omega_{\max}} d\omega\, \mathfrak{s}_{\cal E}(\omega) = Q^2\,,
\end{align}
where $\omega_{\max}$ is the maximum value of the angular coordinate $\omega$.  However, we need to ensure that the spectral function of any event is unique, so that there is a bijection between the detector or celestial sphere representation of an event and its spectral function.  This can be established following a proof from Ref.~\cite{boutin2004reconstructing}.  There, the authors consider a point cloud in $\mathbb{R}^n$, an arbitrary distribution of points in Euclidean space.  They prove that the spectral representation of the point cloud, that is, its distribution of pairwise point distances, is unique for almost all configurations of points.  Here, ``almost all'' means that if there are any degeneracies in which multiple pairs of points have the same distance, then there are multiple point clouds that have the same pairwise distance spectrum.  Note that for distances drawn from a continuous distribution, the probability of a degeneracy is formally 0.  Additionally, this uniqueness is only true up to isometries of the embedding space.  In $\mathbb{R}^n$, for example, you can obviously translate, rotate, or reflect the point cloud and the distribution of pairwise distances is unchanged.  However, this is also a desirable property as the particular coordinate frame that one uses to describe physics is arbitrary.

Thus, the spectral function as motivated from the EEC and the proof that the pairwise distance distribution is almost always unique (up to isometries) proves that the spectral function is a one-dimensional distribution that does indeed encode the energy flow of an event in a unique way.  As such, we can use the one-dimensional $p$-Wasserstein metric definition of Eq.~\ref{eq:wassdist} to define a {\it spectral} energy mover's distance (SEMD).  The details are not especially challenging to work out, so I will not provide a complete derivation here, but instead point interested readers to Refs.~\cite{Larkoski:2023qnv,Gambhir:2024ndc}.  The SEMD as defined by the $(p=2)$-Wasserstein metric between spectral functions $\mathfrak{s}_A$ and $\mathfrak{s}_B$ of events $A$ and $B$, respectively, can be expressed as
\begin{align}
\text{SEMD}(\mathfrak{s}_A,\mathfrak{s}_B) &= d_{p=2}(\mathfrak{s}_A,\mathfrak{s}_B)^2 \\
&= \sum_{i<j\in {\cal E}_A} 2E_iE_j\omega_{ij}^2+ \sum_{i<j\in {\cal E}_B} 2E_iE_j\omega_{ij}^2 - 2\sum_{\substack{n\in {\cal E}_A^2,\ell\in {\cal E}_B^2\\\omega_n< \omega_{n+1}\\\omega_\ell<\omega_{\ell+1}}}\omega_n\omega_\ell\, \text{ReLU}\left(
{\cal S}_{n\ell}
\right)\,.\nonumber
\end{align}
I note that the SEMD is technically defined as the {\it square} of the actual metric, as in this form there is no overall square-root.  Here, the notation $n\in {\cal E}_A^2$, for example, means one of the peaks away from the origin in the spectral function, of which there are ${N\choose 2}$, where $N$ is the number of particles. $\omega_n<\omega_{n+1}$ just means that these peaks have been ordered in increasing angle $\omega$, and $\text{ReLU}(x)=x\,\Theta(x)$ is the rectified linear unit function.\footnote{Interestingly, ReLU is a common activation function in a machine learning architecture, but its appearance here seems completely coincidental.}  Its argument is
\begin{align}
{\cal S}_{n\ell} = \min\left[
\mathfrak{S}_A^+(\omega_n),\mathfrak{S}_B^+(\omega_\ell)
\right] -\max\left[
\mathfrak{S}_A^-(\omega_n),\mathfrak{S}_B^-(\omega_\ell)
\right] \,.
\end{align}
The inclusive and exclusive cumulative spectral functions are defined as
\begin{align}
&\mathfrak{S}^+(\omega_n) = \sum_{i\in {\cal E}}E_i^2 + \sum_{\substack{n\geq m\in {\cal E}^2\\\omega_m<\omega_{m+1}}}(2EE)_m\,,
&\mathfrak{S}^-(\omega_n) = \sum_{i\in {\cal E}}E_i^2 + \sum_{\substack{n> m\in {\cal E}^2\\\omega_m<\omega_{m+1}}}(2EE)_m\,.
\end{align}
The shorthand $(2EE)_m$ means the squared energy weight of the $m$th peak in the spectral function away from the origin.

Whew! Finally in this form, we can identify a very convenient choice for the angular function $\omega_{ij}$.  If we choose 
\begin{align}
\omega_{ij} = \sqrt{1-\cos\theta_{ij}}\,,
\end{align}
then the first two terms simplify to the total invariant mass of events $A$ and $B$:
\begin{align}\label{eq:semdtotfinal}
\text{SEMD}({\cal E}_A,{\cal E}_B) = s_A+s_B - 2\sum_{\substack{n\in {\cal E}_A^2,\ell\in {\cal E}_B^2\\\omega_n< \omega_{n+1}\\\omega_\ell<\omega_{\ell+1}}}\omega_n\omega_\ell\, \text{ReLU}\left(
{\cal S}_{n\ell}
\right)\,.
\end{align}
Here, we have expressed the SEMD as a function of events $A$ and $B$'s energy flows ${\cal E}_A$ and ${\cal E}_B$. It is this form that is most commonly used (for obvious reasons) and what I will consider exclusively in these lecture notes.  I stress, however, that again, these choices are not unique nor even particularly well-motivated, they are merely convenient and ``simple'' for us to work with.  As mentioned above, this is exactly the same situation we find ourselves in whenever we define a metric on a spacetime (or any manifold); the manifold is what it is, the coordinates and hence the metric we use to describe it is purely a choice based on the features that we want to study.

Because of its simple, compact form, in these notes we will use the SEMD as the prototypical metric for studying the manifold of collider events.\footnote{As an illustration of the difficulty of even expressing the EMD metric in a closed form useful for perturbative calculations, see the expression for event isotropy \cite{Cesarotti:2020hwb} in the calculation of Ref.~\cite{Atzori:2026blf}.  There, event isotropy is defined on only 3 particle final states in $e^+e^-\to$ hadrons events, but requires a numerical optimization step.  By contrast, an analogous event isotropy observable can be defined from the SEMD, and its expression on events with arbitrary numbers of particles can be written down exactly.  See Eq.~(4.25) of Ref.~\cite{Gambhir:2024ndc}.}  Even with this restriction, any IRC safe metric on collider events will have similar behavior.  IRC safety requires that any metric treats soft and collinear emissions essentially identically and in QCD, soft and collinear emissions dominate the description of events with jets and hadronic final states.  Further, in some recent studies, it has been explicitly observed that the EMD and SEMD exhibit very similar performance, see, e.g., the study of Ref.~\cite{Doherty:2026fli}.

In these notes, we will only study the SEMD on collinear boosted jets.  In the collinear limit for jets $A$ and $B$ that each have two particles, the SEMD of Eq.~\ref{eq:semdtotfinal} takes the very simple form:
\begin{align}\label{eq:semdjetjet}
\text{SEMD}({\cal J}_A,{\cal J}_B) = s_A+s_B-2\sqrt{s_As_B}\sqrt{\frac{\min\left[z_A(1-z_A),z_B(1-z_B)\right]}{\max\left[z_A(1-z_A),z_B(1-z_B)\right]}}\,.
\end{align}
Here, $s_A$ is the invariant mass of jet $A$, and $z_A$ is the energy fraction of one of the particles of that jet, and we use the notation of ${\cal J}_A$, for example, to represent the energy flow of an individual jet.

\section{Anti-QCD Tagging}\label{sec:antiqcd}

QCD dominates the description of physics at the LHC, both due to the large coupling $\alpha_s$, so QCD radiation is more likely and ubiquitous than QED or photon radiation, and because QCD bound states, protons,  are what is being collided.  This has obvious advantages for studying QCD, but makes it challenging to search for physics beyond QCD, beyond the Standard Model.  Further, QCD is approximately a scale-invariant theory, and as such whatever questions you ask on jets will always return an answer.  Regardless of what scale you probe a jet, there will always be some structure you observe, because scale invariance implies structure at all scales.  This of course has its limits because the QCD coupling runs with energy, quarks are confined, and hadrons have non-zero mass, but at sufficiently high energies scale invariance is the dominant feature of QCD.

Thus in the search for new physics, we need to be certain that what we are seeing is not QCD.  However, essentially by definition, we have no idea what the new physics actually would be (or is, if you're more optimistic).  This puts us in a somewhat ambiguous scenario.  Given known signal and background, Neyman-Pearson says that the optimal observable is the likelihood ratio.  In that case, we still might not know the distributions of signal and background perfectly, but if we can classify what they are, we can make headway considering asymptotic limits, general scaling features, etc.  For identification of ``new physics'', we know significantly less.  Technically, most conservatively, and least biased, all we know about potential new physics is that it is {\it not} QCD.\footnote{This statement is itself biased.  While we have a short-distance Lagrangian for QCD that describes the interactions of quarks and gluons through a non-Abelian gauge theory, this in no way means that we actually know all of its consequences.  We still lack first-principles understandings of confinement, chiral symmetry breaking, the quark-gluon plasma, the phase diagram of QCD at finite density and temperature, etc., etc., etc.  Just because we know the Lagrangian, we do not know all of its consequences \cite{Anderson:1972pca}.}  However, unlike the case of binary discrimination, there is in general no universal optimum for multi-label classification.  One reason for this is that once you consider classification beyond two classes, then the likelihood space is greater than one-dimensional.  Just like the challenges we had discussed earlier with the EMD metric, in dimensions beyond 1, points are not total-ordered, so you have to introduce some ordering, some preference, for how to move around that likelihood space.  Given a choice, there can then be an optimal procedure, but this is not unambiguous.  Similarly, classification into ``QCD'' and ``not QCD'' is too ambiguous to have a universally-optimal observable or procedure.

Ambiguity never stopped anyone before, and this ``QCD'' versus ``not QCD'' problem has seen huge interest in recent years.  This general problem is referred to as {\bf anomaly detection} as the machine is trained to identify anomalies (i.e., signatures of ``not QCD'') in a simulated data set.\footnote{The HEP ML Living Review lists 178 references on anomaly detection \cite{Feickert:2021ajf}.  Needless to say, I will not be referencing them here.}  If the relevant phase space is one-dimensional, anomalies are relatively straightforward to understand by eye, looking at a plot.  For example, in the discovery of the Higgs boson in 2012, the ATLAS and CMS experiments plotted the distribution of the invariant mass of pairs of high-energy photons \cite{ATLAS:2012yve,CMS:2012qbp}.  Without a Higgs boson, the production of high energy photons is a scale-invariant process (so-called ``continuum production'') which can be predicted accurately and measured precisely.  The data are binned, and each bin is independent of all other bins because all events are independent of one another and each event only contributes to a single bin.  As such, the distribution of background events in a given bin in this distribution is Poissonian distributed with a known mean.  This then directly enables a simple quantification of observed deviation from expectation.  In the case of the Higgs, a significant excess of events in data was observed above the continuum expectation, ultimately leading to its discovery.

The case of the Higgs is, however, a case that we already knew where to look.  We knew that the invariant mass distribution of photons is interesting, so we plot that distribution.  With general agnosticism, we have neither a prediction for the signal nor even what observables might be useful, so we have to look everywhere.  The phase space for collision events, or even just for the structure of jets, can be enormous, hundreds of dimensions, so ``looking everywhere'' suffers catastropically from the curse of dimensionality.  If we have a more principled approach, we can evade the curse of dimensionality.  Effectively, we want to avoid explicit binning of events with some fixed coordinate system in some high dimensional phase space, but still need to correlate them in a sensible, controlled way.

With our discussion in the previous section of smearing distributions on the dataspace with an event metric, we have exactly that ability.  Ideas along this line have been studied before, e.g., Refs.~\cite{Park:2022zov,Craig:2024rlv,Cai:2025vxl}, but to my knowledge, this is the first theory study of such a proposal.  We can directly calculate the smeared distribution from QCD about some phase space point $\Phi$, $p_\text{QCD}(\Phi|\epsilon)$.  The number of events in this region of radius $\epsilon$ about point $\Phi$ is Poissonian distributed, with mean $\lambda(\Phi|\epsilon)$ equal to
\begin{align}
\lambda(\Phi|\epsilon) = N_\text{tot}\int d\Phi'\, p_\text{QCD}(\Phi')\,\Theta\left(\epsilon - d(\Phi,\Phi')\right)\,,
\end{align}
where $N_\text{tot}$ is the number of events in the dataset ensemble.  In this same phase space region, one can calculate the number of events in data that are within $\epsilon$ of $\Phi$.  This number of events is
\begin{align}
N_\text{data}(\Phi|\epsilon) = N_\text{tot}\int d\Phi'\, p_\text{data}(\Phi')\,\Theta\left(\epsilon - d(\Phi,\Phi')\right)\,.
\end{align}
One can then compare $N_\text{data}(\Phi|\epsilon)$ to $\lambda(\Phi|\epsilon)$ with the standard statistical measures to establish if there is an anomaly and, if so, how interesting it is.

\subsection{Calculations for an Anti-QCD Jet Tagger}

While we can't calculate much for the properties of the actual measured dataset, we can completely quantify the background from QCD.\footnote{Another approach to anomaly detection using an event metric is to identify the ``least representative'' events in an ensemble \cite{Komiske:2019jim}.  The ``least representative'' events are those that are on average farthest from all other events.}  Here, we will just focus on jets from QCD and look to systematically calculate the statistical properties of the smeared distribution on QCD jets, $p_\text{QCD}(\Phi|\epsilon)$.  In practice, however, this will almost assuredly be done with simulated events that can include complete information about aspects that are not included in a calculation (hadronization, detector effects, etc.).  There is, however, no replacement for an honest, first principles calculation for development of intuition, and calculations are deeply enriching to the soul of a theoretical physicist.  So, in this section, we will beat on, and perform our calculation against the current of blind simulation analysis.  Perhaps we might even learn something from it.

Let's construct the smeared distribution as a perturbative series, through expansion of the fixed-order distribution in orders of $\alpha_s$.  We will just work to lowest non-trivial order here.  We have
\begin{align}
p_\text{QCD}(\Phi|\epsilon) &= \int d\Phi'\, p_\text{QCD}(\Phi')\,\Theta\left(
\epsilon - d(\Phi,\Phi')
\right) \\
&= \int d\Phi'\, p^{(0)}_\text{QCD}(\Phi')\,\Theta\left(
\epsilon - d(\Phi,\Phi')
\right)+ \alpha_s\int d\Phi'\, p^{(1)}_\text{QCD}(\Phi')\,\Theta\left(
\epsilon - d(\Phi,\Phi')
\right)+\cdots\nonumber\,,
\end{align}
where the superscript denotes the contribution at that order in $\alpha_s$.  We will exclusively consider collinear jets here and use the SEMD on those jets as defined in Eq.~\ref{eq:semdjetjet}.  At lowest-order in $\alpha_s$, the jet of interest consists of a single particle, and so its mass is 0: $s' = 0$.  Additionally, the lowest-order jet matrix element is simply 1, so the leading term is very simple:
\begin{align}\label{eq:losmear}
\int d\Phi'\, p^{(0)}_\text{QCD}(\Phi')\,\Theta\left(
\epsilon - d(\Phi,\Phi')
\right) = \Theta(\epsilon^2 - s)\,.
\end{align}
That is, a jet in the data is only ``close'' to a leading-order QCD jet if the invariant mass of the jet in data is sufficiently small (according to the SEMD).

The calculation at the next order will be significantly more challenging.  As we are considering collinear jets, we know the phase space and matrix element (see Sec.~3 in the original notes of Ref.~\cite{Larkoski:2024uoc}).  The distribution on jets at this next-to-leading order can be expressed as
\begin{align}
\alpha_s \,d\Phi\,p_\text{QCD}^{(1)}(\Phi) = \frac{\alpha_s}{2\pi}\, \frac{ds\, dz}{s}\, P(z)\,\Theta\left(z(1-z)E^2R^2-s\right) - \frac{\alpha_s}{2\pi}\,\delta(s')\int \frac{ds'\, dz'}{s}\, P(z')\,\Theta\left(z'(1-z')E^2R^2-s'\right)\,.
\end{align}
The first term describes real collinear emission through the splitting function $P(z)$ while the second term is the effective virtual term, chosen in this case so that the total contribution to the cross section at this order is 0.  I have also included the effect of a finite jet radius $R$ with jet energy $E$, as the particles are only interesting to our analysis if they indeed lie in the jet of interest.  Then, with the expression for the SEMD, the contribution to the smeared distribution at next-to-leading order can be expressed as
\begin{align}
\alpha_s\int d\Phi'\, p^{(1)}_\text{QCD}(\Phi')\,\Theta\left(
\epsilon - d(\Phi,\Phi')
\right) &=\frac{\alpha_s}{2\pi}\int \frac{ds'}{s'}\, dz'\, P(z')\,\Theta\left(z'(1-z')E^2R^2-s'\right)\\
&\hspace{-1cm}\times\left[
\Theta\left(
\epsilon^2 - s - s' + 2\sqrt{ss'}\sqrt{\frac{\min[z(1-z),z'(1-z')]}{\max[z(1-z),z'(1-z')]}}
\right)-\Theta(\epsilon^2-s)
\right]\,.\nonumber
\end{align}

To proceed, we will work with a particularly simple form for the splitting function.  The QCD splitting functions enjoy a supersymmetric relationship \cite{dokshitzer1991basics,Seymour:1997kj} where
\begin{align}
P_{q\to qg}(z)+P_{q\to qg}(1-z) = P_{g\to gg}(z)+P_{g\to gg}(1-z) + P_{g\to q\bar q}(z)+ P_{g\to q\bar q}(1-z)\,,
\end{align}
when $C_F = C_A = n_f$.  Of course, on a jet, we do not know with certainty which particle is a quark or a gluon (and anyway we only measure color-singlet hadrons in experiment), so this symmetrization is natural from a measurement perspective.  We will thus consider the symmetrized splitting function
\begin{align}
P_{q\to qg}(z)+P_{q\to qg}(1-z) = C_J\, \frac{1+z^2}{1-z}+C_J\, \frac{1+(1-z)^2}{z} = C_J\,\frac{2-3z(1-z)}{z(1-z)}\,,
\end{align}
where we just use $C_J$ to represent the jet's color Casimir.  To avoid double counting, we only consider $z\in[0,1/2]$.  We now note that everything in the integrand is symmetric in $z'\to 1-z'$, so we can introduce the new phase space coordinate 
\begin{align}
x \equiv z(1-z)\,.
\end{align}
In this coordinate, the contribution to the smeared distribution can be expressed as\footnote{In principle, this integral can be evaluated in a series in $\epsilon$.  One can write, for example, $s'=s+\Delta s$ and $x'=x+\Delta x$, and expand the integrand in powers of $\Delta s$ and $\Delta x$.  However, there still remain several possible hierarchies; the relative sizes compared to $\epsilon$, the absolute size of $s$ itself, etc., which leads to a proliferation of constraints that gets annoying to track down.  So, for the purposes of summer school lectures, I'll save such a detailed calculation for you.}
\begin{align}
\alpha_s\int d\Phi'\, p^{(1)}_\text{QCD}(\Phi')\,\Theta\left(
\epsilon - d(\Phi,\Phi')
\right) &=\frac{\alpha_s}{2\pi}\int \frac{ds'}{s'}\, \frac{dx'}{\sqrt{1-4x'}}\, \frac{2-3x'}{x'}\,\Theta\left(x'E^2R^2-s'\right)\\
&\hspace{1cm}\times\left[
\Theta\left(
\epsilon^2 - s - s' + 2\sqrt{ss'}\sqrt{\frac{\min[x,x']}{\max[x,x']}}
\right)-\Theta(\epsilon^2-s)
\right]\,.\nonumber
\end{align}

Note that the difference of $\Theta$-functions implies that if both are satisfied, then the integral vanishes.  So, only one can be satisfied for a non-zero integral.  Then, we can express the $\Theta$-functions as
\begin{align}
&\Theta\left(
\epsilon^2 - s - s' + 2\sqrt{ss'}\sqrt{\frac{\min[x,x']}{\max[x,x']}}
\right)-\Theta(\epsilon^2-s) \\
&\hspace{1cm}= \Theta\left(
\epsilon^2 - s - s' + 2\sqrt{ss'}\sqrt{\frac{\min[x,x']}{\max[x,x']}}
\right)\Theta(s-\epsilon^2)-\Theta\left(
s + s' - 2\sqrt{ss'}\sqrt{\frac{\min[x,x']}{\max[x,x']}}
-\epsilon^2\right)\Theta(\epsilon^2-s)
\nonumber\,.
\end{align}
Also, note that the first constraint implies that
\begin{align}
s' < 4s\,\frac{\min[x,x']}{\max[x,x']}\,,
\end{align}
and vice-versa for the second constraint.  To further clean up the integrand, we can rescale all invariant masses by $E^2R^2$, $s\to E^2R^2 s$, for example.  This can then be absorbed into the distance $\epsilon$ by the rescaling $\epsilon \to ER\epsilon$.  With this rescaling, the contribution at next-to-leading order to the smeared distribution can be expressed as
\begin{align}\label{eq:nlosmear}
\alpha_s\int d\Phi'\, p^{(1)}_\text{QCD}(\Phi')\,\Theta\left(
\epsilon - d(\Phi,\Phi')
\right) &=\frac{\alpha_s}{2\pi}\int \frac{ds'}{s'}\, \frac{dx'}{\sqrt{1-4x'}}\, \frac{2-3x'}{x'}\,\Theta\left(x'-s'\right)\\
&\hspace{-5cm}\times\left[
\Theta\left(
\epsilon^2 - s - s' + 2\sqrt{ss'}\sqrt{\frac{\min[x,x']}{\max[x,x']}}
\right)\Theta(s-\epsilon^2)-\Theta\left(
s + s' - 2\sqrt{ss'}\sqrt{\frac{\min[x,x']}{\max[x,x']}}
-\epsilon^2\right)\Theta(\epsilon^2-s)
\right]\,.\nonumber
\end{align}

\begin{figure}[t!]
\begin{center}
\includegraphics[width=0.45\textwidth]{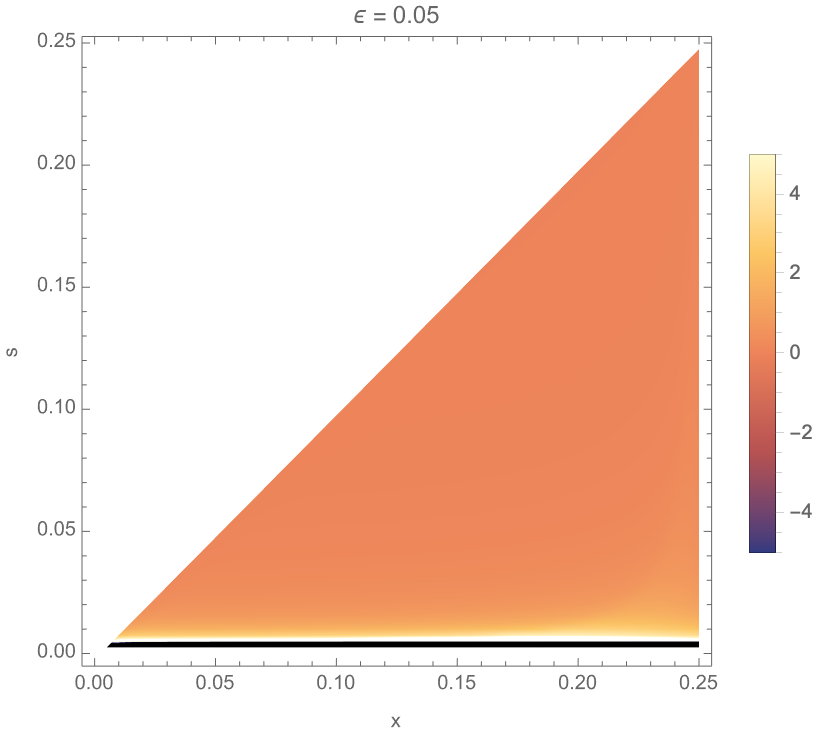}\hspace{0.5cm} \includegraphics[width=0.45\textwidth]{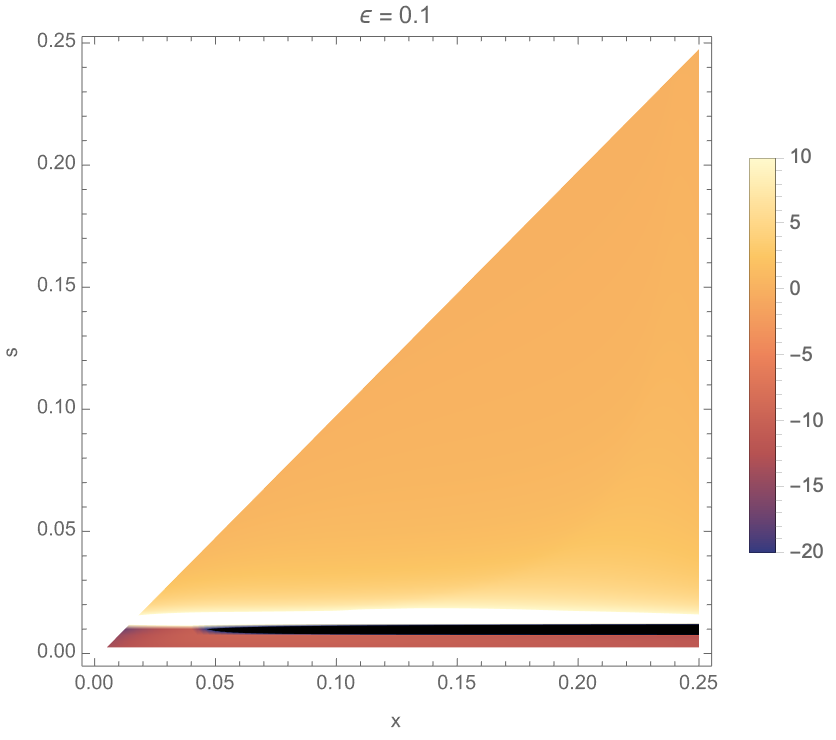}\\
\includegraphics[width=0.45\textwidth]{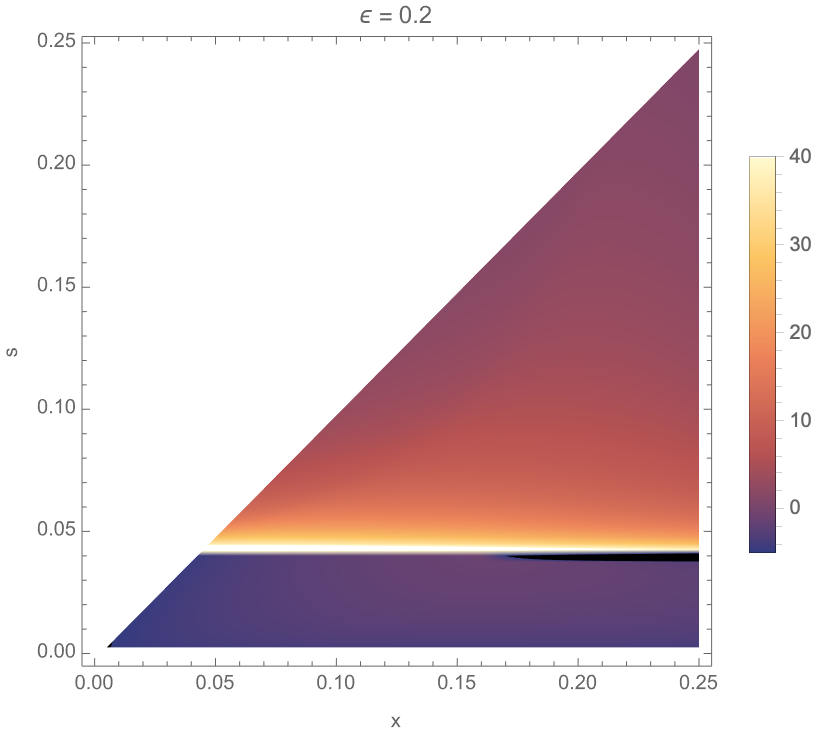}\hspace{0.5cm} \includegraphics[width=0.45\textwidth]{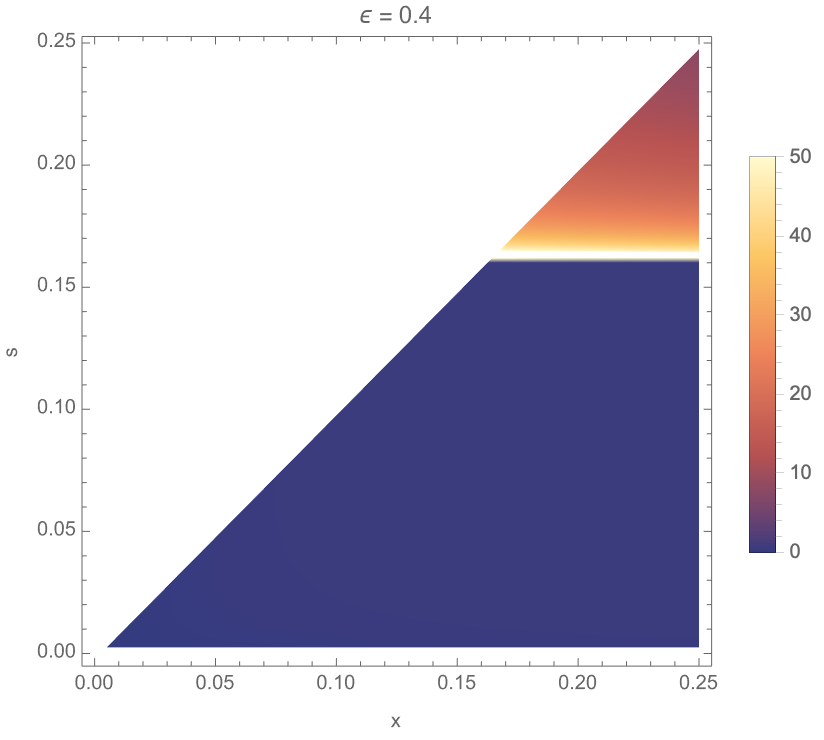}
\caption{\label{fig:smeardists} Next-to-leading order contributions to the SEMD smeared distribution on a jet of Eq.~\ref{eq:nlosmear}, in the phase space defined by $x=z(1-z)$, the product of particle energy fractions, and $s$, the squared invariant mass.  Four different values of the smearing radius are plotted $\epsilon=0.05,0.1,0.2,0.4$.  Black (white) are regions below (above) the range of the legend.}
\end{center}
\end{figure}

This smearing integral can be straightforwardly numerically evaluated on two-particle collinear phase space of $(s,x)$, for a given smearing radius $\epsilon$.  We show some plots of this for smearing radius $\epsilon = 0.05,0.1,0.2,0.4$ in Fig.~\ref{fig:smeardists}.  In these plots, we exclusively show the kinematic dependence of the contribution at next-to-leading order; that is, Eq.~\ref{eq:nlosmear} with no overall factor of $\alpha_s/(2\pi)$.  The complete contribution through next-to-leading order needs this factor, as well as to be added to the leading-order smearing of Eq.~\ref{eq:losmear}.  There are a few things to note in general about these plots.  First, as $\epsilon\to 0$, the smeared distribution vanishes because the effective bin size vanishes as well.  Next, as $\epsilon\to \infty$, the smeared distribution reduces to unity, because every jet is within an infinite distance of any other jet.  On two-body collinear phase space, this can be significantly strengthened.  The maximum distance between two jets on this phase space is $\epsilon_{\max}^2 = s_\text{max} = 1/4$, as this rescaled squared invariant mass is bounded from above by $x$.  So, in this current study, if $\epsilon > 1/2$, then the smeared distribution reduces to unity on all of phase space.

One intriguing thing to note about these plots is that the contribution to smearing at next-to-leading order can be negative in the degenerate phase space region, where $s\to 0$.  This is well-known, because the matrix element for real emissions is positive while the matrix element for virtual or loop diagrams is negative on at least some of phase space.  Such negative contributions aren't immediately problematic for interpretation of the distribution because this is just the next-to-leading order contribution, which must be added to the leading-order contribution.  However, these plots make it clear that there is a region of phase space on which the negative contribution at next-to-leading order becomes very large, so even when multiplying by $\alpha_s/(2\pi)$ and adding to the leading-order term, the distribution may be negative there.  Such negative weights at next-to-leading order are an artifact of degenerate perturbation theory and are necessarily corrected when all-orders are summed (anyway, we never observe negative weight events).  

In practice, however, negative weight events in fixed-order perturbation theory starting at next-to-leading order can be a rather serious problem.  In particular, in the simulation of collider events, one wants the most accurate theoretical prediction possible to compare with data.  This involves matching fixed-order predictions with the parton shower, but doing so means that events in your predicted, simulated ensemble are weighted.  Weighted events in principle converge to the same distributions on phase space, but can dramatically reduce the statistical power of your ensemble, for the same number of events.  A general rule-of-thumb is that if your ensemble contains a fraction $f$ events that have negative weight, the ensemble must be increased by a factor
\begin{align}
\frac{\langle w^2\rangle}{\langle w\rangle^2} = \frac{1}{(1-2f)^2}\,,
\end{align}
to have the same statistics of an ensemble with exclusively unit-weight events.  Here, $\langle w\rangle$ is the mean event weight and $\langle w^2\rangle$ is the second moment.  As an example, if 25\% of your events have negative weight, then you must increase the number of events by a factor of 4 to have the same statistical properties.  This is such a serious degradation that this negative weight problem can be the largest computational bottleneck in some experimental analyses \cite{HSFPhysicsEventGeneratorWG:2020gxw,Campbell:2022qmc,ATLAS:2021yza,CMS:2022psv}.  Procedures for reducing negative weights are now a rather active area of research \cite{Nason:2007vt,Borisyak:2019vbz,Nachman:2020fff,Frederix:2020trv,Andersen:2021mvw,Danziger:2021xvr,Frederix:2023hom,Andersen:2023cku,Andersen:2024mqh,Glazier:2024ogg,Drnevich:2024vfj,CMS:2024jdl,Janssen:2025zke,Shyamsundar:2025nzn,Shyamsundar:2025mfw,vanBeekveld:2025lpz,Palmer:2025jmb,Nachman:2025lid,Gambhir:2025lka,Andersen:2026ppe,Doherty:2026fli,Heimel:2026cxh}, but we won't discuss it more here.

\subsection{Local vs.~Global Significance}\label{sec:locglobsig}

About a given point $\Phi$ on phase space, we can calculate the expected number of events within an SEMD distance $\epsilon$, $\lambda(\Phi|\epsilon)$, and using properties of Poissonian statistics, we can estimate how significant the measured number of events, $N_\text{data}(\Phi|\epsilon)$, in that same region is.  Such a {\bf local significance} requires a point $\Phi$ as input.  However, with no other information, that point $\Phi$ could have been anywhere on phase space.  The probability of an excess anywhere on phase space is much larger because there are now many such regions that could exhibit an excess.  As such, the significance of an excess anywhere on phase space, the {\bf global significance} is smaller.  This effect, that we could have ``looked elsewhere'' for an excess is called the {\bf look-elsewhere effect} and is a general feature in any search for new physics.  We don't have a prior for where that new physics may live, so we have to look everywhere.

Let's imagine that you observe an excess in one bin of your data, and there are a total of $N_\text{bins}$ bins.  Let the local $p$-value of the excess be $p_\text{local}$; that is, the probability that an individual bin has an excess at least as large as the bin of interest is $p_\text{local}$.  Now, that excess could have been anywhere.  The probability that there is at least one excess in all bins as significant as this observed excess is 1 minus the probability that there is no excess this significant.  This global $p$-value is
\begin{align}
p_\text{global} = 1-(1-p_\text{local})^{N_\text{bins}}\,.
\end{align}
As long as the bins are small enough so that $p_\text{local}$ is sufficiently small, this global $p$-value simplifies to
\begin{align}
p_\text{global} \approx N_\text{bins}p_\text{local}\,.
\end{align}
With enough bins, this effect can be rather large.

For our SEMD smearing anomaly detection, we have thus far focused on local significance, determining the properties of individual regions or ``bins'' on phase space.  However, we should properly account for the look-elsewhere effect to have an understanding of the true significance, and if such a purported excess is actually evidence for new physics.  We therefore need to establish the number of bins on phase space given a smearing radius $\epsilon$.  We do have to be a bit careful.  In the analysis above for calculating the look-elsewhere effect, we had assumed that all bins were independent.  That is, an event only ever lives in a single bin.  With our smearing prescription, this is no longer necessarily true, as two events could contribute to the smeared distribution value at different points on phase space, if they are within an SEMD distance of $2\epsilon$.  Thus, distinct points on a smeared distribution are not independent, so some care is needed for proper accounting of statistics.\footnote{In this way, note that our smearing prescription is essentially a moving average that smooths out discontinuous or non-smooth distributions on phase space.}

\begin{figure}[t!]
\begin{center}
\includegraphics[width=0.45\textwidth]{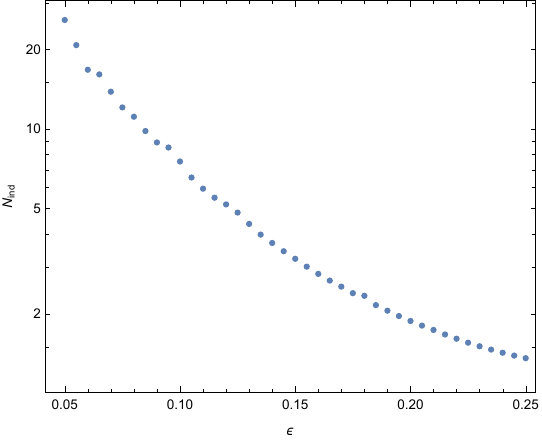} \hspace{0.5cm} \includegraphics[width=0.45\textwidth]{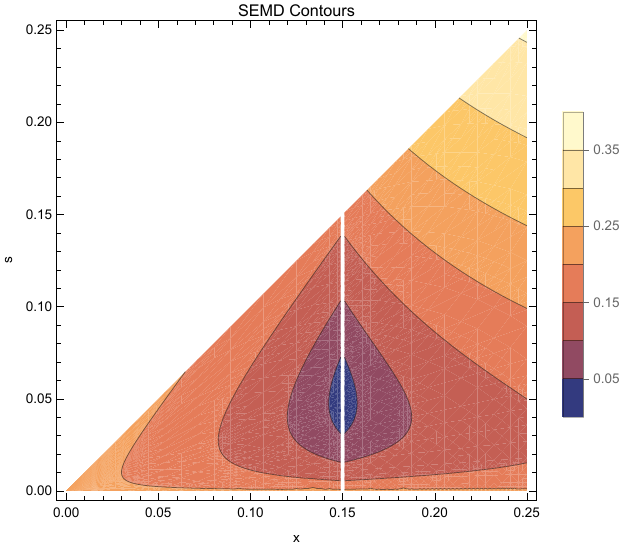}
\caption{\label{fig:nbins} Left: The number of independent regions $N_\text{ind}$ on two-particle jet phase space as a function of SEMD smearing radius $\epsilon$. Right: Plot of contours of fixed SEMD about the point $(x,s) = (0.15,0.05)$ on two-particle jet phase space.}
\end{center}
\end{figure}

Really, what we need to evaluate the look-elsewhere effect and the global significance is the number of independent regions $N_\text{ind}$ of radius $\epsilon$ on phase space.  The centers of these regions are all at least a distance of $2\epsilon$ from one another.  Then, we can take $N_\text{ind}$ as the number of effective bins to establish the size of the look-elsewhere effect.  Calculating this number of independent regions is actually straightforward.  Note that the average volume within a distance $\epsilon$ of any point on phase space is
\begin{align}
\langle \text{Vol}(\epsilon)\rangle = \frac{1}{\int d\Phi}\, \int d\Phi\, d\Phi'\, \Theta\left(
\epsilon - d(\Phi,\Phi')
\right)\,.
\end{align}
The total number of ``average'' regions on phase space is precisely number of independent regions:
\begin{align}
N_\text{ind}(\epsilon) = \frac{\int d\Phi}{\langle \text{Vol}(\epsilon)\rangle} = \frac{\left(\int d\Phi\right)^2}{\int d\Phi\, d\Phi'\, \Theta\left(
\epsilon - d(\Phi,\Phi')
\right)}\,.
\end{align}
We note that this is dependent on the smearing radius $\epsilon$.  The volume of our rescaled two-body jet phase space is simple to calculate:
\begin{align}
\int d\Phi = \int ds\, \frac{dx}{\sqrt{1-4x}}\, \Theta(x-s) = \frac{1}{12}\,.
\end{align}
We then need to calculate the integral over all pairs of phase space points within a distance $\epsilon$ of one another:
\begin{align}
&\int d\Phi\, d\Phi'\, \Theta\left(\epsilon - d(\Phi,\Phi')\right)\\
&\hspace{1cm}= \int ds\, \frac{dx}{\sqrt{1-4x}}\,ds'\, \frac{dx'}{\sqrt{1-4x'}}\, \Theta(x-s)\, \Theta(x'-s')\,\Theta\left(
\epsilon^2 - s - s' + 2\sqrt{ss'}\sqrt{\frac{\min[x,x']}{\max[x,x']}}
\right)\,.\nonumber
\end{align}
While this integral is somewhat easier than the smeared distribution of the previous section, it still has a significant number of constraints.  So again, I leave it to you to exactly evaluate it, and I will just numerically evaluate the integral.  

This number of independent bins on two-particle jet phase space as a function of smearing radius $\epsilon$ is plotted at left in Fig.~\ref{fig:nbins}.  As expected, the number of bins gets large as the smearing radius decreases.  Phase space is two dimensional with volume $1/12$, from the calculation above.  If we simply binned this space with squares of side length $\epsilon$, this na\"ive procedure would have
\begin{align}
N_\text{naive}(\epsilon) = \frac{1}{12\epsilon^2}
\end{align}
bins.  This simple scaling describes the number of independent regions of radius $\epsilon$ on that space with the SEMD rather well.  For example, for $\epsilon = 0.05$, this na\"ive scaling predicts
\begin{align}
N_\text{naive}(\epsilon=0.05) \approx 33
\end{align}
bins.  If $\epsilon = 0.15$, the na\"ive number of bins is instead
\begin{align}
N_\text{naive}(\epsilon=0.15) \approx 4\,.
\end{align}
So, the SEMD binning follows the expectation of bins in two dimensions.  We will see later, however, with a more detailed definition of such a scaling, that the IRC safety of the SEMD implies some rather interesting local geometric structure on phase space.  Some of a hint of this is plotted at right in Fig.~\ref{fig:nbins}, which are the SEMD contours about the point $(x,s)=(0.15,0.05)$ on this phase space.  That is, the contours are constant values of true distance
\begin{align}
\sqrt{\text{SEMD}({\cal J},{\cal J}')} = \sqrt{s+s'-2\sqrt{ss'}\sqrt{\frac{\min[x,x']}{\max[x,x']}}}\,.
\end{align}
IRC safety requires that the contours become parallel as they approach the $s=0$ axis, and this squeezes and warps distances in an unintuitive way.

\section{Scaling Laws}\label{sec:scaling}

Many natural systems exhibit a self-similarity over various length scales; indeed, our introduction to QCD made that self-similarity or scale invariance a fundamental axiom (at least to some first approximation).  Scale invariance, or at least its approximation over a range of length scales, is rather simple to encode in a program or a living system's genetics.  Tree branches, for example, exhibit a self-similar structure because its genes may encode branching as devoting half of available resources to each of the branches produced at the next bifurcation, for example.  As this proceeds, branches of smaller and smaller size are generated as you move from the trunk to the leaves.  

From another perspective, the particular scaling properties of a system encode its intrinsic dimension.  The dimension of a system can be defined extrinsically, for example, through the number of coordinates that are needed to describe its embedding in $\mathbb{R}^n$.  Such a ``degree of freedom'' definition of dimension is strictly integral in value and requires an embedding.  Intrinsic definitions of dimension of some space do not require any explicit coordinate system, they just require a way to measure distances on that space, which dramatically generalizes what a dimension can be.  One definition of an intrinsic dimension is the following.  Let $\epsilon$ be a length with which to measure your system.  With this length, construct characteristic hypercubes on your space.  The number $N_\epsilon$ of these hypercube ``boxes'' is related to the dimension $D$ of your system as:
\begin{align}
N_\epsilon = \epsilon^{-D}\,.
\end{align}
This can be rearranged to solve for dimension $D$, which then defines it through measurements on your system:
\begin{align}
D = -\frac{\log N_\epsilon}{\log\epsilon}\,.
\end{align}
For self-similar systems, the dimension is defined through a limit as your ruler $\epsilon\to 0$:
\begin{align}
D = \lim_{\epsilon\to 0}-\frac{\log N_\epsilon}{\log\epsilon}\,.
\end{align}
In general, this value of the dimension does not need to be integral-valued and is referred to as the box-counting dimension of the space.  This generalized notion of dimension was introduced by Felix Hausdorff \cite{hausdorff1918dimension}.  Standard systems, lines, planes, etc., have (Hausdorff) dimension equal to 1, 2, etc., but systems for which their Hausdorff dimension does not equal their natural embedding dimension are called {\bf fractals}.\footnote{Benoit B.~Mandelbrot is responsible for popularization of fractals through his 1967 paper ``How long is the coastline of Britain?'' \cite{mandelbrot1967long}.  Mandelbrot cites Lewis Richardson's work on thinking about this problem \cite{richardson1961problem}.  Mandelbrot adopted his initials into its own self-similar recursion: the middle ``B.'' stands for ``Benoit B.~Mandelbrot''.  A couple of fun applications of fractal dimension to physics are a calculation of the fractal dimension of the path of a quantum mechanical particle \cite{Abbott:1979bh} and the fractal dimension of the parton shower \cite{Gustafson:1991ru,Bjorken:1991ft}.}

Systems that exhibit a scale invariance therefore define an intrinsic or effective dimensionality of the system.  Conversely, if you observe a scaling law, this implies some underlying order or structure of the system, even if you do not have a description of that structure.  Even if you only observe a scaling law empirically, like in a natural system, you can still use that scaling law and apply to other cases for a prediction.  If the scaling law applies in many seemingly unrelated systems, this suggests that there is a universal, underlying mechanism describing all of them.

Okay, so what does this have to do with machine learning?  Machine learning is driven by two competing quantities: ``compute'', which is just a general term for the cost of resources, and ``loss'' or a measure of accuracy as defined by the value of the loss or objective function obtained after training your machine.  In general, you want to minimize compute while optimizing the loss.  Compute is most practically measured in dollars: the cost of the time to train billions of GPUs at a data farm that uses electricity and cold water from a large river, like the Columbia River near my home in Portland, Oregon, USA.  Correspondingly, if you had a way to predict the loss given finite compute, you would know exactly how much resources (\$\$\$) you would need to invest to make a chatbot that passes the Turing test.  If you observe a scaling law between compute and loss this is exactly the type of relationship that enables such a prediction.

\begin{figure}[t!]
\begin{center}
\includegraphics[width=0.45\textwidth]{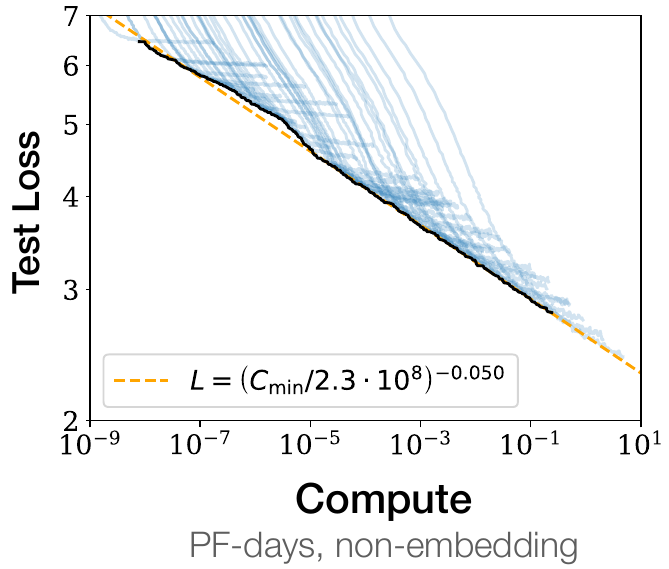}
\caption{\label{fig:powerlaw} Illustration of a power law or scaling relationship between compute and loss in a machine learning study.  Figure from Ref.~\cite{kaplan2020scaling}.}
\end{center}
\end{figure}

In 2020, a group consisting of OpenAI employees and a physics professor, Jared Kaplan, demonstrated that loss versus compute followed a power law functional form over several decades; see Fig.~\ref{fig:powerlaw} from their paper \cite{kaplan2020scaling}.  Scaling laws had been observed or suggested in machine learning by other groups, too, \cite{ahmad1988scaling,cohn1990can,hestness2017deep,rosenfeld2019constructive,henighan2020scaling,rosenfeld2021predictability}, but Ref.~\cite{kaplan2020scaling} changed the playing field and led to the founding of the AI corporation Anthropic.\footnote{Jared Kaplan was a theoretical particle physics professor at Johns Hopkins and worked in the field of amplitudes at the time Anthropic was founded.  Jared and I overlapped at SLAC when I was a student and he was a post-doc. Kaplan's net worth is estimated at \$15 billion as of August 2026.  My net worth is not.}  The tasks that machines were trained on to observe these scaling laws are things like classifying images (cat? dog? Brad Pitt?) or summarizing text (TL;DR, training data was taken from Reddit posts).  The fact that scaling laws are observed suggests an underlying structure to the data, but natural language or the space of images has no foundational theory scaffolding its structure that can be used to truly understand the origin of scaling laws.  Scaling laws in machine learning are extremely powerful for the whole enterprise and in part led to trillion-dollar valuations of multiple AI companies, but they are still just observations, with no first-principles understanding of why they are there or why power law exponents take the values they do.

This is where machine learning on data in physics, and in particular, particle collider physics, is significantly distinct from random data on the internet.  With particle physics data, we can ask the same question of the relationship between compute (number of events in our dataset, for example) and loss (efficacy of classifying jets as ``quark'' or ``gluon'' flavor), but we have an underlying theory of what the jet labels truly mean, and a way to predict the distribution of data on phase space.  Several groups have identified empirical scaling laws in machine learning for particle physics problems, e.g., Refs.~\cite{Batson:2023ohn,ATLAS:2026zdb,Bahl:2026jvt,Vigl:2026ppx,Amram:2026zzv,Uslu:2026ywh,Uslu:2026nln}, but so far little work has been done toward understanding the origin of the scaling laws and predicting scaling exponents up to now.  We will present some first thoughts on this problem, following Ref.~\cite{Larkoski:2025clo}, but this subfield deserves much more theoretical QCD work.

\subsection{Defining ``dimension''}

Getting to calculating scaling laws on machine learning for physics cases requires some set-up so we know what exactly it is we want to calculate in QCD.  As mentioned above, a scaling law is a manifestation of some structure of some (possibly fractal) dimension on your data.  We had introduced the box-counting dimension earlier, but this is a bit too simplistic to be applied on physics data as defined through calculations in QCD.  So we'll need a different, though closely related, definition of dimension.  To motivate this, let's consider some manifold ${\cal M}$ endowed with a metric $d(\cdot,\cdot)$.  About a point $\Pi\in {\cal M}$, the local geometry of manifold is encoded through the volume of the manifold within a ball of radius $\epsilon$:
\begin{align}
\int d\Pi'\, \Theta\left(\epsilon - d(\Pi,\Pi')\right) \propto \epsilon^{\text{dim}(\Pi)}\left(
1-\frac{{\cal S}(\Pi)}{6\left(\text{dim}(\Pi)+2\right)}\,\epsilon^2 +{\cal O}(\epsilon^3)
\right)\,.
\end{align}
That is, the volume of the manifold within a distance $\epsilon$ of some point $\Pi$ scales with the dimension of the manifold at that point, $\text{dim}(\Pi)$.  Corrections to this scaling are suppressed by positive powers of $\epsilon$ and the first correction is proportional to the scalar curvature at the point $\Pi$, ${\cal S}(\Pi)$.

Using the SEMD metric and our prescription for smearing distributions on phase space, this method for extracting a dimension is straightforward.  In our dataset, for example, we can consider a phase space point $\Pi$ as defined by some event that we have measured.  Then, we can count all events that lie within a distance $\epsilon$ of $\Pi$, and the rate at which this scales away as $\epsilon\to 0$ is the dimension about the event of interest.  Note however that as we scale $\epsilon\to 0$, more and more events are required in our dataset to sufficiently fill the event manifold to ensure that a relatively smooth scaling law can be observed.  On any finite dataset, there will be some $\epsilon_{\min}$ below which there are no events within $\epsilon_{\min}$ of $\Pi$.  This is a first hint of a connection between compute (dataset size) and loss (density with which the phase space manifold is sampled).

So far, however, this definition of dimension is defined point-by-point on the manifold and is rather challenging to calculate analytically, especially when considering events with many particles in them.  Further, it fails to provide an overall picture of the effective dimensionality of the entire manifold, unless you calculated the dimensions about many random points on the manifold.  This is naturally easy to correct: we simply average over all points on the manifold to define an average dimension.  We thus can define
\begin{align}
\langle \epsilon^{\text{dim}(\Pi)}\rangle \propto \int d\Pi\, d\Pi'\, \Theta\left(\epsilon - d(\Pi,\Pi')\right)\,.
\end{align}
The average dimension over the manifold at a scale $\epsilon$ can be extracted through an appropriate derivative:
\begin{align}
\text{dim}_\epsilon({\cal M}) \equiv \epsilon \frac{d}{d\epsilon} \log\left[
\int d\Pi\, d\Pi'\, \Theta\left(
\epsilon -d(\Pi,\Pi')
\right)
\right]\,.
\end{align}
This is effectively the correlation dimension \cite{Grassberger:1983zz,theiler1990estimating} of the manifold.  If one is interested in a more detailed description of the dimensionality and how it varies over the manifold, one can consider higher moments of the scaling.  The dimension as we have defined here as a correlation is like the first moment of the dimension on the manifold, while a quantity like
\begin{align}
\langle \epsilon^{2\,\text{dim}(\Pi)}\rangle=\int d\Pi\, d\Pi'\,d\Pi''\,\Theta\left(\epsilon - d(\Pi,\Pi')\right)\Theta\left(\epsilon - d(\Pi,\Pi'')\right)\,,
\end{align}
encodes a second moment of the point-by-point manifold dimension.  This can be useful for quantifying statistical properties of points sampled uniformly on the manifold \cite{Larkoski:2025idq,Larkoski:2025pai}.  In these notes, however, we will stick to the average manifold dimension.

To provide some intuition for this correlation dimension, let's do a simple calculation on a sample of quark jets at high energy.  We can re-express this correlation dimension in our smeared distribution framework with the SEMD metric.  The average volume scaling becomes
\begin{align}
\langle \epsilon^{\text{dim}(\Pi)}\rangle_q = \int d\Phi\, d\Phi'\, p_q(\Phi)\,p_q(\Phi')\,\Theta\left(\epsilon - d(\Phi,\Phi')\right)\,,
\end{align}
where $\Phi$ are points on collinear phase space and $p_q(\Phi)$ is the distribution of a quark jet on that phase space.  We can expand this to order-$\alpha_s$ in perturbation theory and find
\begin{align}
\langle \epsilon^{\text{dim}(\Pi)}\rangle_q = 1+2\cdot\frac{\alpha_s}{2\pi}\int \frac{ds}{s}\,dz\, P_{q\to qg}(z)\,\Theta(z(1-z)E^2R^2-s)\,\Theta\left(\epsilon^2 - s\right)+{\cal O}(\alpha_s^2)\,.
\end{align}
In this expression, we note that the SEMD at lowest order, between two jets each with a single particle, is 0.  So, the leading term is simply 1.  At the next order, one jet has a single particle (order-$\alpha_s^0$) and the other jet has two particles (order-$\alpha_s^1$).  We have also enforced that those two particles have total energy $E$ and lie within a radius $R$ of one another.  The SEMD between such jets is just the mass of the two-particle jet, and either jet could have had two particles, which is the factor of 2 in front of the phase space integral.

We can of course do this integral and take its logarithm, but before that, we can already perform some useful expansions.  Note that
\begin{align}
\log \langle \epsilon^{\text{dim}(\Pi)}\rangle_q = \frac{\alpha_s}{\pi}\int \frac{ds}{s}\,dz\, P_{q\to qg}(z)\,\Theta(z(1-z)E^2R^2-s)\,\Theta\left(\epsilon^2 - s\right)+{\cal O}(\alpha_s^2)\,,
\end{align}
so we can easily take derivatives.  We have
\begin{align}
\epsilon \,\frac{d}{d\epsilon} \log \langle \epsilon^{\text{dim}(\Pi)}\rangle_q &= 2\epsilon^2\frac{\alpha_s}{\pi}\int \frac{ds}{s}\,dz\, P_{q\to qg}(z)\,\Theta(z(1-z)E^2R^2-s)\,\delta\left(\epsilon^2 - s\right)+{\cal O}(\alpha_s^2)\\
&=\frac{2\alpha_s}{\pi}\int dz\, P_{q\to qg}(z)\,\Theta(z(1-z)E^2R^2-\epsilon^2)+{\cal O}(\alpha_s^2)\nonumber\,.
\end{align}
Using the expression for the splitting function,
\begin{align}
P_{q\to qg} = C_F\,\frac{1+(1-z)^2}{z}\,,
\end{align}
we can easily perform this integral.  We find that the dimensionality of the manifold ${\cal M}_q$ of quark jets is
\begin{align}
\text{dim}({\cal M}_q)=\epsilon \,\frac{d}{d\epsilon} \log \langle \epsilon^{\text{dim}(\Pi)}\rangle_q=\frac{2\alpha_s C_F}{\pi}\left(
-2\log\frac{\epsilon^2}{E^2R^2}-\frac{3}{2}
\right)+{\cal O}(\alpha_s^2,\epsilon^2)\,,
\end{align}
where we have ignored corrections suppressed by $\epsilon^2$.  Intriguingly, as one ``looks'' closer and closer at the manifold of quark jets, as $\epsilon\to 0$, the dimension grows without bound.  This feature is a manifestation of the approximate scale invariance of QCD.

\subsection{Scaling Laws for Discrimination}

Let's now connect these various threads, and consider what scaling laws may exist for a binary discrimination problem.  We will start with our very old friend quark versus gluon jet discrimination, but will later study some properties of IRC safe metrics on boosted $H\to b\bar b$ jets, as well.   The set-up of this problem is the following.  Let $p_q(\Phi)$ and $p_g(\Phi)$ be the distributions of quark and gluon jets, respectively, on phase space $\Phi$.  One way to quantify discrimination power is through a measurement of how many quark jets live in the neighborhood of a gluon jet, for example.  For now, we will think about this in the opposite way: what is the minimum radius $\epsilon_{\min}$ of a neighborhood about a gluon jet event such that there is one quark jet event?  If this $\epsilon_{\min}$ is very small for a fixed number of events in our ensemble, discrimination power will be weak, as there is significant overlap of the quark and gluon distributions.

So that is our ``loss'' side of the equation.  For the ``compute'' side, we will consider an ensemble of $n$ quark jet events.  We would like to establish the relationship between $\epsilon_{\min}$ (loss) and $n$ (compute).  A single quark jet from this ensemble in the neighborhood of a gluon jet phase space point $\Phi$ is the requirement that
\begin{align}\label{eq:psdiscloss}
\int d\Phi'\, p_q(\Phi')\,\Theta\left(\epsilon_{\min} - d(\Phi,\Phi')\right) = \frac{1}{n}\,.
\end{align}
That is, the total probability that a quark jet is within $\epsilon_{\min}$ of point $\Phi$ is $1/n$, which is (on average) a single event from our quark jet ensemble.  We have explicit control over the number of events in our ensemble, but the value of $\epsilon_{\min}$ is set by physics.  So, we really want to invert this equation, and determine $\epsilon_{\min}$ as a function of $n$.  As discussed above, evaluating this integral at arbitrary phase space points $\Phi$ is a mess, so to simplify, we will average over the gluon jet distribution on phase space.

That is, we will consider the symmetric, averaged relationship between $\epsilon_{\min}$ and $n$, working from
\begin{align}\label{eq:losscomputescale}
\int d\Phi\,d\Phi'\, p_g(\Phi)\, p_q(\Phi')\,\Theta\left(\epsilon_{\min} - d(\Phi,\Phi')\right) = \frac{1}{n}\,.
\end{align}
All we have done is integrate the distribution of gluon jets $p_g(\Phi)$ over phase space on both sides of Eq.~\ref{eq:psdiscloss} and used the fact that it is normalized to unity at right.  This integral over phase space is something that we can calculate, given the distributions of quark and gluon jets.  We are most interested in the regime in which the number of events in our ensemble is large, $n\gg 1$.  In this limit, a fixed-order expansion on the left is inappropriate.  As we demonstrated above, large logarithms in $\epsilon_\text{min}\to 0$ arise that spoil convergence of the perturbative series.  Instead, we need to resum these large logarithms to all orders and use a distribution for the quark and gluon jets on phase space in which these large logarithms are tamed.

To know what these resummed distributions are, let's remind ourselves what the metric is at lowest order.  The SEMD between a jet with a single particle and a jet with two particles is just the two-particle jet's squared invariant mass $s$.  This actually holds for the SEMD between a one-particle jet and an arbitrary jet: the SEMD is always the squared mass of the more complicated jet.  By the triangle inequality, the SEMD distance between two arbitrary jets is bounded by the sum of their invariant masses.  Specifically, for a quark jet ${\cal J}_q$, a gluon jet ${\cal J}_g$, and a single-particle jet ${\cal J}_1$, their SEMD distance satisfies
\begin{align}
d({\cal J}_q,{\cal J}_g) \leq d({\cal J}_q,{\cal J}_1)+d({\cal J}_g,{\cal J}_1) = \sqrt{s_q} + \sqrt{s_g}\,.
\end{align}
The integral we need to evaluate contains the constraint that $\epsilon_{\min}> d({\cal J}_q,{\cal J}_g)$.  This can be satisfied parametrically by demanding that the invariant masses of both the quark jet and the gluon jet are less than $\epsilon_{\min}$: $\epsilon_{\min}>\sqrt{s_q}, \sqrt{s_g}$.  The triangle inequality demonstrates then that the constraint $\epsilon_{\min}> d({\cal J}_q,{\cal J}_g)$ is at worst violated by a factor of 2.  However, to leading logarithmic accuracy in $\epsilon_{\min}$, this factor of 2 is subleading as its presence does not generate a large logarithm that requires resummation.  

Restricting ourselves to leading logarithmic accuracy, and even simpler to double logarithmic accuracy, we can then very easily evaluate the integral of Eq.~\ref{eq:losscomputescale}.  In Sec.~4.2 of Ref.~\cite{Larkoski:2024uoc}, we had calculated the cumulative distribution or probability that the invariant mass of a jet is bounded from above.  There, we had found the expression that
\begin{align}
\Sigma(\epsilon_{\min}) = e^{-\frac{2\alpha_s C_J}{\pi}\log^2\frac{\epsilon_{\min}}{ER}}\,,
\end{align}
which is the probability that $\epsilon_{\min}^2 > s$, the squared invariant mass of the jet, $ER$ is the product of jet energy $E$ and radius $R$, and $C_J$ is the appropriate quadratic Casimir of the jet.  To demand that both quark and gluon jets have an invariant mass bounded from above means we need to multiply their individual probabilities together.  We then finally have that \cite{Komiske:2022vxg}
\begin{align}
\int d\Phi\,d\Phi'\, p_g(\Phi)\, p_q(\Phi')\,\Theta\left(\epsilon_{\min} - d(\Phi,\Phi')\right) =e^{-\frac{2\alpha_s (C_A+C_F)}{\pi}\log^2\frac{\epsilon_{\min}}{ER}}= \frac{1}{n}\,,
\end{align}
to double logarithmic accuracy.

It is now straightforward to solve for $\epsilon_{\min}$ (loss) in terms of $n$ (compute).  We have
\begin{align}
\frac{\epsilon_{\min}}{ER} = \exp\left(-\sqrt{\frac{\pi\log n}{2\alpha_s(C_A+C_F)}}\right)\,.
\end{align}
This doesn't appear to be a power law at first glance, but let's proceed assuming exactly a power law form, with
\begin{align}
\epsilon_{\min}\propto n^\gamma\,,
\end{align}
for some power $\gamma$.  We can extract $\gamma$ through our usual trick where
\begin{align}
\gamma = \frac{d\log\epsilon_\text{min}}{d\log n} = -\frac{1}{2}\sqrt{\frac{\pi}{2\alpha_s (C_A+C_F)\log n}}\,.
\end{align}
This remains not an exact power law as this exponent varies with ensemble size $n$.  However, this variation is spectacularly slow.  If there is some initial ensemble size $n_0$ on which the exponent is calculated, then the exponent has decreased by half when the ensemble size is $n_1 = n_0^4$.  So even if you started from a very, very small ensemble size of say 1000 jets, you would only observe significant violations of the power law if you had a sample of a trillion jets.  Combined in their entire history, the ATLAS and CMS experiments have only collected about $10^{10}$ or $10^{11}$ jets with transverse momenta greater than 100 GeV.

\begin{figure}[t!]
\begin{center}
\includegraphics[width=0.45\textwidth]{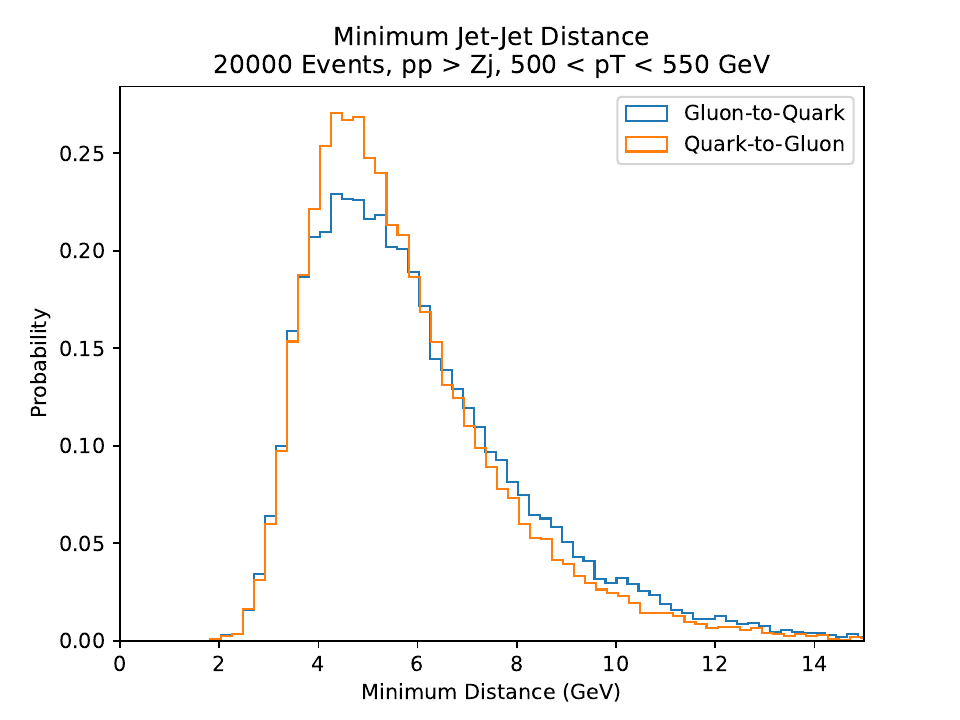}\hspace{0.5cm}\includegraphics[width=0.45\textwidth]{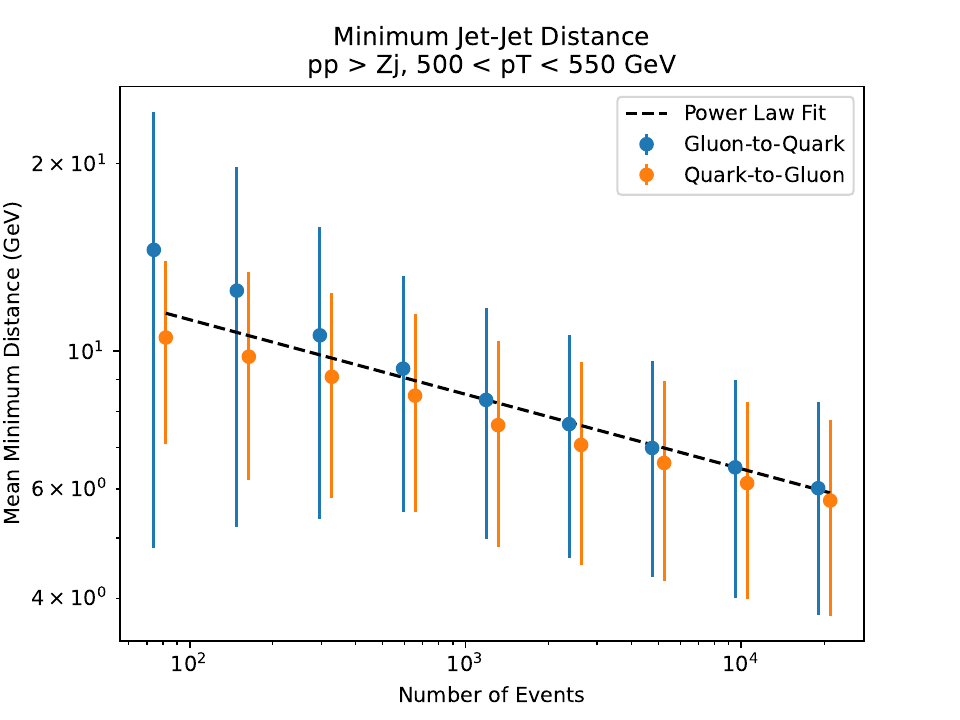}
\caption{\label{fig:scaling} Left: distributions of minimal distances between gluon jets and any quark jet (blue) and quark jets and any gluon jet (yellow) on a simulated ensemble of 20000 jets each.  Right: Plot of the dependence of the mean minimal distances between quark and gluon jets as a function of ensemble size.  The dashed line represents a power law of $\langle \epsilon_{\min}\rangle \propto n^{-0.12}$. Figure from Ref.~\cite{Larkoski:2025clo}.}
\end{center}
\end{figure}

By the way, what is this exponent?  Plugging in the numbers for known quantities, and using $\alpha_s \sim 0.1,0.2$, we have
\begin{align}\label{eq:predictexp}
\gamma = -\frac{1}{2}\sqrt{\frac{\pi}{2\alpha_s (C_A+C_F)\log n}}\approx -\frac{1}{\sqrt{\log n}}\,.
\end{align}
This double logarithmic approximation ignores a lot of physics: running coupling, hard collinear effects, fixed-order corrections, etc., so we aren't necessarily expecting a precision prediction.  However, as a lowest-order approximation, it isn't absolutely terrible.  In Fig.~\ref{fig:scaling}, we show the distribution of the minimum distance between simulated gluon jets and any quark jet (and vice versa) on ensembles with $n=20000$ events, and the scaling of the average minimum distances between cross-sample jets as a function of dataset size.  The mean minimum distance follows a reasonable power law of $\langle \epsilon_{\min}\rangle \propto n^{-0.12}$, which is shallower by about a factor of 2 or so from the prediction of Eq.~\ref{eq:predictexp}.

\subsection{Scaling with IRC safe Metrics}

All of this analysis is predicated on using an IRC safe metric on phase space so that calculations are possible at all in perturbation theory.  As we got a glimpse of in Sec.~\ref{sec:locglobsig}, IRC safe metrics are quite strange objects, as they necessarily must warp phase space in a way that the soft and collinear regions are properly controlled.  As discussed previously, metrics are chosen to emphasize different aspects of the local geometry of your manifold of interest and IRC safe metrics are specifically designed to de-emphasize the degenerate limits.  The question then arises how much of the behavior we identify with the phase space manifold is due to the restriction of an IRC safe metric and how much is due to intrinsic physics?  At some level, this distinction is inseparable; in perturbation theory, IRC safety is a requirement for asking sensible physics questions at all.  However, we can possibly work to tease this out if we have a system with no degeneracies, so IRC safety isn't strictly required (at least not through some order in perturbation theory).

We indeed have such a system, boosted $H\to b\bar b$ decays introduced in Sec.~5 of Ref.~\cite{Larkoski:2024uoc}.  At leading order in perturbation theory, its distribution on phase space is finite and integrable because the mass of the Higgs regulates divergences.  Further, because the mass of the jets is fixed, leading-order phase space is one-dimensional, just the energy fraction $z$ of the bottom quark.  Recall that the normalized leading-order distribution for Higgs decay is simply
\begin{align}
&p_H^{(0)} = 1\,.
\end{align}
With this distribution, we can then calculate the scaling dimension or correlation dimension of the space of Higgs decays to bottom quarks with the SEMD.  This is
\begin{align}
\langle \epsilon^{\text{dim}(\Pi)}\rangle_H&=\int d\Phi\, d\Phi'\, p_H^{(0)}(\Phi)\, p_H^{(0)}(\Phi')\, \Theta\left(
\epsilon - d(\Phi,\Phi')
\right)\\
&=\int dz\, dz'\, \Theta\left(
\epsilon^2 - 2m_H^2+2m_H^2\sqrt{\frac{\min[z(1-z),z'(1-z')]}{\max[z(1-z),z'(1-z')]}}
\right)\,.\nonumber
\end{align}
In the expression of the SEMD, we have replaced the dynamical jet invariant mass $s$ with the fixed Higgs mass, $m_H^2$.

We can evaluate this in the small $\epsilon\to 0$ limit.  In this limit, we necessarily have that $z\sim z'$, so let's introduce the new variable $\Delta z$ where
\begin{align}
z' = z+\Delta z\,.
\end{align}
We can then expand the integrand to linear order in $\Delta z$.  We have
\begin{align}
\frac{\min[z(1-z),(z+\Delta z)(1-z-\Delta z)]}{\max[z(1-z),(z+\Delta z)(1-z-\Delta z)]} &= \frac{\min[z(1-z),z(1-z)+\Delta z(1-2z)]}{\max[z(1-z),z(1-z)+\Delta z(1-2z)]}+{\cal O}(\Delta z^2)\\
&=1-\frac{\left|(1-2z)\Delta z\right|}{z(1-z)}+{\cal O}(\Delta z^2)\,.\nonumber
\end{align}
The argument of the $\Theta$-function reduces to
\begin{align}
\epsilon^2 - 2m_H^2+2m_H^2\sqrt{\frac{\min[z(1-z),z'(1-z')]}{\max[z(1-z),z'(1-z')]}} &=\epsilon^2-m_H^2\frac{\left|(1-2z)\Delta z\right|}{z(1-z)}+{\cal O}(\Delta z^2)\,.
\end{align}
Working with this expansion, we can then evaluate the integral to leading-order in powers of $\epsilon$.  We find
\begin{align}
\langle\epsilon^{\text{dim}(\Pi)}\rangle_H&=\int dz\, dz'\,\Theta\left(\epsilon^2 - 2m_H^2+2m_H^2\sqrt{\frac{\min[z(1-z),z'(1-z')]}{\max[z(1-z),z'(1-z')]}}\right)\\
&=\frac{1}{4}\frac{\epsilon^2}{m_H^2}\left(
1-2\log\frac{\epsilon^2}{2m_H^2}
\right)+{\cal O}(\epsilon^4)\,.\nonumber
\end{align}
From this, it follows that the dimension of the Higgs decay manifold using the SEMD metric is
\begin{align}\label{eq:semddimscale}
\text{dim}({\cal M}_H) = \epsilon\,\frac{d}{d\epsilon}\log\langle\epsilon^{\text{dim}(\Pi)}\rangle_H = \frac{4\log\frac{\epsilon^2}{2m_H^2}+2}{2\log\frac{\epsilon^2}{2m_H^2}-1}\,.
\end{align}
In the limit that $\epsilon\to 0$, this converges to $2$, which is rather interesting in itself because Higgs jets to this order are described by a single quantity on phase space, the energy fraction $z$.  However, the approach to this limit is logarithmically slow, so at small but finite resolution $\epsilon$, the dimension of Higgs decay phase space is some fraction less than 2.

To really see the strangeness of IRC safe metrics, let's contrast this result with that of the absolute simplest metric on this space, just the Euclidean metric.  Let's perform the same calculation of the scaling dimension but with the Euclidean metric.  We'll actually perform this calculation at every point on phase space $z$.  We have
\begin{align}
\int d\Phi'\, p_H^{(0)}(\Phi')\, \Theta\left(\epsilon - d_E(\Phi,\Phi')\right) &= \int_0^1 dz'\,\Theta\left(\epsilon - |z-z'|\right)\\
&=\epsilon\left[
\Theta(z-\epsilon)+\Theta(1-\epsilon-z)
\right]+z\,\Theta(\epsilon-z)+(1-z)\Theta\left(
z-(1-\epsilon)
\right)\,.
\nonumber
\end{align}
We now immediately see that this Euclidean metric results in the expected scaling; as $\epsilon\to 0$, this scales like $\epsilon^1$, demonstrating that the phase space $z\in[0,1]$ is indeed one dimensional.  This is spectacularly different than the scaling with the SEMD in Eq.~\ref{eq:semddimscale}.

On a phase space with a fixed number of particles, one can define a metric through identification of the manifold of phase space.  Broadly, the topology of phase space is that of a hypersphere, because the energy conservation constraint can be expressed through the sum of squared three-momenta \cite{Cox:2018wce,Henning:2019mcv,Henning:2019enq,Larkoski:2020thc,Batson:2021agz}.  The natural metric on a hypersphere is the angle between two points on it, and all expected scaling and dimensionality relationships follow from expectations of geometry in Euclidean space.  The phase space of a jet or event in QCD as defined through an IRC safe metric differs in at least two ways.  First, the number of particles itself fluctuates on an event-by-event level, so there is no sense in which there is a fixed dimension phase space for these events.  This QCD event dataspace ${\cal M}_\text{data}$ is at least the nesting of phase spaces of increasing number of particles or dimension:
\begin{align}
{\cal M}_\text{data} = d\Phi_1\subset d\Phi_2\subset d\Phi_3\subset \cdots\,,
\end{align}
where the subscripts denote the number of particles.  Such a manifold in which phase spaces of increasing dimension are embedded in one another is called a stratified space.

However, this isn't all.  IRC safety enforces particular properties of this stratification.  The boundary of $n$-particle phase space is $(n-1)$-particle phase space and an IRC safe metric connects these phase spaces.  For IRC safety, a metric must return 0 distance between the boundary of $n$-particle phase space and $(n-1)$-particle phase space.  At any given order in perturbation theory, contributions from phase spaces with different numbers of particles exist and matrix element singularities live on the boundaries of phase space of fixed particle number.  In perturbation theory, of course, particle number or multiplicity isn't a well-defined quantity anyway, so the dataspace manifold must have some fluid notion of ``particle'' in general.  This perspective of the general dataspace manifold in a gauge theory like QCD has been studied in very, very little detail; one initial study is Ref.~\cite{Komiske:2020qhg}.  However, to the best of my knowledge, no general construction of the event dataspace manifold in QCD has been presented.  Nevertheless, it is clear that IRC safe metrics must be central to such a construction.  We will present an initial study of the geometry and topology of phase space in Sec.~\ref{sec:geotopops}.

\section{Fundamental Limit of Top Tagging?}\label{sec:toptag}

The problem of identifying the boosted, hadronic decays of top quarks is as old as modern jet substructure, and has proved to be a foundational and inspiring problem for work that followed.\footnote{An early community study that reviewed the state of the field and compared several top tagger algorithms is Ref.~\cite{Abdesselam:2010pt}.}  Top quarks are quarks, and so they have spin-1/2 and carry QCD color charge in the fundamental representation, just like any quark.  Top quarks are also very massive; with a mass of about $m_t \sim 175$ GeV, the top quark is the most massive particle in the Standard Model.  Further, the top quark decays almost instantaneously, on a time scale that is significantly shorter than the time it takes a free quark to hadronize, to fragment into color-singlet hadrons.  Nearly 100\% of the time, the top decays to a $W$ boson and a bottom quark,\footnote{No decay of the top quark to a final state inconsistent with $t\to Wb$ has ever been observed.  Limits on the branching fractions of non-$Wb$ decay modes are bounded from above by about $10^{-3}$ or even less.  See the PDG \cite{ParticleDataGroup:2026aaa} for more details.} and then the $W$ boson further decays, and hadronization occurs.  The $W$ boson decays to quarks about twice as often as to leptons, or about 2/3 of the time.  This follows from a simple counting of final states: the $W$ boson can decay to three possible leptonic final states ($e,\mu,\tau$ paired with their appropriate neutrino), but can decay to six possible quark final states (three colors each of $(u,d)$ or $(c,s)$).  So, about 2/3 of all top quark decays produce three final state quarks, $t\to bq\bar q$, where $qq'$ is either $ud$ or $cs$.

Further, at a hadron collider like the LHC, top quarks are dominantly pair produced as $pp\to t\bar t$, because the top quark is much, much too massive to have any non-zero content in the proton.  So, a top quark has to be produced from the collision of light partons, and protons have a lot of gluons in them.  At high energies, when a pair of top quarks has an invariant mass much larger than $2m_t$ and so each top quark is highly Lorentz boosted in the lab frame, the decay products of these top quarks will become collimated and can be associated into a single jet.  The probability that at least one top quark decays hadronically (that is, to three quarks that hadronize and are observed as hadrons in a detector) is then
\begin{align}
p(t\bar t\to\text{hadrons}) \approx 1-\left(\frac{1}{3}\right)^2\approx 89\%.
\end{align}
So, if we want to understand top quarks at a hadron collider we need to be able to identify them from their hadronic decays.

What's the problem?  While such boosted, hadronic top quark jets have very distinct features, the most prominent of which is a very large invariant mass and a three-prong structure of the individual jets produced from the quarks in decay, the background for such final states is very large.  Jets initiated by light QCD partons are produced in enormous quantities at the LHC.  Further, the mass of such jets can be significantly large.  In the collinear limit, the expectation value for the squared mass of a two-particle jet of radius $R$ is
\begin{align}
\langle m^2_J\rangle = \int d\Phi_\text{coll}^{(N=2)}\, |{\cal M}_\text{coll}|^2\, s\,\Theta_\text{jet} = \frac{\alpha_s}{2\pi}\int ds\, dz\, P_{(ij)\to ij}(z)\, \Theta\left(R^2 - \frac{s}{z(1-z)E^2}\right)\,.
\end{align}
In this expression, $d\Phi_\text{coll}^{(N=2)}$ is two-particle collinear phase space, $|{\cal M}_\text{coll}|^2$ is the collinear matrix element, $s$ is the squared invariant mass, and $\Theta_\text{jet}$ is the jet restriction.  At right, $P_{(ij)\to ij}(z)$ is the appropriate DGLAP splitting function and $E$ is the jet energy. All of these quantities are defined in Secs.~(3.1) and (3.2) of Ref.~\cite{Larkoski:2024uoc}.  Because the splitting functions are purely functions of the energy fraction $z$, we can do the integral over the invariant mass.  We find
\begin{align}
\langle m^2_J\rangle = \frac{\alpha_sR^2E^2}{2\pi}\int dz\, z(1-z)\,P_{(ij)\to ij}(z)\,.
\end{align}
Even without the explicit form of the splitting function, this gives us all the information we need.\footnote{With the expressions for the splitting functions, the integral over $z$ is not challenging.  I point the interested reader to Ref.~\cite{Salam:2010nqg} which completes the calculation and studies the dependence of jet mass on the jet algorithm.  The complete calculations there justify the rule-of-thumb from Ref.~\cite{Ellis:2007ib} that the jet mass is $m_J \sim 0.2 ER$.}  The factor of $z(1-z)$ in the integrand regulates any possible divergences as $z\to 0$ or 1, and so the integral that remains is just some order-1 value.  Just focusing on parametric scaling, let's simply set that integral to 1; the root-mean square jet mass scales like
\begin{align}
\sqrt{\langle m_J^2\rangle} \sim \sqrt{\frac{\alpha_s}{2\pi}}\,RE \approx 0.13 RE\,,
\end{align}
where the approximation at right follows from the value of $\alpha_s = 0.11$, about its value at energies of hundreds of GeV.  This particular value depends on quark and gluon content in the jet ensemble; because of the larger color factor, gluons have a mass about twice that of quarks.  Anyway, for large radius jets, $R\sim 1$ and energies of $E\sim $ TeV, the mean jet mass is close to that of the top mass.  That is, at high enough energies, jets initiated by light QCD partons have masses around that of the top quark, and so simply restricting to jets with masses around 175 GeV is not sufficient to isolate hadronically-decaying top quarks.

\subsection{Defining a ``Top Quark Jet''}

Mass alone is therefore insufficient to identify a boosted top quark jet from the overwhelming QCD background.  What is very different from a jet initiated by a light QCD parton is how that mass is generated.  In a light QCD jet, mass is generated from approximately scale-invariant soft and collinear emissions, while a top quark's mass is encoded in an on-shell decay to three particles.  Because top quarks have an intrinsic mass, the decay occurs in its proper frame, and in the lab frame the three decay products are boosted into three, relatively high energy, subjets of the top quark jet.  In a light QCD jet, because of approximate scale invariance, soft or collinear emissions occur effectively for ``free'', or with probability 1.  By contrast, hard emissions at relatively wide angles are each suppressed by a power of $\alpha_s\sim 0.1$.  So, if we have a way to identify three hard prongs in a jet, a top quark jet always passes, while the fraction of light QCD jets that satisfy this requirement is suppressed by two powers of $\alpha_s$ (two hard emissions are needed off the originator).

Now, not only does a top quark jet have three hard prongs, but also two of those prongs are produced from on-shell $W$ boson decay, and so their invariant mass is constrained to be around that of the $W$ mass, $m_W \sim 80$ GeV.  This identification is a bit more subtle, because of the three prongs, there are ${3\choose 2}=3$ possible pairs of prongs that may have a mass around that of the $W$ boson.  In practice, what is typically done is to find the pair with the mass closest to that of the $W$ boson, and demand that its invariant mass is within some window $dm_W$ to  $m_W = 80$ GeV to qualify as a top quark.  That is, we would require that there is one prong pair invariant mass $m_\text{pair}$ with $m_\text{pair}\in[m_W - dm_W,m_W+dm_W]$.

With more pairs of prongs in a jet, there is a larger probability that any one pair has a mass around that of the $W$ boson.  We can estimate the rate that at least one pair of prongs in a random jet has a mass around the $W$ boson, given that it has three hard prongs and a total mass of $m_t$.  For simplicity, we will assume that the squared invariant masses of pairs of prongs are uniformly distributed and are fixed by the sum rule.  The probability that at least one pair has a mass around the $W$ mass is
\begin{align}
p(W~\text{mass prongs}) &= 3\cdot \frac{2}{m_t^4}\int ds_{12}\, ds_{13}\, ds_{23}\,\delta(m_t^2-s_{12}-s_{13}-s_{23})\\
&\hspace{3cm}\times\Theta\left(
(m_W^2-s_{13})^2-(m_W^2-s_{12})^2
\right)\Theta\left(
(m_W^2-s_{23})^2-(m_W^2-s_{12})^2
\right)\nonumber\\
&\hspace{3cm}\times\Theta\left(
(m_W+dm_W)^2-s_{12}
\right)\Theta\left(
s_{12}-(m_W-dm_W)^2
\right)\nonumber\,.
\end{align}
On the first line, the 3 accounts for any of the three pairs being nearest the $W$ mass and $2/m_t^4$ is the normalization factor.  On the second line, these $\Theta$-functions enforce that $s_{12}$ is closest to the $W$ mass, and the third line enforces that $s_{12}$ is sufficiently close to the $W$ mass.  In general, the mass window width is relatively small compared to the $W$ mass, $dm_W \ll m_W$, so we can expand the window function on the final line to lowest order in $dm_W$.  We have
\begin{align}
&\Theta\left(
(m_W+dm_W)^2-s_{12}
\right)\Theta\left(
s_{12}-(m_W-dm_W)^2
\right) \\
&\hspace{4cm}= \left[
(m_W+dm_W)^2-(m_W-dm_W)^2
\right]\, \delta(s_{12} - m_W^2)+ {\cal O}(dm_W^2)\nonumber\\
&\hspace{4cm}=4 m_W\, dm_W\, \delta(s_{12} - m_W^2)+ {\cal O}(dm_W^2)\nonumber\,.
\end{align}
With this simplification, the probability that at least one pair of prongs has a mass around the $W$ boson mass is 
\begin{align}
p(W~\text{mass prongs}) &= 3\cdot \frac{2}{m_t^4}\int ds_{12}\, ds_{13}\, ds_{23}\,\delta(m_t^2-s_{12}-s_{13}-s_{23})\\
&\hspace{3cm}\times\Theta\left(
(m_W^2-s_{13})^2-(m_W^2-s_{12})^2
\right)\Theta\left(
(m_W^2-s_{23})^2-(m_W^2-s_{12})^2
\right)\nonumber\\
&\hspace{3cm}\times 4 m_W\, dm_W\, \delta(s_{12} - m_W^2)+{\cal O}(dm_W^2)\nonumber\\
&=24\,\frac{dm_W}{m_W}\, \frac{m_W^2(m_t^2-m_W^2)}{m_t^4} + {\cal O}(dm_W^2)\,.\nonumber
\end{align}
Plugging in values for the top and $W$ masses, this probability is approximately
\begin{align}
p(W~\text{mass prongs}) \approx 4\frac{dm_W}{m_W}\,.
\end{align}
The window about the $W$ mass is typically around $dm_W \sim 10$ GeV, so the probability that just a completely random pair of prongs has a mass near the $W$ can be about 1/2 itself.

These considerations suggest that simply identification of three hard prongs in a jet is the dominant discrimination mechanism.  Historically, a large number of observables were constructed to identify three-prong structure in jets, e.g., Refs.~\cite{Seymour:1993mx,Gerbush:2007fe,Brooijmans:2008zza,Kaplan:2008ie,Thaler:2008ju,Almeida:2008yp,Plehn:2009rk,Almeida:2008tp,CMS:2009lxa,ATLAS:2010rhr,Plehn:2010st,Thaler:2010tr,Almeida:2010pa,Thaler:2011gf,Jankowiak:2011qa,Soper:2012pb,Larkoski:2013eya,Larkoski:2014zma,Dasgupta:2018emf}, and one can refine this further.  For example, because the $W$ boson is a color singlet, the quarks from its decay will dominantly radiate in the region between them, and this can be exploited to tease out a bit more discrimination power \cite{Gallicchio:2010sw,Hook:2011cq}.  However, at some point, the phase space for what a top quark jet can be versus a light QCD jet gets too large and unwieldy to encapsulate into a compact observable that a human can easily understand.  It would only be natural, then, to generate huge ensembles of top quark jets and light QCD jets and from them train a machine to classify jets \cite{Kasieczka:2019dbj}.

\subsection{Top Quark Jets from Na\"ive Dimensional Analysis}

Not surprisingly, this is indeed the direction that the vast majority of the field has turned.\footnote{The HEP ML Living Review lists 58 references under its ``top quark tagging'' heading, with 55 of them since 2020 \cite{Feickert:2021ajf}.}  Standard simulated datasets have been produced and are available publicly \cite{Qu:2022mxj,qu_2022_6619768} for anyone to train their favorite machine and then to compete to be the most performant tagging architecture.  Obviously, the performance cannot improve without bound because the optimal discriminant is known (the likelihood by Neyman-Pearson) and eventually a machine will be powerful enough to approximate the likelihood to arbitrary accuracy.  So why don't people just go out and attempt to identify the likelihood directly, as extracted from the standard simulated datasets?

\begin{figure}[t!]
\begin{center}
\includegraphics[width=0.45\textwidth]{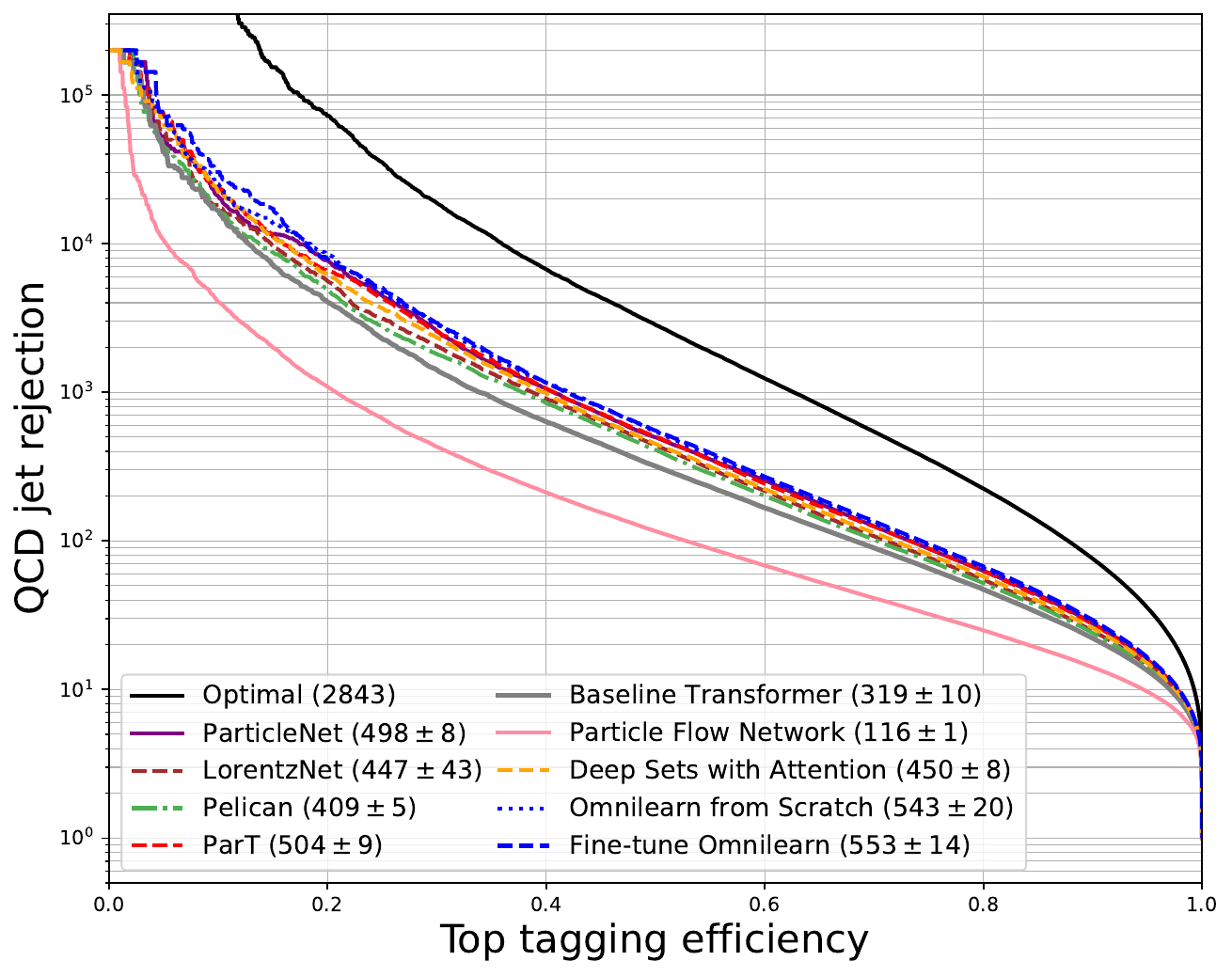}\hspace{1cm} 
\includegraphics[width=0.38\textwidth]{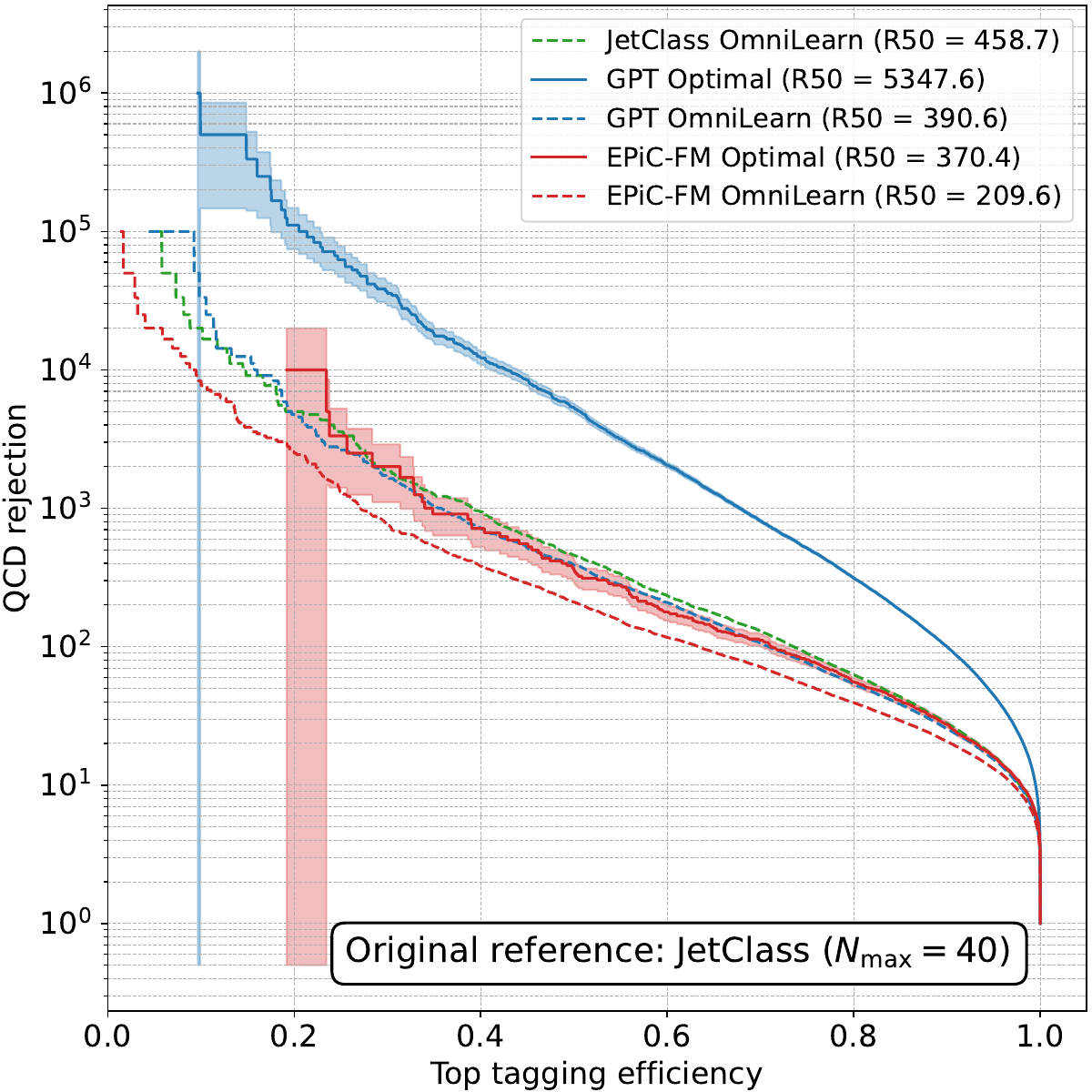}
\caption{\label{fig:fundtop} Left: ROC curve from Ref.~\cite{Geuskens:2024tfo} that demonstrates large separation between claimed optimal QCD jet rejection rate versus top tagging efficiency (solid black) versus several machine learning architectures for this problem (all other curves).  Right: The corresponding ROC curve from Ref.~\cite{Pang:2025lbs} in which their claimed optimal rejection rate (solid red) is close to the performance of other machines (dashed), and far below the over-optimal curve from Ref.~\cite{Geuskens:2024tfo} (solid blue).}
\end{center}
\end{figure}

That is indeed what has been done, but of course, not without controversy.  Ref.~\cite{Geuskens:2024tfo} attempted to construct the likelihood exactly from simulated datasets using a transformer architecture (very similar to that behind ChatGPT, Claude, Gemini, etc.) and discretizing particle momenta.  In this framework, then, the system can be cast into a natural language problem, with all particle probabilities in the dataset well-defined and finite, and the architecture can regress to the true likelihood with arbitrary accuracy.  (See Ref.~\cite{Geuskens:2024tfo} for many more details as to what they did!)  With the likelihood generated, this group then compared that ``optimal'' discrimination power to that approximated by a slew of published models.  Intriguingly, all the models they tested had very similar discrimination power, but the ``optimal'' discriminant performed roughly 10 times better.  See the left plot of Fig.~\ref{fig:fundtop} for their central plot of the receiver operating characteristic (ROC) curve showing this.  However, about a year later another group challenged these claims \cite{Pang:2025lbs}, and demonstrated that the architecture used in Ref.~\cite{Geuskens:2024tfo} actually strongly overestimated the performance of the likelihood.  Ref.~\cite{Pang:2025lbs} claimed that if you correctly implemented the ``optimal'' likelihood, that performance was already very close to that of many models that already exist.  See the right plot of Fig.~\ref{fig:fundtop} for this group's corresponding ROC curve.

I will not attempt to divine who is right or who is wrong in this debate (and that will anyway likely require more discussion in the community). However, using the na\"ive powers of dimensional analysis, we can get an estimate for the scale of the QCD rejection rate as a function of top jet tagging rate, which can hopefully provide a sanity check or benchmark for maximum possible performance from a machine.  The calculations in this section are based on Ref.~\cite{Larkoski:2024hfe}.

The very first thing we must establish is to define precisely what signal and background jets are.  Signal jets are straightforward: they are hadronically-decaying top quarks at large Lorentz boosts.  We have already discussed their properties in some detail.  For background jets, this actually requires knowing what events are in the simulated data samples used to train the models in Refs.~\cite{,Geuskens:2024tfo,Pang:2025lbs}.  These jets are very simple, just high energy jets initiated by light QCD partons, with no constraint on their internal kinematics whatsoever.  No mass cuts on the jet have been imposed, no restrictions on pronginess, etc.; just QCD jets. While these samples are generic, inclusive QCD jets, we will use some knowledge from the experimental realm to help constrain the background.  The ability of experiments to identify and tag a jet that contains a long-lived bottom hadron (correspondingly interpreted as initiated by a bottom quark) is essentially perfect \cite{ATLAS:2025dkv,CMS:2026mee}.  So, as the top quark decays to a bottom quark, we will assume that all of the nominal jets in our sample contain a bottom quark.  That is, the QCD background we consider are bottom jets.\footnote{Robustly defining the partonic flavor of a jet is subtle in perturbation theory, and can be practically impossible in experiment for light parton species.  There have recently been several definitions of infrared safe jet flavor \cite{Banfi:2006hf,Caletti:2022hnc,Caletti:2022glq,Czakon:2022wam,Gauld:2022lem,Caola:2023wpj,Larkoski:2025afg,Generet:2025gdy} and a recent community Les Houches study \cite{Behring:2025ilo}.}

Our goal is them to calculate the probability that a background jet passes all the cuts to look exactly like a top quark jet, given that the jet has a bottom quark in it, $p(\text{passes cuts}|b)$.  Schematically, this can be expressed as the product of the relevant phase space volume and a matrix element:
\begin{align}
p(\text{passes cuts}|b) = d\Phi\, |{\cal M}|^2\,.
\end{align}
As we are considering jets, we will work in the highly-boosted, collinear limit, so we can simplify phase space and the matrix element accordingly.  The top quark decay proceeds as $t\to b W\to bqq'$, a sequence of two $1\to 2$ decays.  If the $W$ boson mass was large compared to the top mass, then the $W$ would be produced almost at rest in the top rest frame.  On the other hand, if the $W$ boson's mass is small, then it is itself highly boosted in the top rest frame.  The boost factor is controlled by the ratio of squared masses, and we note that
\begin{align}
\frac{m_t^2}{m_W^2} \approx 4.8\,,
\end{align}
which is ``very'' large.  This large boost means that the decay can be approximated as sequential $1\to 2$ collinear splittings.  To ensure that a bottom quark is in the QCD jet, it must be initiated by a bottom quark.  So, we take this probability to have the form:
\begin{align}\label{eq:toptagbkgmaster}
p(\text{passes cuts}|b) \approx d\Phi_2^{(bg)}\,|{\cal M}_{b\to bg}|^2\,d\Phi_2^{(ij)}|{\cal M}_{g\to ij}|^2\,,
\end{align}
where $d\Phi_2$ is collinear phase space
\begin{align}
d\Phi_2 = \frac{dz\, ds}{16\pi^2}\,.
\end{align}
The two collinear QCD splittings that ultimately produce the three-particle final state that mimics the top decay is $b\to bg\to bij$, where $ij$ are either $(gg)$ or $(q\bar q)$.  We can now approximate every factor in the expression of Eq.~\ref{eq:toptagbkgmaster}.

Let's start with the initial splitting, $d\Phi_2^{(bg)}\,|{\cal M}_{b\to bg}|^2$.  This factor is analogous to the decay $t\to Wb$, so let's establish the kinematics of this decay.  We start in the top rest frame in which the four-momenta of the bottom quark and $W$ boson can be expressed as
\begin{align}
&p_b = (p,0,0,p)\,, &p_W = \left(
\sqrt{p^2+m_W^2},0,0,-p
\right)\,.
\end{align}
We have assumed that the bottom quark is massless.  The momentum $p$ is fixed by demanding that the invariant mass of these particles is the top mass:
\begin{align}
(p_b+p_W)^2 = m_t^2=m_W^2 + 2p\sqrt{p^2+m_W^2}+2p^2\,,
\end{align}
so that it follows that
\begin{align}
p = \frac{m_t^2-m_W^2}{2m_t}\,.
\end{align}

Let's now move out of the rest frame, to the lab frame.  To do this, we must perform a Lorentz boost.  Assuming that the total energy of the top jet is $E$ after boosting in a direction an angle $\theta$ from the direction of the bottom quark, the energy of the bottom quark becomes
\begin{align}
E_b = \gamma p\left(1 +\beta\cos\theta\right) = \frac{E}{m_t}p\left(
1+\sqrt{1-\frac{m_t^2}{E^2}}\cos\theta
\right)\,.
\end{align}
For unpolarized top production, the direction of the boost is uniform on the celestial sphere; that is, the distribution of $\theta$ is uniform in $\cos\theta$.  The energy fraction of the bottom quark in the lab frame is
\begin{align}
z = \frac{E_b}{E} = \frac{p}{m_t}\left(
1+\sqrt{1-\frac{m_t^2}{E^2}}\cos\theta
\right)\,.
\end{align}
We can now safely take the large boost $E\to \infty$ limit, in which the energy fraction simplifies to 
\begin{align}
z \to \frac{m_t^2-m_W^2}{2m_t^2}\left(
1+\cos\theta
\right)\,.
\end{align}
While the minimum energy fraction of the bottom quark is 0, even in this infinite boost limit, the maximum energy fraction of the bottom quark is:
\begin{align}
z_{\max} = \frac{m_t^2-m_W^2}{m_t^2}\,,
\end{align}
which is less than 1 because the $W$ boson is massive.

Now, I note that the matrix element or splitting function for $b\to bg$ takes the form
\begin{align}
|{\cal M}_{b\to bg}|^2 \approx \frac{8\pi\alpha_s C_F}{m_t^2}\, f(z)\,,
\end{align}
where $m_t$ is the invariant mass of the jet and $f(z)$ is some smooth function of the energy fraction $z$, regulated by the constraint that the $W$ boson (faked here by the gluon) is massive.  As we have done before, because this function is smooth on phase space, we will just set it to unity for simplicity (as we'll only care about parametric scaling).  Putting these pieces together, the product of phase space and this matrix element is approximately
\begin{align}
d\Phi_2^{(bg)}\,|{\cal M}_{b\to bg}|^2 \approx \frac{m_t^2-m_W^2}{m_t^2}\,\frac{dm_t}{m_t}\,\frac{\alpha_s C_F}{\pi}\,,
\end{align}
after integrating over the energy fraction of the bottom quark, $z$, and assuming that the mass of the jet is within $dm_t$ of the top mass, $m_t$.\footnote{Note that we have defined $dm_t$ as the width of the window about the invariant mass of the jet, $m_t$.  However, the coordinate on phase space is the squared invariant mass $s$, so there is a Jacobian that we've accounted for in this expression.}  This is the probability for the splitting $b\to bg$ to fake the decay $t\to bW$.

We then need to do the same analysis of the subsequent $W$ decay.  However, we assume that its decay products are massless, so the kinematics is significantly simplified and can be recycled from the top quark decay analysis.  For the matrix element, the gluon can split to two gluons (with probability proportional to the adjoint Casimir $C_A$) or to a quark--anti-quark pair (with probability proportional to the number of relevant quark flavors and the Killing form normalization $T_R$).  The matrix element for gluon splitting takes the schematic form
\begin{align}
|{\cal M}_{g\to ij}|^2 \approx \frac{8\pi\alpha_s}{m_W^2}\, \left[
C_A f_A(z) + (n_f-1)T_R f_T(z)
\right]\,,
\end{align}
where $f_A(z)$ and $f_T(z)$ are two smooth functions of energy fraction $z$, regulated by the non-zero $W$ mass.  The number of active or massless quark flavors $n_f$ is 5 (as we assume all quarks up through the bottom are massless).  However, the $W$ boson's decay to the bottom quark is strongly suppressed by off-diagonal elements in the CKM matrix, so we will ignore it.  Setting these functions to unity for isolation of parametrics, the probability that a $W$-mass gluon splits scales like
\begin{align}
d\Phi_2^{(ij)}|{\cal M}_{g\to ij}|^2 \approx \frac{dm_W}{m_W}\,\frac{\alpha_s}{\pi}\left(
(n_f-1)T_R + C_A
\right)\,.
\end{align}

Let's now review.  We have defined a top quark jet as a jet with the following properties:
\begin{enumerate}
\item contains a bottom quark,
\item has a total mass within a window width $dm_t$ of the top quark mass $m_t$,
\item has three hard prongs corresponding to the three partonic decay products, and
\item the mass of the non-bottom quark pair of those decay products is within a window width $dm_W$ of the $W$ mass $m_W$.
\end{enumerate}
By construction within our framework and assumptions, a boosted, hadronically-decaying top quark satisfies these properties with probability 1.  (Of course, further emissions in the parton shower, hadronization, detector effects, etc., will smear this out, but this is the simplest place to start.)  By contrast from our calculation above, a background QCD jet only satisfies these requirements with probability:
\begin{align}
p(\text{passes cuts}|b) \approx \frac{m_t^2-m_W^2}{m_t^2}\,\frac{dm_t}{m_t}\,\frac{dm_W}{m_W}\,\left(\frac{\alpha_s }{\pi}\right)^2 C_F\left(
(n_f-1)T_R + C_A
\right)\,.
\end{align}

Let's now plug in some numbers to get a sense of the scale of this probability.  The Casimirs are $C_F = 4/3$ and $C_A = 3$ in QCD.  Including the bottom quark, $n_f=5$ and $T_R$ = 1/2.  At high energies, $\alpha_s \sim 0.1$, so we find that the probability is approximately
\begin{align}
p(\text{passes cuts}|b) \approx 6\times 10^{-3}\,\frac{dm_t}{m_t}\,\frac{dm_W}{m_W}\,.
\end{align}
The effective width of the windows about the top and $W$ masses are in practice a not-too-small fraction of the mass, so this probability is naturally at the scale of $p(\text{passes cuts}|b) \approx 10^{-3}\,,10^{-4}$ range.  The corresponding QCD rejection rate would be the inverse of this, and from Fig.~\ref{fig:fundtop}, this is indeed the general scale of the QCD jet rejection rate achieved by modern machine architectures.  Machines of course are sensitive to many other physics effects that we have ignored here.  I leave it to you to determine if this simple analysis sheds insight into the problem of the gap between machines and the claimed optimal likelihood.

\section{Regression with Momentum Conservation}\label{sec:rambo}

The blessing of machine learning, as opposed to simple fitting of a parametrized functional form, is that the nonparametric nature of the output has the ability describe a huge number of systems, and describe them in a way that is not a way that humans would do it.  The curse of machine learning, as opposed to fitting a function constrained by known symmetries, is that the nonparametric nature of the output can produce anything, and in particular can badly violate known symmetries.  In particle physics, of course the systems we study enjoy Poincar\'e invariance, which has the consequence of energy, momentum, and angular momentum conservation that is exhibited in every collider event.  If your machine is trained on particle physics data, given a large enough training sample and enough compute, the machine will indeed learn these conservation laws, by the universal approximation theorem \cite{cybenko1989approximation,hornik1991approximation,leshno1993multilayer,Bogatskiy:2022czk}.  However, all training sets and compute resources are finite, and so in practice, these conservation laws will essentially always be broken at some level.  Conservation laws are a tricky beast: there is no sense in which a conservation law that corresponds to a true symmetry of Nature can be ``slightly'' broken.  Every particle collision event conserves energy, momentum, and angular momentum {\it exactly}, and any violation of these conservation laws is a significant problem for how to understand and validate the output of a machine.

Naturally, then, development of architectures that incorporate or can learn symmetries in the data have seen significant development in the past few years \cite{Dolan:2020qkr,Craven:2021ems,Gong:2022lye,Bogatskiy:2022hub,Hao:2022zns,Forestano:2023fpj,Buhmann:2023pmh,Forestano:2023qcy,Bogatskiy:2023nnw,Bright-Thonney:2023gdl,Bressler:2024wzc,Chatterjee:2024pbp,Bhardwaj:2024djv,Sahu:2024sts,Spinner:2024hjm,Maitre:2024hzp,Brehmer:2024yqw,Woodward:2024dxb,Nabat:2024nce,Sanz:2025sld,Spinner:2025prg,Favaro:2025pgz,Hebbar:2025adf,Petitjean:2025zjf,Breso-Pla:2026tlz,Abasov:2026jed,Kato:2026txd}.  This includes equivariant architectures that faithfully represent the Lorentz symmetry of the data, or architectures that can learn and implement exact symmetries, or modifications of loss functions to penalize conservation law failure.  While these recent developments are significant, the issue of symmetries and representation of particle collider events extends back decades, to the first Monte Carlo computer program implementations that were created to sample $n$-particle Lorentz-invariant phase space, the space on which collider events live, efficiently \cite{Kleiss:1985gy}.  Even further back, to the beginning of understanding relativity itself, one is confronted with questions of how two observers can communicate with each other in a way that is universal and frame-independent.  Two observers, for example, in general disagree with their measurement of the lifetime of the muon.  However, they always agree on the lifetime that the muon itself would measure in its proper (or, as we are in France), {\it propre}, frame.

If we have a definition of a {\it propre} frame and a prescription for transforming to and from it, we can easily guarantee exact Lorentz invariance, Poincar\'e invariance, momentum conservation, etc., of our architecture at an event-by-event level.  While the presentation in this section follows from Ref.~\cite{Bogorad:2026oxa}, in the machine learning context there were some ideas about this earlier \cite{Favaro:2025pgz}, and, as we will see, it was actually essentially completely known 40 years ago \cite{Kleiss:1985gy}.  This starts with understanding the symmetries of massless Lorentz invariant phase space.

\subsection{Symmetries of Massless Phase Space}

Recall that differential, massless, Lorentz-invariant phase space for $n$ particles is
\begin{align}
d\Phi_n &= \prod_{i=1}^n \left[\frac{d^4p_i}{(2\pi)^4}\,2\pi\,\delta(p_i^2) \,\Theta(p_i^0)\right](2\pi)^4\, \delta^{(4)}\left(Q-\sum_{i=1}^n p_i\right)\,,
\end{align}
where $p_i$ is the $i$th particle four-momentum and $Q$ is the net four-vector of the system.  Here, note that we explicitly impose that the energy or $0$th component of momentum is non-negative with $\Theta(p_i^0)$.  If $\Lambda$ is an element of the proper orthochronous Lorentz group $\text{SO}(3,1)^+$ (that is, a Lorentz transformation that preserves positive energies), then we can transform every momentum with this Lorentz transformation to move to a new frame, $p_i\to \Lambda p_i$. Under this transformation, phase space transforms exceptionally simply:
\begin{align}
d\Phi_n &\to \prod_{i=1}^n \left[\frac{d^4p_i}{(2\pi)^4}\,2\pi\,\delta(p_i^2) \,\Theta(p_i^0)\right](2\pi)^4\, \delta^{(4)}\left(\Lambda^{-1}Q-\sum_{i=1}^n p_i\right)\,.
\end{align}
Here, we have used the fact that the Jacobian of the transformation is 1 because
\begin{align}
\det\Lambda = 1\,,
\end{align}
by definition.  Because the Lorentz group is a group, the inverse transformation $\Lambda^{-1}$ is still a proper orthochronous Lorentz transformation, so all that has happened to phase space is that the net momentum $Q$ has been transformed, $Q\to \Lambda^{-1}Q$.  This is just an extremely roundabout way to say that Lorentz invariant phase space is, well, Lorentz invariant.

Massless phase space is not only Lorentz invariant, but also scale invariant.  Under a scale transformation, the four-momentum transforms as $p_i\to \lambda p_i$, where $\lambda$ is a positive real number, $\lambda \in \mathbb{R}^+$.  Phase space transforms homogeneously under a rescaling:
\begin{align}
d\Phi_n &\to \prod_{i=1}^n \left[\frac{\lambda^4\,d^4p_i}{(2\pi)^4}\,2\pi\,\delta(\lambda\,p_i^2) \,\Theta(\lambda\, p_i^0)\right](2\pi)^4\, \delta^{(4)}\left(Q-\sum_{i=1}^n \lambda\,p_i\right)\\
&=\lambda^{2n-4}\prod_{i=1}^n \left[\frac{d^4p_i}{(2\pi)^4}\,2\pi\,\delta(p_i^2) \,\Theta(p_i^0)\right](2\pi)^4\, \delta^{(4)}\left(\frac{Q}{\lambda}-\sum_{i=1}^n \,p_i\right)
\nonumber\,.
\end{align}
That is, under a scale transformation, the structure of massless phase space is unchanged, but the total momentum $Q$ has transformed to $Q\to Q/\lambda$.  Note that, unlike Lorentz symmetry, this scale invariance requires exclusively massless particles.  A particle's mass is not a dynamical quantity and so would not transform under such a scaling.  In that case, a scale transformation would break the structure of massive phase space, effectively modifying the masses of all particles as $m_i \to m_i/\lambda$.  If the mass of a particle changes, then the particle changes, as the mass is a Casimir of the Poincar\'e group and so defines the irreducible representation of the particle.

The relationship between Lorentz invariance, scale invariance, and conformal invariance has long been very intriguing in the study of general quantum field theories.  It is known that two-dimensional quantum field theories that enjoy Lorentz and scale invariance are fully conformal invariant \cite{Zamolodchikov:1986gt}, but in higher dimensions the problem remains open.  A sensible quantum field theory is unitary, and the interplay of unitarity with spacetime symmetries is extremely subtle and complex.  For more information about this fascinating research program, see, e.g., Refs.~\cite{Polchinski:1987dy,Nakayama:2013is,Dymarsky:2013pqa}.  For the case at hand, however, this Lorentz invariance and scale invariance is much more pedestrian; it is merely the symmetry of phase space, and not a fundamental quantum field theory.  Anyway, massless particles in QCD are an approximation anyway, valid at sufficiently high energies when we can ignore the masses of quarks or hadrons. There is only one particle that we have ever detected that we believe is truly massless: the photon.  

\subsection{Property of Maximal Entropy}

Let's now shelve our knowledge that massless phase space is both Lorentz and scale invariant and make another, seemingly orthogonal, interesting observation.  Because there is a fixed net four momentum $Q$ and particle energies must be non-negative, phase space is compact and has a finite volume (which we will calculate shortly).  Also, just considering phase space itself, it is ``flat'': there are no correlations between particles beyond energy-momentum conservation.  Dynamical correlations would be encoded in a matrix element on that phase space, which in turn represents the underlying quantum field theory that describes the interactions between particles.  Phase space alone essentially assumes that the matrix element is unity.

Such a uniform distribution on a compact space maximizes entropy; given a finite region and some particles, the distribution that is maximally mixed is when the particles are uniformly distributed on that space.  This is straightforward to prove for one-dimensional distributions, and can easily be generalized to higher dimensions.  In one dimension, consider the distribution $p(x)$ defined in some finite domain $x\in {\cal R}$.  This distribution is normalized
\begin{align}
1 = \int_{\cal R} dx\, p(x)\,.
\end{align}
To determine the distribution that maximizes entropy given this fixed normalization, we turn to our old friend Lagrange multipliers.  We introduce the Lagrange multiplier $\alpha$ to enforce the normalization and add to it the entropy of the distribution to define the Lagrangian:
\begin{align}
{\cal L}\left[p(x),\alpha\right] = \alpha\left[\int_{\cal R} dx\, p(x) -1\right]- \int_{\cal R} dx\, p(x)\log p(x)\,.
\end{align}
Note that the entropy, the term at right, is the continuous generalization of the more familiar discrete definition of entropy $S$,
\begin{align}
S = -\sum_i p_i\log p_i\,,
\end{align}
where $p_i$ is the probability of state $i$.\footnote{In the continuous generalization, $p(x)$ is a probability {\it density}, and not a probability, so continuous entropy is not invariant to coordinate transformations.}

We can then take the distributional derivative of the Lagrangian with respect to $p(x)$ and demand that it vanishes to extremize entropy:
\begin{align}
\frac{\delta{\cal L}}{\delta p(x)} = 0 = - \int_{\cal R} dx\left[ \log p(x)+(1-\alpha)\right]\,.
\end{align}
For this to vanish in general, the probability distribution $p(x)$ must be constant.  In terms of the Lagrange multiplier it is
\begin{align}
p(x) = e^{\alpha-1}\,.
\end{align}
$\alpha$ is then fixed by demanding that the distribution be normalized.  Again, this argument can be generalized to higher dimensions, but we won't do that here.  Essentially, all that is needed to do that proof is to define the general volume measure on the space $dV$ on which the entropy is evaluated. In general, however, the uniform distribution on a compact manifold maximizes entropy with a fixed normalization.  This implies that massless phase space itself is a maximal entropy distribution.

Let's again tuck away this maximal entropy property of phase space along with its symmetries.  Let's establish the maximal entropy distribution with other constraints.  Let's now consider a normalized distribution on $x\in[0,\infty)$ that also has a fixed mean $\mu$.  What is the maximal entropy distribution with these constraints?  We introduce two Lagrange multipliers $\alpha$ and $\beta$ to enforce the normalization and mean, and then add to them the entropy:
\begin{align}
{\cal L}\left[p(x),\alpha,\beta\right] = \alpha\left[\int_0^\infty dx\, p(x) -1\right]+\beta\left[\int_0^\infty dx\, x\,p(x) -\mu\right]- \int_0^\infty dx\, p(x)\log p(x)\,.
\end{align}
Taking the distributional derivative with respect to $p(x)$ and extremizing, we have the requirement that
\begin{align}
\frac{\delta{\cal L}}{\delta p(x)} = 0 = - \int_0^\infty dx\left[ \log p(x)+(1-\alpha)-\beta x\right]\,.
\end{align}
For this to vanish completely generally, this requires $x$ to be exponentially distributed:
\begin{align}
p(x) \propto e^{\beta x}\,,
\end{align}
where $\beta$ is set by $\mu$.  You are of course very familiar with this result from statistical mechanics, as this is the derivation of the Boltzmann factor when particles in a thermal system have an average energy.

\subsection{RAMBO Algorithm}

Let's summarize the observations that we've made thus far.  Massless, Lorentz invariant phase space enjoys Lorentz and dilation (scaling) symmetries.  Uniform on phase space is also the maximal entropy distribution on that compact manifold.  The maximal entropy distribution on the semi-infinite domain $x\in[0,\infty)$ with a fixed mean is the exponential distribution.  Note also that two exponential distributions can be transformed into one another by precisely a scale transformation.  Let's consider the exponential distribution unit normalized with unit mean:
\begin{align}
dx\, p(x) = dx\, e^{-x}\,.
\end{align}
Now, rescale $x\to \lambda x$.  The distribution transforms to
\begin{align}
dx\, p(x) \to \lambda \, dx\, e^{-\lambda x}\,.
\end{align}
This is still unit normalized and a scale transformation maps the domain to itself: $\mathbb{R}^+ \to \lambda \mathbb{R}^+ = \mathbb{R}^+$.  However, now the mean is $\mu\to 1/\lambda$.

These results provide us with a very powerful algorithm for sampling phase space.  In general, sampling a space with universal constraints, like energy-momentum conservation constraints on phase space, in a predictable way is significantly challenging.  However, the symmetries and maximal entropy properties enable us to uniformly sample massless phase space with a Monte Carlo algorithm.  The idea is as follows.  Consider the conjugate space $d\widetilde\Phi_n$ with no universal constraints, but with individual particle energies exponentially distributed.  This differential space is
\begin{align}
d\widetilde\Phi_n = \prod_{i=1}^n\left[
d^4 q_i\, \delta(q_i^2)\,\Theta(q_i^0)\, e^{-k\cdot q_i}
\right] = \prod_{i=1}^n\left[
\frac{d^3 q_i}{2|\vec q_i|}\, e^{-|\vec q_i|}
\right] = \prod_{i=1}^n\left[
\frac{|\vec q_i|\, e^{-|\vec q_i|}\,d |\vec q_i|}{2}\, d\Omega_i
\right]\,.
\end{align}
In this expression, we have chosen the reference four-vector $k = (1,0,0,0)$. Normalization can always be fixed up later; we really only care about the functional form of the distribution.  Here, $d\Omega_i$ is the angular measure on the two-sphere for particle $i$.

Because every particle is independent of every other particle on this space, we can easily sample all corresponding $q$-momenta.  Sampling points on the two-sphere is easy: azimuthal angles are uniform on $\phi\in[0,2\pi)$ and polar angles are uniform in $\cos\theta\in[-1,1]$.  The method for sampling energies with this distribution follows from a simple observation.  The chi-squared probability distribution
\begin{align}
P_n(x) = \frac{x^{n-1}e^{-x}}{\Gamma(n)}\,,
\end{align}
can be sampled by taking 
\begin{align}
x = -\log\left(
\prod_{i=1}^n \xi_i
\right)\,,
\end{align}
where $\xi_i\in[0,1]$ is uniformly distributed.  This can be proved by induction.  First, with $n=1$, we have
\begin{align}
P_1(x) = \int d\xi_1\,\delta\left(
x+\log\xi_1
\right)=\int d\xi_1\, \xi_1\,\delta\left(\xi_1-e^{-x}\right) = e^{-x}\,.
\end{align}
Now, we assume $P_n(x)$.  $P_{n+1}(x)$ is
\begin{align}
P_{n+1}(x) &=\int \prod_{i=1}^{n+1} d\xi_i\, \delta\left(
x+\sum_{i=1}^{n+1}\log\xi_i
\right) =\int_0^1 d\xi \,P_n(x+\log\xi)\\
&=\frac{e^{-x}}{\Gamma(n)}\int_0^1 \frac{d\xi}{\xi}\,(x+\log\xi)^{n-1} \,\Theta(x+\log\xi)
\nonumber\\
&=\frac{x^n e^{-x}}{n\,\Gamma(n)}\,,\nonumber
\end{align}
completing the proof.  Step one of the algorithm is generating the $n$ momenta in this $q$-space.

The next step exploits the maximal entropy properties and symmetries.  Given this set of $n$ momenta $\{q_i\}_{i=1}$, they have some net momentum
\begin{align}
\widetilde Q = \sum_{i=1}^n q_i\,.
\end{align}
By Lorentz invariance, we are free to boost this to any other frame we want; for simplicity let's boost to the center-of-mass (COM) frame in which the three-momentum is 0.  As Lorentz transformations are a symmetry, the system maintains its maximal entropy property.  However, after this Lorentz boost, the net energy is still rather unconstrained.  Never fear; we still have scale invariance at our disposal.  We then rescale all particle momenta so that the total energy is unity (say).  Again, scale transformations are a symmetry, so the distribution retains its maximal entropy property.  But now, we have generated the maximal entropy distribution of $n$ on-shell, massless four momenta with 0 net three-momentum and unit total energy.  This is now {\it exactly} the uniform on phase space distribution we were seeking!

This algorithm for uniform sampling of massless phase space is called RAMBO \cite{Kleiss:1985gy}.\footnote{If you are of a younger generation or not so familiar with American films of the 1980s, you likely do not know the origin of this name.  Rambo was the main character from the 1972 novel {\it First Blood} by David Morrell and a 1982 film starring Sylvester Stallone.  Rambo was a Vietnam War veteran troubled by his war experience.  While that first film from 1982 was a rather serious portrait of a veteran's PTSD, the five films later in the series became increasingly campy and ``Rambo'' became a general term for someone randomly shooting a machine gun in all directions.  This point was parodied in the 1989 film {\it UHF} starring ``Weird Al'' Yankovic.  The RAMBO algorithm samples uniformly on massless phase space by Monte Carlo ``shooting in all directions'', hence the name.}  That original paper has all the details about the precise Lorentz and scale transformations necessary for a rather simple algorithm that can be easily implemented in your favorite code.  The compact representation of the algorithm is
\begin{align}
d\widetilde \Phi_n \xrightarrow{\text{boost to COM frame; rescale total energy}} d\Phi_n\,.
\end{align}
Interestingly, the original authors did not seem to know about the maximal entropy argument employed here, but simply happened upon the exponential energy distribution in the conjugate $q$-space as it had desired properties.  This then connects back with our original goal.  To ensure exact energy and momentum conservation of our machine learning architecture, all we need to do is perform the learning or regression in the conjugate $q$-space.  With that result, we can then Lorentz boost and scale our events back to the desired frame.  While this does not work with massive particles (because scale invariance is lost), one can introduce weighted events that account for mass effects \cite{Kleiss:1985gy}.  For more details on algorithmic implementation of such a $q$-space regression see Ref.~\cite{Bogorad:2026oxa}.

\subsection{The Volume of Phase Space}

To actually calculate distributions on massless phase space, it is useful to know its total volume.  With the total volume of phase space, we can correspondingly define a Monte Carlo sampling procedure of a distribution of interest on phase space from the (normalized) uniform distribution provided by RAMBO.  To calculate the volume of phase space, we will work with the slightly cleaner form for differential phase space
\begin{align}
d\Pi_n &= \prod_{i=1}^n \left[d^4p_i\,\delta(p_i^2) \,\Theta(p_i^0)\right] \delta^{(4)}\left(Q-\sum_{i=1}^n p_i\right)\,,
\end{align}
where we have just removed all explicit factors of $2\pi$.  These can of course be restored at the end, if necessary.  By Lorentz invariance and dimensional analysis, the total volume of phase space can be expressed as
\begin{align}
\int d\Pi_n = a_n \left(Q^2\right)^{n-2}\,,
\end{align}
where $Q^2$ is the total squared invariant mass of the event and $a_n$ is some dimensionless constant that depends on the number of particles, $n$.

We will now work recursively, and work with an expression that constructs a meta-massive particle from the first $n-1$ particles:
\begin{align}
\int d\Pi_n &= \int \prod_{i=1}^n \left[d^4p_i\,\delta(p_i^2) \,\Theta(p_i^0)\right] \delta^{(4)}\left(Q-\sum_{i=1}^n p_i\right)\int d^4q\,\delta^{(4)}\left(q-\sum_{i=1}^{n-1}p_i\right)\, dw^2\, \delta(w^2-q^2)\\
&=\int d^4 p_n\,\delta(p_n^2)\, \Theta(p_n^0)\,d^4 q\,\delta(q^2-w^2)\, \Theta(q^0) \,\delta^{(4)}(Q-p_n-q)\times a_{n-1}(w^2)^{(n-1)-2}\, dw^2\,.\nonumber
\end{align}
At right on the first line, we have multiplied by 1:
\begin{align}
1 = \int d^4q\,\delta^{(4)}\left(q-\sum_{i=1}^{n-1}p_i\right)\, dw^2\, \delta(w^2-q^2)\,,
\end{align}
and then simplified factors on the second line through integration over particles 1 through $n-1$.  Note the appearance of the coefficient $a_{n-1}$ and the energy weight $(w^2)^{(n-1)-2}$, whose product is the volume of $(n-1)$-body massless phase space with total invariant mass $w^2$.

Now, note that
\begin{align}
\int d^4 p_n\,\delta(p_n^2)\, \Theta(p_n^0)\,d^4 q\,\delta(q^2-w^2)\, \Theta(q^0)\,\delta^{(4)}(Q-p_n-q)
\end{align}
is just the volume of two-body phase space with one massive and one massless particle with total four-momentum $Q$.  The volume of two-body phase space is (see, e.g., Appendix A.5 of Peskin and Schroeder \cite{Peskin:1995ev})
\begin{align}
\int d^4 p_n\,\delta(p_n^2)\, \Theta(p_n^0)\,d^4 q\,\delta(q^2-w^2)\, \Theta(q^0)\,\delta^{(4)}(Q-p_n-q) = \frac{\pi}{2}\frac{2|\vec p|}{\sqrt{Q^2}}\,,
\end{align}
where $|\vec p|$ is the magnitude of the three-momentum of either final state particle in the center-of-mass frame.  To evaluate this, note that the energy of the massless particle is $E_p = |\vec p|$, while the energy of the massive particle is $E_q = \sqrt{w^2+|\vec p|^2}$.  Their sum is the total energy in the center-of-mass frame:
\begin{align}
E_p+E_q = \sqrt{Q^2} = |\vec p|+\sqrt{w^2+|\vec p|^2}\,.
\end{align}
The solution is
\begin{align}
|\vec p| = \frac{Q^2-w^2}{2\sqrt{Q^2}}\,.
\end{align}
Using this, we then have
\begin{align}
\int d^4 p_n\,\delta(p_n^2)\, \Theta(p_n^0)\,d^4 q\,\delta(q^2-w^2)\, \Theta(q^0) \,\delta^{(4)}(Q-p_n-q)= \frac{\pi}{2}\left(
1-\frac{w^2}{Q^2}
\right)\,.
\end{align}

With this result, the total volume of phase space is
\begin{align}
\int d\Pi_n = a_n(Q^2)^{n-2} =& \frac{\pi}{2}a_{n-1}\int_0^{Q^2}dw^2\left(
1-\frac{w^2}{Q^2}
\right)\, (w^2)^{n-3}\\
&=\frac{\pi}{2}a_{n-1}\frac{(Q^2)^{n-2}}{(n-1)(n-2)}\,.\nonumber
\end{align}
This recursion relation can easily be solved and we find
\begin{align}
\int d\Pi_n = \left(
\frac{\pi}{2}
\right)^{n-1}\frac{(Q^2)^{n-2}}{(n-1)!(n-2)!}\,.
\end{align}

\subsection{Geometry and Topology of Phase Space}\label{sec:geotopops}

We can approach the evaluation of the volume of phase space from a very different perspective, which also provides some insight into the local geometry and global topology of fixed $n$-body phase space.  Again, removing factors of $2\pi$, differential phase space is typically expressed as
\begin{align}
d\Pi_n &= \prod_{i=1}^n \left[d^4p_i\,\delta(p_i^2) \,\Theta(p_i^0)\right] \delta^{(4)}\left(Q-\sum_{i=1}^n p_i\right)\,.
\end{align}
While this volume form on phase space is expressed in ``natural'' coordinates of collider experiment, as we directly measure particle momenta, it is less convenient for establishing properties of the geometry of phase space.  Let's work in the center-of-mass frame where $Q = (Q,0,0,0)$ is both the total energy and net four momentum of the event.  The $\delta$-function constraints in Cartesian coordinates aren't the most convenient, so let's introduce a parametrization that naturally imposes the on-shell condition, $\delta(p^2)$.  We take inspiration from the spinor-helicity formalism of amplitudes, see, e.g., Refs.~\cite{Mangano:1990by,Dixon:1996wi}, in which we encode an on-shell, massless four-vector $p$ through the complex-valued two-component spinors $\lambda^1,\lambda^2$.  This is accomplished by dotting the four momentum with the Pauli matrices\footnote{We've moved Lorentz indices of the momentum $p$ lower so that the notation that follows is a bit simpler.  The presentation in this section follows Ref.~\cite{Larkoski:2020thc}.}
\begin{align}\label{eq:momspinor}
(p\cdot \sigma)^{\alpha\dot\alpha} = \left(
\begin{array}{cc}
p_0-p_3 & -p_1+ip_2\\
 -p_1-ip_2 & p_0+p_3 
\end{array}
\right)^{\alpha\dot\alpha} = \left(
\begin{array}{cc}
\lambda^1\tilde\lambda^{\dot 1} & \lambda^1\tilde\lambda^{\dot 2}\\
\lambda^2\tilde\lambda^{\dot 1} & \lambda^2\tilde\lambda^{\dot 2} 
\end{array}
\right)^{\alpha\dot\alpha}\,.
\end{align}
Here, the Pauli matrices are
\begin{align}
\sigma^\mu = (\mathbb{I},\vec \sigma)^\mu\,,
\end{align}
where $\mathbb{I}$ is the $2\times 2$ identity matrix, and $\alpha$ and $\dot\alpha$ are spinor indices.  This momentum matrix is Hermitian because four-momentum is real-valued, and so the spinors are related by complex conjugation: $\lambda^* = \tilde\lambda$.  In this formalism, the on-shell condition is automatic, through the fact that the determinant of this matrix is 0.

Additionally, the spinors enjoy a freedom due to invariance under the little group.  A massless particle travels at the speed of light, and so there is no Lorentz transformation that one can perform to ``stop'' or ``turn around'' a massless particle.  Thus, the axis along the direction of motion of a massless particle is itself a Lorentz invariant, and therefore a good axis about which to quantize angular momentum.  Along a single axis, angular momentum can only point one of two ways, and this helicity is therefore Lorentz invariant.  Helicity quantized along the direction of massless particle momentum enjoys a freedom because we can rotate about that axis which leaves the particle unchanged.  A rotation about a single axis in three spatial dimensions forms a group, the group U(1), one-dimensional unitary rotations.  This U(1) little group manifests itself in spinor-helicity notation because we can freely multiply the spinors by a complex phase, and the momentum matrix remains unchanged.  That is, under the transformation
\begin{align}
&\lambda \to e^{i\phi} \lambda\,, &\tilde \lambda \to e^{-i\phi}\tilde \lambda\,,
\end{align}
Eq.~\ref{eq:momspinor} is unchanged.

With this observation, we can therefore express on-shell differential phase space (with no momentum conservation yet) as
\begin{align}
d^4p\, \delta(p^2)\, \Theta(p_0) = \frac{d^2\lambda^1\, d^2\lambda^2}{\text{U}(1)}\,.
\end{align}
Here, the division by U(1) represents an implicit restriction to one element of the little group, or, equivalently, division by the volume of the group U(1).  I will leave it as an exercise to you to perform the change of variables using Eq.~\ref{eq:momspinor} to verify this relationship.  To clean up this expression, we will introduce the dimensionless and complex-valued vectors $\vec u$ and $\vec v$, where
\begin{align}
&\vec u = \frac{1}{\sqrt{Q}}\left(
\lambda^1_1\ \ \lambda^1_2\  \ \cdots \  \ \lambda^1_n
\right)\,,
&\vec v = \frac{1}{\sqrt{Q}}\left(
\lambda^2_1\ \ \lambda^2_2\ \ \cdots \ \  \lambda^2_n
\right)\,,
\end{align}
where the subscript denotes the particle.  With these coordinates, momentum conservation can be simply expressed.  First, note that in the center-of-mass frame, the momentum conserving $\delta$-functions can be written in the form
\begin{align}
\delta^{(4)}\left(Q-\sum_{i=1}^n p_i\right) = \delta\left(
Q - \sum_{i=1}^n(p_{i,0}+p_{i,3})
\right)\delta\left(
Q - \sum_{i=1}^n(p_{i,0}-p_{i,3})
\right)\delta\left(
\sum_{i=1}^n \vec p_{i,1}
\right)\delta\left(
\sum_{i=1}^n \vec p_{i,2}
\right)\,.
\end{align}
This is similar to the light-cone coordinates introduced in Sec.~3.1.1 of Ref.~\cite{Larkoski:2024uoc}, with a different choice of normalization.  In the coordinates $\vec u$ and $\vec v$, this simplifies extremely compactly to
\begin{align}
\delta^{(4)}\left(Q-\sum_{i=1}^n p_i\right) = Q^{-4}\delta\left(1-|\vec u|^2\right)\delta\left(1-|\vec v|^2\right)\delta^{(2)}\left(\vec u^\dagger \vec v\right)\,,
\end{align}
where $\dagger$ is the Hermitian conjugate.  That is, vectors $\vec u$ and $\vec v$ are orthogonal and unit normalized.  Phase space in spinor-helicity coordinates is then
\begin{align}
d\Pi = Q^{2n-4}\,\frac{d^n\vec u\, d^n\vec v}{\text{U}(1)^n}\,\delta\left(1-|\vec u|^2\right)\delta\left(1-|\vec v|^2\right)\delta^{(2)}\left(\vec u^\dagger \vec v\right)\,.
\end{align}

Now, let's clean this up a bit.  Let's simplify the phase space of $\vec u$ by introducing polar coordinates for each component of $\vec u$.  With
\begin{align}
u_j= r_j e^{i\phi_j}\,,
\end{align}
for the $j$th component of vector $\vec u$, with $r_j$ the radius and $\phi_j$ the phase, we have
\begin{align}
\frac{d^n\vec u}{\text{U}(1)^n}\,\delta\left(1-|\vec u|^2\right) = \frac{\prod_{j=1}^n r_j\, dr_j\, d\phi_j}{\text{U}(1)^n}\,\delta\left(1-\sum_{j=1}^n r_j^2\right)\,.
\end{align}
The phase $\phi_j$ is precisely the little group rotation that we can freely integrate over, as it does not represent a dynamical degree of freedom.  So we can integrate over this angle using little group freedom, and we also note that the volume of the group U(1) is $2\pi$, which exactly cancels the integral over the phases $\phi_j$.  Also, let's call
\begin{align}
\rho_j \equiv r_j^2\,,
\end{align}
so that this part of phase space simplifies to
\begin{align}
\frac{d^n\vec u}{\text{U}(1)^n}\,\delta\left(1-|\vec u|^2\right) = \frac{1}{2^n}\prod_{j=1}^n d\rho_j\,\delta\left(1-\sum_{j=1}^n \rho_j\right)\,.
\end{align}
This describes an $(n-1)$-simplex and we note that the volume of such a simplex is
\begin{align}
\int  \prod_{j=1}^n d\rho_j \, \delta\left(1-\sum_{j=1}^n \rho_j\right) = \frac{1}{(n-1)!}\,.
\end{align}

Let's next turn to what remains of the differential phase space.  Phase space of the $\vec v$ vector can be simplified by integrating over the $n$th component with the $\delta$-function that enforces $\vec u^\dagger \vec v = 0$:
\begin{align}\label{eq:vdiffform}
d^n\vec v\,\delta\left(1-|\vec v|^2\right)\delta^{(2)}\left(\vec u^\dagger \vec v\right) = \frac{d^{n-1}\vec v}{\rho_n}\,\delta\left(
1-\sum_{i=1}^{n-1}|v_i|^2-|v_n|^2
\right)\,.
\end{align}
Here, of course the component $v_n$ is now a function of all other components of $\vec u$ and $\vec v$.  In particular, the $\delta$-function sets
\begin{align}
 u_n v_n = -\sum_{i=1}^{n-1}u_i v_i\,.
\end{align}
The inverse Jacobian of this transformation is then the square of component $u_n$, $J^{-1} = u_n^2 = \rho_n$.  We also note that we can choose all components of $\vec u$ to be real and positive, by little group invariance.

We still have that unit normalization $\delta$-function, and it would be very convenient if we could find a coordinate transformation $\vec v\to \vec v'$ such that
\begin{align}
\sum_{i=1}^{n-1}|v_i|^2+ |v_n|^2 = \sum_{i=1}^{n-1}|v'_i|^2 = \vec v^\dagger \mathbb{A}^\dagger\mathbb{A}\vec v\,,
\end{align}
for some matrix $\mathbb{A}$.  If we can find such a transformation matrix $\mathbb{A}$ that is real, so that real and imaginary parts of $\vec v$ are not mixed, and that induces the correct Jacobian to exactly cancel the $1/\rho_n$ factor of Eq.~\ref{eq:vdiffform}, we can have an extremely nice form for the differential phase space.  That is, we would desire that
\begin{align}
\frac{d^{n-1}\vec v}{\rho_n}\,\delta\left(
1-\sum_{i=1}^{n-1}|v_i|^2-|v_n|^2
\right) = d^{n-1}\vec v'\, \delta\left(1-|\vec v'|^2\right)\,,
\end{align}
then the measure at right is the measure of the $(2n-3)$-sphere, $S^{2n-3}$.  I won't provide the details here, but such a matrix $\mathbb{A}$ does exist and can be found.  This matrix has components \cite{Cai:2024xnt}
\begin{align}
\mathbb{A}_{ij} = \frac{1}{u_n}\left(
\frac{u_iu_j}{1+u_n}+u_n\,\delta_{ij}
\right)\,,
\end{align}
with $i,j\in\{1,\dotsc,n-1\}$.  I will leave it as an exercise to you to verify that this matrix does indeed produce the correct Jacobian, and what the form of the vector $\vec v'$ is.

With this transformation, we can then express $n$-particle massless, on-shell phase space in the extremely compact (though, to be fair, very abstract!) form of
\begin{align}
d\Pi_n = \frac{Q^{2n-4}}{2^n} \prod_{i=1}^n d\rho_i\, \delta\left(
1-\sum_{i=1}^n\rho_i
\right)\, d^{n-1}\vec v' \,\delta\left(1-|\vec v'|^2\right)\,.
\end{align}
This volume form describes the product of an $(n-1)$-simplex and a $(2n-3)$-sphere.  It's now extremely easy to calculate the volume of phase space.  As mentioned above, the volume of the simplex is
\begin{align}
\int \prod_{i=1}^n d\rho_i\, \delta\left(
1-\sum_{i=1}^n\rho_i
\right) = \frac{1}{(n-1)!}\,.
\end{align}
The volume of the $(2n-3)$-sphere is
\begin{align}
\int d^{n-1}\vec v' \,\delta\left(1-|\vec v'|^2\right) = \frac{2\pi^{n-1}}{(n-2)!}\,.
\end{align}
The total volume of phase space is thus
\begin{align}
\int d\Pi_n = \left(
\frac{\pi}{2}
\right)^{n-1}\frac{Q^{2n-4}}{(n-1)!(n-2)!}\,,
\end{align}
exactly as we had found in the previous section.  This construction also makes the spherical topology of phase space manifest.

\subsection{Phase Space in the Large-$n$ Limit}

Phase space has rather interesting simplifications in the limit that the number of particles $n$ gets large, $n\to\infty$.  A very general type of observable to study is the correlation of $m$ particles.  We had seen this before in the energy-energy correlator, which quantifies two-particle correlations.  However, in taking moments of many IRC safe observables, $m$ particle correlations arise.  For example, consider the two-point energy correlation function observable \cite{Banfi:2004yd,Larkoski:2013eya,Komiske:2017aww}
\begin{align}
e_\beta = \frac{1}{Q^2}\sum_{i,j}E_iE_j\left(
\frac{1-\cos\theta_{ij}}{2}
\right)^\beta\,.
\end{align}
where the sum runs over all pairs of particles in the event and IRC safety requires that the angular exponent $\beta > 0$.  We also assume that we are working in the center-of-mass frame so $Q$ is the total energy of the events. The mean of this observable is the sum over two-particle correlations:
\begin{align}
\langle e_\beta\rangle = \frac{1}{Q^2}\sum_{i,j} \left\langle E_iE_j\left(
\frac{1-\cos\theta_{ij}}{2}
\right)^\beta \right\rangle\,,
\end{align}
where $\langle \cdot\rangle$ means the expectation value on the appropriate phase space and matrix element, if applicable.  The second moment of the energy correlation function involves four-particle correlations:
\begin{align}
\langle e_\beta^2\rangle = \frac{1}{Q^4}\sum_{i,j,k,l} \left\langle E_iE_jE_kE_l\left(
\frac{1-\cos\theta_{ij}}{2}
\right)^\beta\left(
\frac{1-\cos\theta_{kl}}{2}
\right)^\beta \right\rangle\,,
\end{align}
and continuing for higher moments.

If one is only interested in relatively low moments, then we can meaningfully consider the limit in which the number $m$ of correlated particles is significantly less than the total number of particles $n$, $m\ll n$.  Considering evaluation of an $m$-particle correlation on uniform phase space, we can integrate over $n-m$ particles to produce a reduced phase space for $m$ particles that still respects energy-momentum conservation.  From the analysis above, we have
\begin{align}
&\frac{1}{\int d\Pi_n}\int \frac{d\Pi_n}{d\Pi_{m}} \\
&\hspace{0.5cm}= \left(
\frac{\pi}{2}
\right)^{-m}\frac{(n-1)!(n-2)!(Q^2)^{-n+2}}{{(n-m-1)!(n-m-2)!}}\prod_{i=1}^m \left[d^4p_i\, \delta(p_i^2)\,\Theta(p^0_i)\right]\,\left[
Q^2-2Q\sum_{i=1}^m E_i+\sum_{i,j=1}^m p_i\cdot p_j
\right]^{n-m-2}\,.\nonumber
\end{align}
We denote integration over $n-m$ particles through the schematic differential phase space $d\Pi_n/d\Pi_{m}$, where we have ``modded out'' $n$-body phase space by $m$-body phase space.  The factor in square brackets at right is the invariant mass of the $n-m$ particle subsystem that was integrated over.  Note also that we have normalized by the total volume of phase space.

Now, considering the $n\to \infty$ limit with $m$ fixed, we can simplify this invariant mass.  Note that the sum of all particle energies is $Q$:
\begin{align}
\sum_{i=1}^n E_i = Q\,.
\end{align}
The average particle energy is therefore the total energy divided by $n$:
\begin{align}
\langle E\rangle = \frac{Q}{n}\,.
\end{align}
All particles are treated identically and there is no matrix element that enforces a hierarchy (unlike in QCD), so any individual particle will have an energy that is reasonably close the average.  We can then power count the particle energies to scale like $1/n\ll 1$ in the large-$n$ limit.

On the other hand, the sum of pairs of particles' four-vector dot products sums to the total squared invariant mass:
\begin{align}
\sum_{i,j=1}^n p_i\cdot p_j = Q^2\,.
\end{align}
Assuming that $i\neq j$, the average dot product is
\begin{align}
\langle p_i\cdot p_j\rangle = \frac{Q^2}{n(n-1)}\,,
\end{align}
because there are a total of $n(n-1)$ pairs of non-identical particles $(i,j)$.  With the same assumptions as for energies, particle dot products take values around this average, and can be power counted as scaling like $1/n^2$ in the large-$n$ limit.  This scaling is parametrically smaller than that for the sum of energies,
\begin{align}
p_i\cdot p_j \sim\frac{1}{n^2}\ll QE_i \sim \frac{1}{n}\,,
\end{align}
and so the dot products can be ignored at leading order in the large-$n$ limit.  That is, expanding to lowest order in $1/n$, the invariant mass of the $n-m$ integrated particles simplifies to
\begin{align}
\left[
Q^2-2Q\sum_{i=1}^m E_i+\sum_{i,j=1}^m p_i\cdot p_j
\right]^{n-m-2} &= (Q^2)^{n-m-2}\,\exp\left[
(n-m-2)\log\left(
1-\frac{2}{Q}\sum_{i=1}^m E_i+\frac{1}{Q^2}\sum_{i,j=1}^m p_i\cdot p_j
\right)
\right]\nonumber\\
&=(Q^2)^{n-m-2} \, e^{-\frac{2n}{Q}\sum_{i=1}^m E_i}+{\cal O}(n^{-1})\,.
\end{align}
As required by the assumptions of statistical mechanics, the exponential Boltzmann factor appears in this limit.

Then, using this result, the large-$n$ limit of phase space simplifies to
\begin{align}
&\lim_{n\to \infty}\frac{1}{\int d\Pi_n}\int \frac{d\Pi_n}{d\Pi_{m}} = \left(
\frac{\pi}{2}
\right)^{-m}\frac{(n-1)!(n-2)!(Q^2)^{-m}}{{(n-m-1)!(n-m-2)!}}\prod_{i=1}^m \left[d^4p_i\, \delta(p_i^2)\,\Theta(p^0_i)\, e^{-\frac{2nE_i}{Q}}\right]\,.
\end{align}
We can still do some cleaning up.  The large-$n$ limit of the factorial factors is
\begin{align}
\lim_{n\to\infty}\frac{(n-1)!(n-2)!}{{(n-m-1)!(n-m-2)!}} = n^{2m}\,,
\end{align}
because $n!\sim n^n$ in the large-$n$ limit by Stirling's approximation.  We can also perform integrals using the on-shell $\delta$-function constraints.  We then find
\begin{align}
&\lim_{n\to \infty}\frac{1}{\int d\Pi_n}\int \frac{d\Pi_n}{d\Pi_{m}} =\prod_{i=1}^m \left[\frac{(2n)^2E_i\, dE_i}{Q^2}\, \frac{d\Omega_i}{4\pi}\, e^{-\frac{2nE_i}{Q}}\right]\,,
\end{align}
where $d\Omega_i$ is the differential area of particle $i$ on the celestial sphere.  This reduced phase space is completely factorized and normalized to integrate to unity.  It is also trivial to verify that the average value of a single particle's energy is
\begin{align}
\langle E\rangle = \frac{Q}{n}\,,
\end{align}
as required.

\section{Conclusion}\label{sec:concs}

AI agents are getting better, more accurate, sounding more human, solving more interesting open mathematics problems, and their capabilities will only improve in the future as the scaling laws that seem to govern their performance are showing no signs of tapering off.  Unfortunately, that also means that it is easier and easier for anyone to produce results that feel right or have the right ``vibes'', but are completely vacuous.  This issue is rather serious now for machine learning in particle collider physics as AI agents' coding capabilities are beyond all but the very best human programmers, so more and more construction of machine learning architecture is being outsourced to AI.  While it has always been rather difficult to reproduce a group's machine learning claims (do you have exactly the same machine? did you modify the loss function slightly?  did they forget to mention how they modified the activation function?), this is becoming more of an issue and the barrier to performing machine learning analyses is dropping lower and lower.

It is therefore only becoming more important than ever to have human-interpretable, validated ground truth that can be used to justify claims from machine learning studies.  In these notes and in Ref.~\cite{Larkoski:2024uoc}, I have attempted to share almost everything that I know about this theory/machine learning boundary, its established lore, its foundational laws, and its open problems.  I hope you can help solve some of them, and find new directions to explore.

\section*{Acknowledgements}

I thank Fran\c cois Arleo and Paul Caucal for the invitation to return to Saint-Jacut-de-la-Mer and to all of the students for their comments, questions, and curiosity throughout the school.  I thank Nathaniel Craig, Rikab Gambhir, Yoni Kahn, Simone Marzani, Tom Melia, and Jesse Thaler for collaboration and discussion on many of the topics presented here.

\bibliography{refs}

\begin{thebibliography}{206}%
\makeatletter
\providecommand \@ifxundefined [1]{%
 \@ifx{#1\undefined}
}%
\providecommand \@ifnum [1]{%
 \ifnum #1\expandafter \@firstoftwo
 \else \expandafter \@secondoftwo
 \fi
}%
\providecommand \@ifx [1]{%
 \ifx #1\expandafter \@firstoftwo
 \else \expandafter \@secondoftwo
 \fi
}%
\providecommand \natexlab [1]{#1}%
\providecommand \enquote  [1]{``#1''}%
\providecommand \bibnamefont  [1]{#1}%
\providecommand \bibfnamefont [1]{#1}%
\providecommand \citenamefont [1]{#1}%
\providecommand \href@noop [0]{\@secondoftwo}%
\providecommand \href [0]{\begingroup \@sanitize@url \@href}%
\providecommand \@href[1]{\@@startlink{#1}\@@href}%
\providecommand \@@href[1]{\endgroup#1\@@endlink}%
\providecommand \@sanitize@url [0]{\catcode `\\12\catcode `\$12\catcode
  `\&12\catcode `\#12\catcode `\^12\catcode `\_12\catcode `\%12\relax}%
\providecommand \@@startlink[1]{}%
\providecommand \@@endlink[0]{}%
\providecommand \url  [0]{\begingroup\@sanitize@url \@url }%
\providecommand \@url [1]{\endgroup\@href {#1}{\urlprefix }}%
\providecommand \urlprefix  [0]{URL }%
\providecommand \Eprint [0]{\href }%
\providecommand \doibase [0]{https://doi.org/}%
\providecommand \selectlanguage [0]{\@gobble}%
\providecommand \bibinfo  [0]{\@secondoftwo}%
\providecommand \bibfield  [0]{\@secondoftwo}%
\providecommand \translation [1]{[#1]}%
\providecommand \BibitemOpen [0]{}%
\providecommand \bibitemStop [0]{}%
\providecommand \bibitemNoStop [0]{.\EOS\space}%
\providecommand \EOS [0]{\spacefactor3000\relax}%
\providecommand \BibitemShut  [1]{\csname bibitem#1\endcsname}%
\let\auto@bib@innerbib\@empty
\bibitem [{\citenamefont {Larkoski}(2024{\natexlab{a}})}]{Larkoski:2024uoc}%
  \BibitemOpen
  \bibfield  {author} {\bibinfo {author} {\bibfnamefont {A.~J.}\ \bibnamefont
  {Larkoski}},\ }\bibfield  {title} {\bibinfo {title} {{QCD masterclass
  lectures on jet physics and machine learning}},\ }\href
  {https://doi.org/10.1140/epjc/s10052-024-13341-0} {\bibfield  {journal}
  {\bibinfo  {journal} {Eur. Phys. J. C}\ }\textbf {\bibinfo {volume} {84}},\
  \bibinfo {pages} {1117} (\bibinfo {year} {2024}{\natexlab{a}})},\ \Eprint
  {https://arxiv.org/abs/2407.04897} {arXiv:2407.04897 [hep-ph]} \BibitemShut
  {NoStop}%
\bibitem [{\citenamefont {Guevara}\ \emph {et~al.}(2026)\citenamefont
  {Guevara}, \citenamefont {Lupsasca}, \citenamefont {Skinner}, \citenamefont
  {Strominger},\ and\ \citenamefont {Weil}}]{Guevara:2026qzd}%
  \BibitemOpen
  \bibfield  {author} {\bibinfo {author} {\bibfnamefont {A.}~\bibnamefont
  {Guevara}}, \bibinfo {author} {\bibfnamefont {A.}~\bibnamefont {Lupsasca}},
  \bibinfo {author} {\bibfnamefont {D.}~\bibnamefont {Skinner}}, \bibinfo
  {author} {\bibfnamefont {A.}~\bibnamefont {Strominger}},\ and\ \bibinfo
  {author} {\bibfnamefont {K.}~\bibnamefont {Weil}},\ }\bibfield  {title}
  {\bibinfo {title} {{Single-minus gluon tree amplitudes are nonzero}},\
  }\href@noop {} {\  (\bibinfo {year} {2026})},\ \Eprint
  {https://arxiv.org/abs/2602.12176} {arXiv:2602.12176 [hep-th]} \BibitemShut
  {NoStop}%
\bibitem [{\citenamefont {Schwartz}(2026)}]{Schwartz:2026ekw}%
  \BibitemOpen
  \bibfield  {author} {\bibinfo {author} {\bibfnamefont {M.~D.}\ \bibnamefont
  {Schwartz}},\ }\bibfield  {title} {\bibinfo {title} {{Resummation of the
  C-Parameter Sudakov Shoulder Using Effective Field Theory}},\ }\href@noop {}
  {\  (\bibinfo {year} {2026})},\ \Eprint {https://arxiv.org/abs/2601.02484}
  {arXiv:2601.02484 [hep-ph]} \BibitemShut {NoStop}%
\bibitem [{\citenamefont {Shih}(2026)}]{Shih:2026lmy}%
  \BibitemOpen
  \bibfield  {author} {\bibinfo {author} {\bibfnamefont {D.}~\bibnamefont
  {Shih}},\ }\bibfield  {title} {\bibinfo {title} {{Learning to Unscramble:
  Simplifying Symbolic Expressions via Self-Supervised Oracle Trajectories}},\
  }\href@noop {} {\  (\bibinfo {year} {2026})},\ \Eprint
  {https://arxiv.org/abs/2603.11164} {arXiv:2603.11164 [hep-th]} \BibitemShut
  {NoStop}%
\bibitem [{\citenamefont {Moreno}\ \emph {et~al.}(2026)\citenamefont {Moreno},
  \citenamefont {Bright-Thonney}, \citenamefont {Novak}, \citenamefont
  {Garcia},\ and\ \citenamefont {Harris}}]{Moreno:2026mqk}%
  \BibitemOpen
  \bibfield  {author} {\bibinfo {author} {\bibfnamefont {E.~A.}\ \bibnamefont
  {Moreno}}, \bibinfo {author} {\bibfnamefont {S.}~\bibnamefont
  {Bright-Thonney}}, \bibinfo {author} {\bibfnamefont {A.}~\bibnamefont
  {Novak}}, \bibinfo {author} {\bibfnamefont {D.}~\bibnamefont {Garcia}},\ and\
  \bibinfo {author} {\bibfnamefont {P.}~\bibnamefont {Harris}},\ }\bibfield
  {title} {\bibinfo {title} {{AI Agents Can Already Autonomously Perform
  Experimental High Energy Physics}},\ }\href@noop {} {\  (\bibinfo {year}
  {2026})},\ \Eprint {https://arxiv.org/abs/2603.20179} {arXiv:2603.20179
  [hep-ex]} \BibitemShut {NoStop}%
\bibitem [{\citenamefont {Sohl-Dickstein}()}]{sohlant}%
  \BibitemOpen
  \bibfield  {author} {\bibinfo {author} {\bibfnamefont {J.}~\bibnamefont
  {Sohl-Dickstein}},\ }\href@noop {} {\bibinfo {title} {Advice for a (young)
  investigator in the first and last days of the anthropocene}},\ \bibinfo
  {howpublished}
  {\url{https://drive.google.com/file/d/1Z15cWqTkK2dWIb1vOK-bDgLx5RnTv4fI}}\BibitemShut
  {NoStop}%
\bibitem [{\citenamefont {Brown}()}]{brownsand}%
  \BibitemOpen
  \bibfield  {author} {\bibinfo {author} {\bibfnamefont {A.}~\bibnamefont
  {Brown}},\ }\href@noop {} {\bibinfo {title} {Training sand to think:
  Artificial general intelligence \& future of physics}},\ \bibinfo
  {howpublished}
  {\url{https://www.youtube.com/watch?v=Mw60FH5iflI}}\BibitemShut {NoStop}%
\bibitem [{\citenamefont {Schwartz}()}]{schwartz_anthro}%
  \BibitemOpen
  \bibfield  {author} {\bibinfo {author} {\bibfnamefont {M.}~\bibnamefont
  {Schwartz}},\ }\href@noop {} {\bibinfo {title} {Vibe physics: The ai grad
  student}},\ \bibinfo {howpublished}
  {\url{https://www.anthropic.com/research/vibe-physics}}\BibitemShut {NoStop}%
\bibitem [{\citenamefont {OpenAI}({\natexlab{a}})}]{openairesearch}%
  \BibitemOpen
  \bibfield  {author} {\bibinfo {author} {\bibnamefont {OpenAI}},\ }\href@noop
  {} {\bibinfo {title} {Accelerating scientific discovery with chatgpt for
  academic researchers}},\ \bibinfo {howpublished}
  {\url{https://openai.com/index/chatgpt-for-academic-researchers/}}
  ({\natexlab{a}})\BibitemShut {NoStop}%
\bibitem [{\citenamefont {OpenAI}({\natexlab{b}})}]{openainavier}%
  \BibitemOpen
  \bibfield  {author} {\bibinfo {author} {\bibnamefont {OpenAI}},\ }\href@noop
  {} {\bibinfo {title} {{On the Navier–Stokes Millennium Prize Problem}}},\
  \bibinfo {howpublished}
  {\url{https://openai.com/index/navier-stokes-solution/}}
  ({\natexlab{b}})\BibitemShut {NoStop}%
\bibitem [{\citenamefont {Buckmaster}()}]{buckmaster_state}%
  \BibitemOpen
  \bibfield  {author} {\bibinfo {author} {\bibfnamefont {T.}~\bibnamefont
  {Buckmaster}},\ }\href@noop {} {\bibinfo {title} {{Statement}}},\ \bibinfo
  {howpublished}
  {\url{https://cims.nyu.edu/~tristanb/statement.pdf}}\BibitemShut {NoStop}%
\bibitem [{\citenamefont {Alper}\ \emph {et~al.}(2026)\citenamefont {Alper},
  \citenamefont {Barany}, \citenamefont {Chavarri~Villarello}, \citenamefont
  {Dahmen}, \citenamefont {Dean}, \citenamefont {Ganapathy}, \citenamefont
  {Harris}, \citenamefont {Holmes}, \citenamefont {Jamnik}, \citenamefont
  {Kelk}, \citenamefont {Kra}, \citenamefont {Martin}, \citenamefont
  {Naskrecki}, \citenamefont {Ochigame}, \citenamefont {Portegies},\ and\
  \citenamefont {Schmitt}}]{alper_2026_20302944}%
  \BibitemOpen
  \bibfield  {author} {\bibinfo {author} {\bibfnamefont {J.}~\bibnamefont
  {Alper}}, \bibinfo {author} {\bibfnamefont {M.}~\bibnamefont {Barany}},
  \bibinfo {author} {\bibfnamefont {A.}~\bibnamefont {Chavarri~Villarello}},
  \bibinfo {author} {\bibfnamefont {S.}~\bibnamefont {Dahmen}}, \bibinfo
  {author} {\bibfnamefont {W.}~\bibnamefont {Dean}}, \bibinfo {author}
  {\bibfnamefont {K.}~\bibnamefont {Ganapathy}}, \bibinfo {author}
  {\bibfnamefont {M.}~\bibnamefont {Harris}}, \bibinfo {author} {\bibfnamefont
  {D.}~\bibnamefont {Holmes}}, \bibinfo {author} {\bibfnamefont
  {M.}~\bibnamefont {Jamnik}}, \bibinfo {author} {\bibfnamefont
  {S.}~\bibnamefont {Kelk}}, \bibinfo {author} {\bibfnamefont {B.}~\bibnamefont
  {Kra}}, \bibinfo {author} {\bibfnamefont {U.}~\bibnamefont {Martin}},
  \bibinfo {author} {\bibfnamefont {B.}~\bibnamefont {Naskrecki}}, \bibinfo
  {author} {\bibfnamefont {R.}~\bibnamefont {Ochigame}}, \bibinfo {author}
  {\bibfnamefont {J.}~\bibnamefont {Portegies}},\ and\ \bibinfo {author}
  {\bibfnamefont {J.}~\bibnamefont {Schmitt}},\ }\href
  {https://doi.org/10.5281/zenodo.20302944} {\bibinfo {title} {Leiden
  declaration on artificial intelligence and mathematics}} (\bibinfo {year}
  {2026})\BibitemShut {NoStop}%
\bibitem [{\citenamefont {Gowers}()}]{gowersleiden}%
  \BibitemOpen
  \bibfield  {author} {\bibinfo {author} {\bibfnamefont {T.}~\bibnamefont
  {Gowers}},\ }\href@noop {} {\bibinfo {title} {{Thoughts about the Leiden
  Declaration}}},\ \bibinfo {howpublished}
  {\url{https://gowers.wordpress.com/2026/07/26/thoughts-about-the-leiden-declaration/}}\BibitemShut
  {NoStop}%
\bibitem [{\citenamefont {Tao}()}]{taomathinai}%
  \BibitemOpen
  \bibfield  {author} {\bibinfo {author} {\bibfnamefont {T.}~\bibnamefont
  {Tao}},\ }\href@noop {} {\bibinfo {title} {{Mathematics in the age of AI}}},\
  \bibinfo {howpublished}
  {\url{https://teorth.github.io/tao-web/slides/age-of-ai-icm-2026.pdf}}\BibitemShut
  {NoStop}%
\bibitem [{\citenamefont {Neyman}\ and\ \citenamefont
  {Pearson}(1933)}]{Neyman:1933wgr}%
  \BibitemOpen
  \bibfield  {author} {\bibinfo {author} {\bibfnamefont {J.}~\bibnamefont
  {Neyman}}\ and\ \bibinfo {author} {\bibfnamefont {E.~S.}\ \bibnamefont
  {Pearson}},\ }\bibfield  {title} {\bibinfo {title} {{On the Problem of the
  Most Efficient Tests of Statistical Hypotheses}},\ }\href
  {https://doi.org/10.1098/rsta.1933.0009} {\bibfield  {journal} {\bibinfo
  {journal} {Phil. Trans. Roy. Soc. Lond. A}\ }\textbf {\bibinfo {volume}
  {231}},\ \bibinfo {pages} {289} (\bibinfo {year} {1933})}\BibitemShut
  {NoStop}%
\bibitem [{\citenamefont {Kaplan}\ \emph {et~al.}(2020)\citenamefont {Kaplan},
  \citenamefont {McCandlish}, \citenamefont {Henighan}, \citenamefont {Brown},
  \citenamefont {Chess}, \citenamefont {Child}, \citenamefont {Gray},
  \citenamefont {Radford}, \citenamefont {Wu},\ and\ \citenamefont
  {Amodei}}]{kaplan2020scaling}%
  \BibitemOpen
  \bibfield  {author} {\bibinfo {author} {\bibfnamefont {J.}~\bibnamefont
  {Kaplan}}, \bibinfo {author} {\bibfnamefont {S.}~\bibnamefont {McCandlish}},
  \bibinfo {author} {\bibfnamefont {T.}~\bibnamefont {Henighan}}, \bibinfo
  {author} {\bibfnamefont {T.~B.}\ \bibnamefont {Brown}}, \bibinfo {author}
  {\bibfnamefont {B.}~\bibnamefont {Chess}}, \bibinfo {author} {\bibfnamefont
  {R.}~\bibnamefont {Child}}, \bibinfo {author} {\bibfnamefont
  {S.}~\bibnamefont {Gray}}, \bibinfo {author} {\bibfnamefont {A.}~\bibnamefont
  {Radford}}, \bibinfo {author} {\bibfnamefont {J.}~\bibnamefont {Wu}},\ and\
  \bibinfo {author} {\bibfnamefont {D.}~\bibnamefont {Amodei}},\ }\bibfield
  {title} {\bibinfo {title} {Scaling laws for neural language models},\
  }\href@noop {} {\bibfield  {journal} {\bibinfo  {journal} {arXiv preprint
  arXiv:2001.08361}\ } (\bibinfo {year} {2020})}\BibitemShut {NoStop}%
\bibitem [{\citenamefont {Seymour}(1994)}]{Seymour:1993mx}%
  \BibitemOpen
  \bibfield  {author} {\bibinfo {author} {\bibfnamefont {M.~H.}\ \bibnamefont
  {Seymour}},\ }\bibfield  {title} {\bibinfo {title} {{Searches for new
  particles using cone and cluster jet algorithms: A Comparative study}},\
  }\href {https://doi.org/10.1007/BF01559532} {\bibfield  {journal} {\bibinfo
  {journal} {Z. Phys. C}\ }\textbf {\bibinfo {volume} {62}},\ \bibinfo {pages}
  {127} (\bibinfo {year} {1994})}\BibitemShut {NoStop}%
\bibitem [{\citenamefont {Geuskens}\ \emph {et~al.}(2025)\citenamefont
  {Geuskens}, \citenamefont {Gite}, \citenamefont {Kr{\"a}mer}, \citenamefont
  {Mikuni}, \citenamefont {M{\"u}ck}, \citenamefont {Nachman},\ and\
  \citenamefont {Reyes-Gonz{\'a}lez}}]{Geuskens:2024tfo}%
  \BibitemOpen
  \bibfield  {author} {\bibinfo {author} {\bibfnamefont {J.}~\bibnamefont
  {Geuskens}}, \bibinfo {author} {\bibfnamefont {N.}~\bibnamefont {Gite}},
  \bibinfo {author} {\bibfnamefont {M.}~\bibnamefont {Kr{\"a}mer}}, \bibinfo
  {author} {\bibfnamefont {V.}~\bibnamefont {Mikuni}}, \bibinfo {author}
  {\bibfnamefont {A.}~\bibnamefont {M{\"u}ck}}, \bibinfo {author}
  {\bibfnamefont {B.}~\bibnamefont {Nachman}},\ and\ \bibinfo {author}
  {\bibfnamefont {H.}~\bibnamefont {Reyes-Gonz{\'a}lez}},\ }\bibfield  {title}
  {\bibinfo {title} {{Fundamental limit of jet tagging}},\ }\href
  {https://doi.org/10.1103/mjj1-w2b1} {\bibfield  {journal} {\bibinfo
  {journal} {Phys. Rev. D}\ }\textbf {\bibinfo {volume} {112}},\ \bibinfo
  {pages} {L091901} (\bibinfo {year} {2025})},\ \Eprint
  {https://arxiv.org/abs/2411.02628} {arXiv:2411.02628 [hep-ph]} \BibitemShut
  {NoStop}%
\bibitem [{\citenamefont {Pang}\ \emph {et~al.}(2025)\citenamefont {Pang},
  \citenamefont {Faroughy}, \citenamefont {Shih}, \citenamefont {Das},\ and\
  \citenamefont {Kasieczka}}]{Pang:2025lbs}%
  \BibitemOpen
  \bibfield  {author} {\bibinfo {author} {\bibfnamefont {I.}~\bibnamefont
  {Pang}}, \bibinfo {author} {\bibfnamefont {D.~A.}\ \bibnamefont {Faroughy}},
  \bibinfo {author} {\bibfnamefont {D.}~\bibnamefont {Shih}}, \bibinfo {author}
  {\bibfnamefont {R.}~\bibnamefont {Das}},\ and\ \bibinfo {author}
  {\bibfnamefont {G.}~\bibnamefont {Kasieczka}},\ }\bibfield  {title} {\bibinfo
  {title} {{SURFing to the Fundamental Limit of Jet Tagging}},\ }\href@noop {}
  {\  (\bibinfo {year} {2025})},\ \Eprint {https://arxiv.org/abs/2511.15779}
  {arXiv:2511.15779 [hep-ph]} \BibitemShut {NoStop}%
\bibitem [{\citenamefont {Rosenblatt}(1956)}]{rosenblatt1956remarks}%
  \BibitemOpen
  \bibfield  {author} {\bibinfo {author} {\bibfnamefont {M.}~\bibnamefont
  {Rosenblatt}},\ }\bibfield  {title} {\bibinfo {title} {Remarks on some
  nonparametric estimates of a density function},\ }\href@noop {} {\bibfield
  {journal} {\bibinfo  {journal} {The Annals of Mathematical Statistics}\
  }\textbf {\bibinfo {volume} {27}},\ \bibinfo {pages} {832} (\bibinfo {year}
  {1956})}\BibitemShut {NoStop}%
\bibitem [{\citenamefont {Parzen}(1962)}]{parzen1962estimation}%
  \BibitemOpen
  \bibfield  {author} {\bibinfo {author} {\bibfnamefont {E.}~\bibnamefont
  {Parzen}},\ }\bibfield  {title} {\bibinfo {title} {On estimation of a
  probability density function and mode},\ }\href@noop {} {\bibfield  {journal}
  {\bibinfo  {journal} {The annals of mathematical statistics}\ }\textbf
  {\bibinfo {volume} {33}},\ \bibinfo {pages} {1065} (\bibinfo {year}
  {1962})}\BibitemShut {NoStop}%
\bibitem [{\citenamefont {Komiske}\ \emph {et~al.}(2019)\citenamefont
  {Komiske}, \citenamefont {Metodiev},\ and\ \citenamefont
  {Thaler}}]{Komiske:2019fks}%
  \BibitemOpen
  \bibfield  {author} {\bibinfo {author} {\bibfnamefont {P.~T.}\ \bibnamefont
  {Komiske}}, \bibinfo {author} {\bibfnamefont {E.~M.}\ \bibnamefont
  {Metodiev}},\ and\ \bibinfo {author} {\bibfnamefont {J.}~\bibnamefont
  {Thaler}},\ }\bibfield  {title} {\bibinfo {title} {{Metric Space of Collider
  Events}},\ }\href {https://doi.org/10.1103/PhysRevLett.123.041801} {\bibfield
   {journal} {\bibinfo  {journal} {Phys. Rev. Lett.}\ }\textbf {\bibinfo
  {volume} {123}},\ \bibinfo {pages} {041801} (\bibinfo {year} {2019})},\
  \Eprint {https://arxiv.org/abs/1902.02346} {arXiv:1902.02346 [hep-ph]}
  \BibitemShut {NoStop}%
\bibitem [{\citenamefont {Mullin}\ \emph {et~al.}(2021)\citenamefont {Mullin},
  \citenamefont {Nicholls}, \citenamefont {Pacey}, \citenamefont {Parker},
  \citenamefont {White},\ and\ \citenamefont {Williams}}]{Mullin:2019mmh}%
  \BibitemOpen
  \bibfield  {author} {\bibinfo {author} {\bibfnamefont {A.}~\bibnamefont
  {Mullin}}, \bibinfo {author} {\bibfnamefont {S.}~\bibnamefont {Nicholls}},
  \bibinfo {author} {\bibfnamefont {H.}~\bibnamefont {Pacey}}, \bibinfo
  {author} {\bibfnamefont {M.}~\bibnamefont {Parker}}, \bibinfo {author}
  {\bibfnamefont {M.}~\bibnamefont {White}},\ and\ \bibinfo {author}
  {\bibfnamefont {S.}~\bibnamefont {Williams}},\ }\bibfield  {title} {\bibinfo
  {title} {{Does SUSY have friends? A new approach for LHC event analysis}},\
  }\href {https://doi.org/10.1007/JHEP02(2021)160} {\bibfield  {journal}
  {\bibinfo  {journal} {JHEP}\ }\textbf {\bibinfo {volume} {02}},\ \bibinfo
  {pages} {160}},\ \Eprint {https://arxiv.org/abs/1912.10625} {arXiv:1912.10625
  [hep-ph]} \BibitemShut {NoStop}%
\bibitem [{\citenamefont {Komiske}\ \emph
  {et~al.}(2020{\natexlab{a}})\citenamefont {Komiske}, \citenamefont
  {Metodiev},\ and\ \citenamefont {Thaler}}]{Komiske:2020qhg}%
  \BibitemOpen
  \bibfield  {author} {\bibinfo {author} {\bibfnamefont {P.~T.}\ \bibnamefont
  {Komiske}}, \bibinfo {author} {\bibfnamefont {E.~M.}\ \bibnamefont
  {Metodiev}},\ and\ \bibinfo {author} {\bibfnamefont {J.}~\bibnamefont
  {Thaler}},\ }\bibfield  {title} {\bibinfo {title} {{The Hidden Geometry of
  Particle Collisions}},\ }\href {https://doi.org/10.1007/JHEP07(2020)006}
  {\bibfield  {journal} {\bibinfo  {journal} {JHEP}\ }\textbf {\bibinfo
  {volume} {07}},\ \bibinfo {pages} {006}},\ \Eprint
  {https://arxiv.org/abs/2004.04159} {arXiv:2004.04159 [hep-ph]} \BibitemShut
  {NoStop}%
\bibitem [{\citenamefont {Crispim~Rom{\~a}o}\ \emph {et~al.}(2021)\citenamefont
  {Crispim~Rom{\~a}o}, \citenamefont {Castro}, \citenamefont {Milhano},
  \citenamefont {Pedro},\ and\ \citenamefont {Vale}}]{CrispimRomao:2020ejk}%
  \BibitemOpen
  \bibfield  {author} {\bibinfo {author} {\bibfnamefont {M.}~\bibnamefont
  {Crispim~Rom{\~a}o}}, \bibinfo {author} {\bibfnamefont {N.~F.}\ \bibnamefont
  {Castro}}, \bibinfo {author} {\bibfnamefont {J.~G.}\ \bibnamefont {Milhano}},
  \bibinfo {author} {\bibfnamefont {R.}~\bibnamefont {Pedro}},\ and\ \bibinfo
  {author} {\bibfnamefont {T.}~\bibnamefont {Vale}},\ }\bibfield  {title}
  {\bibinfo {title} {{Use of a generalized energy Mover{\textquoteright}s
  distance in the search for rare phenomena at colliders}},\ }\href
  {https://doi.org/10.1140/epjc/s10052-021-08891-6} {\bibfield  {journal}
  {\bibinfo  {journal} {Eur. Phys. J. C}\ }\textbf {\bibinfo {volume} {81}},\
  \bibinfo {pages} {192} (\bibinfo {year} {2021})},\ \Eprint
  {https://arxiv.org/abs/2004.09360} {arXiv:2004.09360 [hep-ph]} \BibitemShut
  {NoStop}%
\bibitem [{\citenamefont {Cai}\ \emph {et~al.}(2020)\citenamefont {Cai},
  \citenamefont {Cheng}, \citenamefont {Craig},\ and\ \citenamefont
  {Craig}}]{Cai:2020vzx}%
  \BibitemOpen
  \bibfield  {author} {\bibinfo {author} {\bibfnamefont {T.}~\bibnamefont
  {Cai}}, \bibinfo {author} {\bibfnamefont {J.}~\bibnamefont {Cheng}}, \bibinfo
  {author} {\bibfnamefont {N.}~\bibnamefont {Craig}},\ and\ \bibinfo {author}
  {\bibfnamefont {K.}~\bibnamefont {Craig}},\ }\bibfield  {title} {\bibinfo
  {title} {{Linearized optimal transport for collider events}},\ }\href
  {https://doi.org/10.1103/PhysRevD.102.116019} {\bibfield  {journal} {\bibinfo
   {journal} {Phys. Rev. D}\ }\textbf {\bibinfo {volume} {102}},\ \bibinfo
  {pages} {116019} (\bibinfo {year} {2020})},\ \Eprint
  {https://arxiv.org/abs/2008.08604} {arXiv:2008.08604 [hep-ph]} \BibitemShut
  {NoStop}%
\bibitem [{\citenamefont {Larkoski}\ and\ \citenamefont
  {Melia}(2020)}]{Larkoski:2020thc}%
  \BibitemOpen
  \bibfield  {author} {\bibinfo {author} {\bibfnamefont {A.~J.}\ \bibnamefont
  {Larkoski}}\ and\ \bibinfo {author} {\bibfnamefont {T.}~\bibnamefont
  {Melia}},\ }\bibfield  {title} {\bibinfo {title} {{Covariantizing phase
  space}},\ }\href {https://doi.org/10.1103/PhysRevD.102.094014} {\bibfield
  {journal} {\bibinfo  {journal} {Phys. Rev. D}\ }\textbf {\bibinfo {volume}
  {102}},\ \bibinfo {pages} {094014} (\bibinfo {year} {2020})},\ \Eprint
  {https://arxiv.org/abs/2008.06508} {arXiv:2008.06508 [hep-ph]} \BibitemShut
  {NoStop}%
\bibitem [{\citenamefont {Tsan}\ \emph {et~al.}(2021)\citenamefont {Tsan},
  \citenamefont {Kansal}, \citenamefont {Aportela}, \citenamefont {Diaz},
  \citenamefont {Duarte}, \citenamefont {Krishna}, \citenamefont {Mokhtar},
  \citenamefont {Vlimant},\ and\ \citenamefont {Pierini}}]{Tsan:2021brw}%
  \BibitemOpen
  \bibfield  {author} {\bibinfo {author} {\bibfnamefont {S.}~\bibnamefont
  {Tsan}}, \bibinfo {author} {\bibfnamefont {R.}~\bibnamefont {Kansal}},
  \bibinfo {author} {\bibfnamefont {A.}~\bibnamefont {Aportela}}, \bibinfo
  {author} {\bibfnamefont {D.}~\bibnamefont {Diaz}}, \bibinfo {author}
  {\bibfnamefont {J.}~\bibnamefont {Duarte}}, \bibinfo {author} {\bibfnamefont
  {S.}~\bibnamefont {Krishna}}, \bibinfo {author} {\bibfnamefont
  {F.}~\bibnamefont {Mokhtar}}, \bibinfo {author} {\bibfnamefont {J.-R.}\
  \bibnamefont {Vlimant}},\ and\ \bibinfo {author} {\bibfnamefont
  {M.}~\bibnamefont {Pierini}},\ }\bibfield  {title} {\bibinfo {title}
  {{Particle Graph Autoencoders and Differentiable, Learned Energy Mover's
  Distance}},\ }in\ \href@noop {} {\emph {\bibinfo {booktitle} {{35th
  Conference on Neural Information Processing Systems}}}}\ (\bibinfo {year}
  {2021})\ \Eprint {https://arxiv.org/abs/2111.12849} {arXiv:2111.12849
  [physics.data-an]} \BibitemShut {NoStop}%
\bibitem [{\citenamefont {Cai}\ \emph {et~al.}(2022)\citenamefont {Cai},
  \citenamefont {Cheng}, \citenamefont {Craig},\ and\ \citenamefont
  {Craig}}]{Cai:2021hnn}%
  \BibitemOpen
  \bibfield  {author} {\bibinfo {author} {\bibfnamefont {T.}~\bibnamefont
  {Cai}}, \bibinfo {author} {\bibfnamefont {J.}~\bibnamefont {Cheng}}, \bibinfo
  {author} {\bibfnamefont {K.}~\bibnamefont {Craig}},\ and\ \bibinfo {author}
  {\bibfnamefont {N.}~\bibnamefont {Craig}},\ }\bibfield  {title} {\bibinfo
  {title} {{Which metric on the space of collider events?}},\ }\href
  {https://doi.org/10.1103/PhysRevD.105.076003} {\bibfield  {journal} {\bibinfo
   {journal} {Phys. Rev. D}\ }\textbf {\bibinfo {volume} {105}},\ \bibinfo
  {pages} {076003} (\bibinfo {year} {2022})},\ \Eprint
  {https://arxiv.org/abs/2111.03670} {arXiv:2111.03670 [hep-ph]} \BibitemShut
  {NoStop}%
\bibitem [{\citenamefont {Andersen}\ and\ \citenamefont
  {Maier}(2022)}]{Andersen:2021mvw}%
  \BibitemOpen
  \bibfield  {author} {\bibinfo {author} {\bibfnamefont {J.~R.}\ \bibnamefont
  {Andersen}}\ and\ \bibinfo {author} {\bibfnamefont {A.}~\bibnamefont
  {Maier}},\ }\bibfield  {title} {\bibinfo {title} {{Unbiased elimination of
  negative weights in Monte Carlo samples}},\ }\href
  {https://doi.org/10.1140/epjc/s10052-022-10372-3} {\bibfield  {journal}
  {\bibinfo  {journal} {Eur. Phys. J. C}\ }\textbf {\bibinfo {volume} {82}},\
  \bibinfo {pages} {433} (\bibinfo {year} {2022})},\ \Eprint
  {https://arxiv.org/abs/2109.07851} {arXiv:2109.07851 [hep-ph]} \BibitemShut
  {NoStop}%
\bibitem [{\citenamefont {Kitouni}\ \emph {et~al.}(2022)\citenamefont
  {Kitouni}, \citenamefont {Nolte},\ and\ \citenamefont
  {Williams}}]{Kitouni:2022qyr}%
  \BibitemOpen
  \bibfield  {author} {\bibinfo {author} {\bibfnamefont {O.}~\bibnamefont
  {Kitouni}}, \bibinfo {author} {\bibfnamefont {N.}~\bibnamefont {Nolte}},\
  and\ \bibinfo {author} {\bibfnamefont {M.}~\bibnamefont {Williams}},\
  }\bibfield  {title} {\bibinfo {title} {{Finding NEEMo: Geometric Fitting
  using Neural Estimation of the Energy Mover's Distance}},\ }\href@noop {} {\
  (\bibinfo {year} {2022})},\ \Eprint {https://arxiv.org/abs/2209.15624}
  {arXiv:2209.15624 [stat.ML]} \BibitemShut {NoStop}%
\bibitem [{\citenamefont {Alipour-Fard}\ \emph {et~al.}(2023)\citenamefont
  {Alipour-Fard}, \citenamefont {Komiske}, \citenamefont {Metodiev},\ and\
  \citenamefont {Thaler}}]{Alipour-Fard:2023prj}%
  \BibitemOpen
  \bibfield  {author} {\bibinfo {author} {\bibfnamefont {S.}~\bibnamefont
  {Alipour-Fard}}, \bibinfo {author} {\bibfnamefont {P.~T.}\ \bibnamefont
  {Komiske}}, \bibinfo {author} {\bibfnamefont {E.~M.}\ \bibnamefont
  {Metodiev}},\ and\ \bibinfo {author} {\bibfnamefont {J.}~\bibnamefont
  {Thaler}},\ }\bibfield  {title} {\bibinfo {title} {{Pileup and Infrared
  Radiation Annihilation (PIRANHA): a paradigm for continuous jet grooming}},\
  }\href {https://doi.org/10.1007/JHEP09(2023)157} {\bibfield  {journal}
  {\bibinfo  {journal} {JHEP}\ }\textbf {\bibinfo {volume} {09}},\ \bibinfo
  {pages} {157}},\ \Eprint {https://arxiv.org/abs/2305.00989} {arXiv:2305.00989
  [hep-ph]} \BibitemShut {NoStop}%
\bibitem [{\citenamefont {Larkoski}\ and\ \citenamefont
  {Thaler}(2023)}]{Larkoski:2023qnv}%
  \BibitemOpen
  \bibfield  {author} {\bibinfo {author} {\bibfnamefont {A.~J.}\ \bibnamefont
  {Larkoski}}\ and\ \bibinfo {author} {\bibfnamefont {J.}~\bibnamefont
  {Thaler}},\ }\bibfield  {title} {\bibinfo {title} {{A spectral metric for
  collider geometry}},\ }\href {https://doi.org/10.1007/JHEP08(2023)107}
  {\bibfield  {journal} {\bibinfo  {journal} {JHEP}\ }\textbf {\bibinfo
  {volume} {08}},\ \bibinfo {pages} {107}},\ \Eprint
  {https://arxiv.org/abs/2305.03751} {arXiv:2305.03751 [hep-ph]} \BibitemShut
  {NoStop}%
\bibitem [{\citenamefont {Davis}\ \emph {et~al.}(2023)\citenamefont {Davis},
  \citenamefont {Menzo}, \citenamefont {Youssef},\ and\ \citenamefont
  {Zupan}}]{Davis:2023lxq}%
  \BibitemOpen
  \bibfield  {author} {\bibinfo {author} {\bibfnamefont {A.}~\bibnamefont
  {Davis}}, \bibinfo {author} {\bibfnamefont {T.}~\bibnamefont {Menzo}},
  \bibinfo {author} {\bibfnamefont {A.}~\bibnamefont {Youssef}},\ and\ \bibinfo
  {author} {\bibfnamefont {J.}~\bibnamefont {Zupan}},\ }\bibfield  {title}
  {\bibinfo {title} {{Earth mover{\textquoteright}s distance as a measure of CP
  violation}},\ }\href {https://doi.org/10.1007/JHEP06(2023)098} {\bibfield
  {journal} {\bibinfo  {journal} {JHEP}\ }\textbf {\bibinfo {volume} {06}},\
  \bibinfo {pages} {098}},\ \Eprint {https://arxiv.org/abs/2301.13211}
  {arXiv:2301.13211 [hep-ph]} \BibitemShut {NoStop}%
\bibitem [{\citenamefont {Ba}\ \emph {et~al.}(2023)\citenamefont {Ba},
  \citenamefont {Dogra}, \citenamefont {Gambhir}, \citenamefont {Tasissa},\
  and\ \citenamefont {Thaler}}]{Ba:2023hix}%
  \BibitemOpen
  \bibfield  {author} {\bibinfo {author} {\bibfnamefont {D.}~\bibnamefont
  {Ba}}, \bibinfo {author} {\bibfnamefont {A.~S.}\ \bibnamefont {Dogra}},
  \bibinfo {author} {\bibfnamefont {R.}~\bibnamefont {Gambhir}}, \bibinfo
  {author} {\bibfnamefont {A.}~\bibnamefont {Tasissa}},\ and\ \bibinfo {author}
  {\bibfnamefont {J.}~\bibnamefont {Thaler}},\ }\bibfield  {title} {\bibinfo
  {title} {{SHAPER: can you hear the shape of a jet?}},\ }\href
  {https://doi.org/10.1007/JHEP06(2023)195} {\bibfield  {journal} {\bibinfo
  {journal} {JHEP}\ }\textbf {\bibinfo {volume} {06}},\ \bibinfo {pages}
  {195}},\ \Eprint {https://arxiv.org/abs/2302.12266} {arXiv:2302.12266
  [hep-ph]} \BibitemShut {NoStop}%
\bibitem [{\citenamefont {Craig}\ \emph {et~al.}(2024)\citenamefont {Craig},
  \citenamefont {Howard},\ and\ \citenamefont {Li}}]{Craig:2024rlv}%
  \BibitemOpen
  \bibfield  {author} {\bibinfo {author} {\bibfnamefont {N.}~\bibnamefont
  {Craig}}, \bibinfo {author} {\bibfnamefont {J.~N.}\ \bibnamefont {Howard}},\
  and\ \bibinfo {author} {\bibfnamefont {H.}~\bibnamefont {Li}},\ }\bibfield
  {title} {\bibinfo {title} {{Exploring Optimal Transport for Event-Level
  Anomaly Detection at the Large Hadron Collider}},\ }\href@noop {} {\
  (\bibinfo {year} {2024})},\ \Eprint {https://arxiv.org/abs/2401.15542}
  {arXiv:2401.15542 [hep-ph]} \BibitemShut {NoStop}%
\bibitem [{\citenamefont {Cai}\ \emph {et~al.}(2024)\citenamefont {Cai},
  \citenamefont {Cheng}, \citenamefont {Craig}, \citenamefont {Koszegi},\ and\
  \citenamefont {Larkoski}}]{Cai:2024xnt}%
  \BibitemOpen
  \bibfield  {author} {\bibinfo {author} {\bibfnamefont {T.}~\bibnamefont
  {Cai}}, \bibinfo {author} {\bibfnamefont {J.}~\bibnamefont {Cheng}}, \bibinfo
  {author} {\bibfnamefont {N.}~\bibnamefont {Craig}}, \bibinfo {author}
  {\bibfnamefont {G.}~\bibnamefont {Koszegi}},\ and\ \bibinfo {author}
  {\bibfnamefont {A.~J.}\ \bibnamefont {Larkoski}},\ }\bibfield  {title}
  {\bibinfo {title} {{The Phase Space Distance Between Collider Events}},\
  }\href@noop {} {\  (\bibinfo {year} {2024})},\ \Eprint
  {https://arxiv.org/abs/2405.16698} {arXiv:2405.16698 [hep-ph]} \BibitemShut
  {NoStop}%
\bibitem [{\citenamefont {Gambhir}\ \emph {et~al.}(2025)\citenamefont
  {Gambhir}, \citenamefont {Larkoski},\ and\ \citenamefont
  {Thaler}}]{Gambhir:2024ndc}%
  \BibitemOpen
  \bibfield  {author} {\bibinfo {author} {\bibfnamefont {R.}~\bibnamefont
  {Gambhir}}, \bibinfo {author} {\bibfnamefont {A.~J.}\ \bibnamefont
  {Larkoski}},\ and\ \bibinfo {author} {\bibfnamefont {J.}~\bibnamefont
  {Thaler}},\ }\bibfield  {title} {\bibinfo {title} {{SPECTER: efficient
  evaluation of the spectral EMD}},\ }\href
  {https://doi.org/10.1007/JHEP12(2024)219} {\bibfield  {journal} {\bibinfo
  {journal} {JHEP}\ }\textbf {\bibinfo {volume} {12}},\ \bibinfo {pages}
  {219}},\ \Eprint {https://arxiv.org/abs/2410.05379} {arXiv:2410.05379
  [hep-ph]} \BibitemShut {NoStop}%
\bibitem [{\citenamefont {Cai}\ \emph {et~al.}(2025)\citenamefont {Cai},
  \citenamefont {Craig}, \citenamefont {Craig},\ and\ \citenamefont
  {Lin}}]{Cai:2025fyw}%
  \BibitemOpen
  \bibfield  {author} {\bibinfo {author} {\bibfnamefont {T.}~\bibnamefont
  {Cai}}, \bibinfo {author} {\bibfnamefont {N.}~\bibnamefont {Craig}}, \bibinfo
  {author} {\bibfnamefont {K.}~\bibnamefont {Craig}},\ and\ \bibinfo {author}
  {\bibfnamefont {X.}~\bibnamefont {Lin}},\ }\bibfield  {title} {\bibinfo
  {title} {{Multiscale optimal transport for complete collider events}},\
  }\href {https://doi.org/10.1103/7nhv-hndt} {\bibfield  {journal} {\bibinfo
  {journal} {Phys. Rev. D}\ }\textbf {\bibinfo {volume} {112}},\ \bibinfo
  {pages} {036021} (\bibinfo {year} {2025})},\ \Eprint
  {https://arxiv.org/abs/2501.10681} {arXiv:2501.10681 [hep-ph]} \BibitemShut
  {NoStop}%
\bibitem [{\citenamefont {D'Agnolo}\ \emph {et~al.}(2025)\citenamefont
  {D'Agnolo}, \citenamefont {Glioti}, \citenamefont {Rigo},\ and\ \citenamefont
  {Valenti}}]{DAgnolo:2025qqr}%
  \BibitemOpen
  \bibfield  {author} {\bibinfo {author} {\bibfnamefont {R.~T.}\ \bibnamefont
  {D'Agnolo}}, \bibinfo {author} {\bibfnamefont {A.}~\bibnamefont {Glioti}},
  \bibinfo {author} {\bibfnamefont {G.}~\bibnamefont {Rigo}},\ and\ \bibinfo
  {author} {\bibfnamefont {A.}~\bibnamefont {Valenti}},\ }\bibfield  {title}
  {\bibinfo {title} {{The Intrinsic Dimension of Collider Events and
  Model-Independent Searches in 100 Dimensions}},\ }\href@noop {} {\  (\bibinfo
  {year} {2025})},\ \Eprint {https://arxiv.org/abs/2511.20760}
  {arXiv:2511.20760 [hep-ph]} \BibitemShut {NoStop}%
\bibitem [{\citenamefont {Andersen}\ \emph {et~al.}(2026)\citenamefont
  {Andersen}, \citenamefont {Cole},\ and\ \citenamefont
  {Maier}}]{Andersen:2026ppe}%
  \BibitemOpen
  \bibfield  {author} {\bibinfo {author} {\bibfnamefont {J.~R.}\ \bibnamefont
  {Andersen}}, \bibinfo {author} {\bibfnamefont {E.}~\bibnamefont {Cole}},\
  and\ \bibinfo {author} {\bibfnamefont {A.}~\bibnamefont {Maier}},\ }\bibfield
   {title} {\bibinfo {title} {{Precision Cell Resampling with a Relative and
  Resonant Aware Metric}},\ }\href@noop {} {\  (\bibinfo {year} {2026})},\
  \Eprint {https://arxiv.org/abs/2605.13237} {arXiv:2605.13237 [hep-ph]}
  \BibitemShut {NoStop}%
\bibitem [{\citenamefont {Monge}(1781)}]{monge1781memoire}%
  \BibitemOpen
  \bibfield  {author} {\bibinfo {author} {\bibfnamefont {G.}~\bibnamefont
  {Monge}},\ }\bibfield  {title} {\bibinfo {title} {M{\'e}moire sur la
  th{\'e}orie des d{\'e}blais et des remblais},\ }\href@noop {} {\bibfield
  {journal} {\bibinfo  {journal} {Mem. Math. Phys. Acad. Royale Sci.}\ ,\
  \bibinfo {pages} {666}} (\bibinfo {year} {1781})}\BibitemShut {NoStop}%
\bibitem [{\citenamefont {Kantorovich}(1939)}]{kantorovich1939mathematical}%
  \BibitemOpen
  \bibfield  {author} {\bibinfo {author} {\bibfnamefont {L.~V.}\ \bibnamefont
  {Kantorovich}},\ }\bibfield  {title} {\bibinfo {title} {The mathematical
  method of production planning and organization},\ }\href@noop {} {\bibfield
  {journal} {\bibinfo  {journal} {Management Science}\ }\textbf {\bibinfo
  {volume} {6}},\ \bibinfo {pages} {363} (\bibinfo {year} {1939})}\BibitemShut
  {NoStop}%
\bibitem [{\citenamefont {Vaserstein}(1969)}]{vaserstein1969markov}%
  \BibitemOpen
  \bibfield  {author} {\bibinfo {author} {\bibfnamefont {L.~N.}\ \bibnamefont
  {Vaserstein}},\ }\bibfield  {title} {\bibinfo {title} {Markov processes over
  denumerable products of spaces, describing large systems of automata},\
  }\href@noop {} {\bibfield  {journal} {\bibinfo  {journal} {Problemy Peredachi
  Informatsii}\ }\textbf {\bibinfo {volume} {5}},\ \bibinfo {pages} {64}
  (\bibinfo {year} {1969})}\BibitemShut {NoStop}%
\bibitem [{\citenamefont {Dobrushin}(1970)}]{dobrushin1970prescribing}%
  \BibitemOpen
  \bibfield  {author} {\bibinfo {author} {\bibfnamefont {R.~L.}\ \bibnamefont
  {Dobrushin}},\ }\bibfield  {title} {\bibinfo {title} {Prescribing a system of
  random variables by conditional distributions},\ }\href@noop {} {\bibfield
  {journal} {\bibinfo  {journal} {Theory of Probability \& Its Applications}\
  }\textbf {\bibinfo {volume} {15}},\ \bibinfo {pages} {458} (\bibinfo {year}
  {1970})}\BibitemShut {NoStop}%
\bibitem [{\citenamefont {Lott}\ and\ \citenamefont
  {Villani}(2009)}]{lott2009ricci}%
  \BibitemOpen
  \bibfield  {author} {\bibinfo {author} {\bibfnamefont {J.}~\bibnamefont
  {Lott}}\ and\ \bibinfo {author} {\bibfnamefont {C.}~\bibnamefont {Villani}},\
  }\bibfield  {title} {\bibinfo {title} {Ricci curvature for metric-measure
  spaces via optimal transport},\ }\href@noop {} {\bibfield  {journal}
  {\bibinfo  {journal} {Annals of Mathematics}\ ,\ \bibinfo {pages} {903}}
  (\bibinfo {year} {2009})}\BibitemShut {NoStop}%
\bibitem [{\citenamefont {Villani}\ \emph {et~al.}(2009)\citenamefont {Villani}
  \emph {et~al.}}]{villani2009optimal}%
  \BibitemOpen
  \bibfield  {author} {\bibinfo {author} {\bibfnamefont {C.}~\bibnamefont
  {Villani}} \emph {et~al.},\ }\href@noop {} {\emph {\bibinfo {title} {Optimal
  transport: old and new}}},\ Vol.\ \bibinfo {volume} {338}\ (\bibinfo
  {publisher} {Springer},\ \bibinfo {year} {2009})\BibitemShut {NoStop}%
\bibitem [{\citenamefont {Kuhn}(1955)}]{kuhn1955hungarian}%
  \BibitemOpen
  \bibfield  {author} {\bibinfo {author} {\bibfnamefont {H.~W.}\ \bibnamefont
  {Kuhn}},\ }\bibfield  {title} {\bibinfo {title} {The hungarian method for the
  assignment problem},\ }\href@noop {} {\bibfield  {journal} {\bibinfo
  {journal} {Naval research logistics quarterly}\ }\textbf {\bibinfo {volume}
  {2}},\ \bibinfo {pages} {83} (\bibinfo {year} {1955})}\BibitemShut {NoStop}%
\bibitem [{\citenamefont {Ollivier}(2009)}]{ollivier2009reduction}%
  \BibitemOpen
  \bibfield  {author} {\bibinfo {author} {\bibfnamefont {F.}~\bibnamefont
  {Ollivier}},\ }\bibfield  {title} {\bibinfo {title} {The reduction to normal
  form of a non-normal system of differential equations: De {\ae}quationum
  differentialium systemate non normali ad formam normalem revocando},\
  }\href@noop {} {\bibfield  {journal} {\bibinfo  {journal} {Applicable Algebra
  in Engineering, Communication and Computing}\ }\textbf {\bibinfo {volume}
  {20}},\ \bibinfo {pages} {33} (\bibinfo {year} {2009})}\BibitemShut {NoStop}%
\bibitem [{\citenamefont {Flamary}\ \emph {et~al.}(2021)\citenamefont
  {Flamary}, \citenamefont {Courty}, \citenamefont {Gramfort}, \citenamefont
  {Alaya}, \citenamefont {Boisbunon}, \citenamefont {Chambon}, \citenamefont
  {Chapel}, \citenamefont {Corenflos}, \citenamefont {Fatras}, \citenamefont
  {Fournier}, \citenamefont {Gautheron}, \citenamefont {Gayraud}, \citenamefont
  {Janati}, \citenamefont {Rakotomamonjy}, \citenamefont {Redko}, \citenamefont
  {Rolet}, \citenamefont {Schutz}, \citenamefont {Seguy}, \citenamefont
  {Sutherland}, \citenamefont {Tavenard}, \citenamefont {Tong},\ and\
  \citenamefont {Vayer}}]{flamary2021pot}%
  \BibitemOpen
  \bibfield  {author} {\bibinfo {author} {\bibfnamefont {R.}~\bibnamefont
  {Flamary}}, \bibinfo {author} {\bibfnamefont {N.}~\bibnamefont {Courty}},
  \bibinfo {author} {\bibfnamefont {A.}~\bibnamefont {Gramfort}}, \bibinfo
  {author} {\bibfnamefont {M.~Z.}\ \bibnamefont {Alaya}}, \bibinfo {author}
  {\bibfnamefont {A.}~\bibnamefont {Boisbunon}}, \bibinfo {author}
  {\bibfnamefont {S.}~\bibnamefont {Chambon}}, \bibinfo {author} {\bibfnamefont
  {L.}~\bibnamefont {Chapel}}, \bibinfo {author} {\bibfnamefont
  {A.}~\bibnamefont {Corenflos}}, \bibinfo {author} {\bibfnamefont
  {K.}~\bibnamefont {Fatras}}, \bibinfo {author} {\bibfnamefont
  {N.}~\bibnamefont {Fournier}}, \bibinfo {author} {\bibfnamefont
  {L.}~\bibnamefont {Gautheron}}, \bibinfo {author} {\bibfnamefont {N.~T.}\
  \bibnamefont {Gayraud}}, \bibinfo {author} {\bibfnamefont {H.}~\bibnamefont
  {Janati}}, \bibinfo {author} {\bibfnamefont {A.}~\bibnamefont
  {Rakotomamonjy}}, \bibinfo {author} {\bibfnamefont {I.}~\bibnamefont
  {Redko}}, \bibinfo {author} {\bibfnamefont {A.}~\bibnamefont {Rolet}},
  \bibinfo {author} {\bibfnamefont {A.}~\bibnamefont {Schutz}}, \bibinfo
  {author} {\bibfnamefont {V.}~\bibnamefont {Seguy}}, \bibinfo {author}
  {\bibfnamefont {D.~J.}\ \bibnamefont {Sutherland}}, \bibinfo {author}
  {\bibfnamefont {R.}~\bibnamefont {Tavenard}}, \bibinfo {author}
  {\bibfnamefont {A.}~\bibnamefont {Tong}},\ and\ \bibinfo {author}
  {\bibfnamefont {T.}~\bibnamefont {Vayer}},\ }\bibfield  {title} {\bibinfo
  {title} {Pot: Python optimal transport},\ }\href
  {http://jmlr.org/papers/v22/20-451.html} {\bibfield  {journal} {\bibinfo
  {journal} {Journal of Machine Learning Research}\ }\textbf {\bibinfo {volume}
  {22}},\ \bibinfo {pages} {1} (\bibinfo {year} {2021})}\BibitemShut {NoStop}%
\bibitem [{\citenamefont {Flamary}\ \emph {et~al.}(2026)\citenamefont
  {Flamary}, \citenamefont {Vincent-Cuaz}, \citenamefont {Courty},
  \citenamefont {Gramfort}, \citenamefont {Kachaiev}, \citenamefont
  {Quang~Tran}, \citenamefont {David}, \citenamefont {Bonet}, \citenamefont
  {Cassereau}, \citenamefont {Gnassounou}, \citenamefont {Tanguy},
  \citenamefont {Delon}, \citenamefont {Collas}, \citenamefont {Mazelet},
  \citenamefont {Chapel}, \citenamefont {Kerdoncuff}, \citenamefont {Yu},
  \citenamefont {Feickert}, \citenamefont {Krzakala}, \citenamefont {Liu},
  \citenamefont {Fernandes~Montesuma}, \citenamefont {Neike}, \citenamefont
  {Genest}, \citenamefont {Coeurjolly}, \citenamefont {Germain}, \citenamefont
  {O'Shea}, \citenamefont {Corneli},\ and\ \citenamefont
  {Genans}}]{flamary2026pot}%
  \BibitemOpen
  \bibfield  {author} {\bibinfo {author} {\bibfnamefont {R.}~\bibnamefont
  {Flamary}}, \bibinfo {author} {\bibfnamefont {C.}~\bibnamefont
  {Vincent-Cuaz}}, \bibinfo {author} {\bibfnamefont {N.}~\bibnamefont
  {Courty}}, \bibinfo {author} {\bibfnamefont {A.}~\bibnamefont {Gramfort}},
  \bibinfo {author} {\bibfnamefont {O.}~\bibnamefont {Kachaiev}}, \bibinfo
  {author} {\bibfnamefont {H.}~\bibnamefont {Quang~Tran}}, \bibinfo {author}
  {\bibfnamefont {L.}~\bibnamefont {David}}, \bibinfo {author} {\bibfnamefont
  {C.}~\bibnamefont {Bonet}}, \bibinfo {author} {\bibfnamefont
  {N.}~\bibnamefont {Cassereau}}, \bibinfo {author} {\bibfnamefont
  {T.}~\bibnamefont {Gnassounou}}, \bibinfo {author} {\bibfnamefont
  {E.}~\bibnamefont {Tanguy}}, \bibinfo {author} {\bibfnamefont
  {J.}~\bibnamefont {Delon}}, \bibinfo {author} {\bibfnamefont
  {A.}~\bibnamefont {Collas}}, \bibinfo {author} {\bibfnamefont
  {S.}~\bibnamefont {Mazelet}}, \bibinfo {author} {\bibfnamefont
  {L.}~\bibnamefont {Chapel}}, \bibinfo {author} {\bibfnamefont
  {T.}~\bibnamefont {Kerdoncuff}}, \bibinfo {author} {\bibfnamefont
  {X.}~\bibnamefont {Yu}}, \bibinfo {author} {\bibfnamefont {M.}~\bibnamefont
  {Feickert}}, \bibinfo {author} {\bibfnamefont {P.}~\bibnamefont {Krzakala}},
  \bibinfo {author} {\bibfnamefont {T.}~\bibnamefont {Liu}}, \bibinfo {author}
  {\bibfnamefont {E.}~\bibnamefont {Fernandes~Montesuma}}, \bibinfo {author}
  {\bibfnamefont {N.}~\bibnamefont {Neike}}, \bibinfo {author} {\bibfnamefont
  {B.}~\bibnamefont {Genest}}, \bibinfo {author} {\bibfnamefont
  {D.}~\bibnamefont {Coeurjolly}}, \bibinfo {author} {\bibfnamefont
  {T.}~\bibnamefont {Germain}}, \bibinfo {author} {\bibfnamefont
  {S.}~\bibnamefont {O'Shea}}, \bibinfo {author} {\bibfnamefont
  {M.}~\bibnamefont {Corneli}},\ and\ \bibinfo {author} {\bibfnamefont
  {F.}~\bibnamefont {Genans}},\ }\href
  {https://doi.org/10.5281/zenodo.17161062} {\bibinfo {title} {Pot python
  optimal transport}} (\bibinfo {year} {2026})\BibitemShut {NoStop}%
\bibitem [{\citenamefont {Basham}\ \emph {et~al.}(1978)\citenamefont {Basham},
  \citenamefont {Brown}, \citenamefont {Ellis},\ and\ \citenamefont
  {Love}}]{Basham:1978bw}%
  \BibitemOpen
  \bibfield  {author} {\bibinfo {author} {\bibfnamefont {C.~L.}\ \bibnamefont
  {Basham}}, \bibinfo {author} {\bibfnamefont {L.~S.}\ \bibnamefont {Brown}},
  \bibinfo {author} {\bibfnamefont {S.~D.}\ \bibnamefont {Ellis}},\ and\
  \bibinfo {author} {\bibfnamefont {S.~T.}\ \bibnamefont {Love}},\ }\bibfield
  {title} {\bibinfo {title} {{Energy Correlations in electron - Positron
  Annihilation: Testing QCD}},\ }\href
  {https://doi.org/10.1103/PhysRevLett.41.1585} {\bibfield  {journal} {\bibinfo
   {journal} {Phys. Rev. Lett.}\ }\textbf {\bibinfo {volume} {41}},\ \bibinfo
  {pages} {1585} (\bibinfo {year} {1978})}\BibitemShut {NoStop}%
\bibitem [{\citenamefont {Hofman}\ and\ \citenamefont
  {Maldacena}(2008)}]{Hofman:2008ar}%
  \BibitemOpen
  \bibfield  {author} {\bibinfo {author} {\bibfnamefont {D.~M.}\ \bibnamefont
  {Hofman}}\ and\ \bibinfo {author} {\bibfnamefont {J.}~\bibnamefont
  {Maldacena}},\ }\bibfield  {title} {\bibinfo {title} {{Conformal collider
  physics: Energy and charge correlations}},\ }\href
  {https://doi.org/10.1088/1126-6708/2008/05/012} {\bibfield  {journal}
  {\bibinfo  {journal} {JHEP}\ }\textbf {\bibinfo {volume} {05}},\ \bibinfo
  {pages} {012}},\ \Eprint {https://arxiv.org/abs/0803.1467} {arXiv:0803.1467
  [hep-th]} \BibitemShut {NoStop}%
\bibitem [{\citenamefont {Dixon}\ \emph {et~al.}(2018)\citenamefont {Dixon},
  \citenamefont {Luo}, \citenamefont {Shtabovenko}, \citenamefont {Yang},\ and\
  \citenamefont {Zhu}}]{Dixon:2018qgp}%
  \BibitemOpen
  \bibfield  {author} {\bibinfo {author} {\bibfnamefont {L.~J.}\ \bibnamefont
  {Dixon}}, \bibinfo {author} {\bibfnamefont {M.-X.}\ \bibnamefont {Luo}},
  \bibinfo {author} {\bibfnamefont {V.}~\bibnamefont {Shtabovenko}}, \bibinfo
  {author} {\bibfnamefont {T.-Z.}\ \bibnamefont {Yang}},\ and\ \bibinfo
  {author} {\bibfnamefont {H.~X.}\ \bibnamefont {Zhu}},\ }\bibfield  {title}
  {\bibinfo {title} {{Analytical Computation of Energy-Energy Correlation at
  Next-to-Leading Order in QCD}},\ }\href
  {https://doi.org/10.1103/PhysRevLett.120.102001} {\bibfield  {journal}
  {\bibinfo  {journal} {Phys. Rev. Lett.}\ }\textbf {\bibinfo {volume} {120}},\
  \bibinfo {pages} {102001} (\bibinfo {year} {2018})},\ \Eprint
  {https://arxiv.org/abs/1801.03219} {arXiv:1801.03219 [hep-ph]} \BibitemShut
  {NoStop}%
\bibitem [{\citenamefont {Chen}\ \emph {et~al.}(2020)\citenamefont {Chen},
  \citenamefont {Moult}, \citenamefont {Zhang},\ and\ \citenamefont
  {Zhu}}]{Chen:2020vvp}%
  \BibitemOpen
  \bibfield  {author} {\bibinfo {author} {\bibfnamefont {H.}~\bibnamefont
  {Chen}}, \bibinfo {author} {\bibfnamefont {I.}~\bibnamefont {Moult}},
  \bibinfo {author} {\bibfnamefont {X.}~\bibnamefont {Zhang}},\ and\ \bibinfo
  {author} {\bibfnamefont {H.~X.}\ \bibnamefont {Zhu}},\ }\bibfield  {title}
  {\bibinfo {title} {{Rethinking jets with energy correlators: Tracks,
  resummation, and analytic continuation}},\ }\href
  {https://doi.org/10.1103/PhysRevD.102.054012} {\bibfield  {journal} {\bibinfo
   {journal} {Phys. Rev. D}\ }\textbf {\bibinfo {volume} {102}},\ \bibinfo
  {pages} {054012} (\bibinfo {year} {2020})},\ \Eprint
  {https://arxiv.org/abs/2004.11381} {arXiv:2004.11381 [hep-ph]} \BibitemShut
  {NoStop}%
\bibitem [{\citenamefont {Moult}\ and\ \citenamefont
  {Zhu}(2025)}]{Moult:2025nhu}%
  \BibitemOpen
  \bibfield  {author} {\bibinfo {author} {\bibfnamefont {I.}~\bibnamefont
  {Moult}}\ and\ \bibinfo {author} {\bibfnamefont {H.~X.}\ \bibnamefont
  {Zhu}},\ }\bibfield  {title} {\bibinfo {title} {{Energy Correlators: A
  Journey From Theory to Experiment}},\ }\href@noop {} {\  (\bibinfo {year}
  {2025})},\ \Eprint {https://arxiv.org/abs/2506.09119} {arXiv:2506.09119
  [hep-ph]} \BibitemShut {NoStop}%
\bibitem [{\citenamefont {Jankowiak}\ and\ \citenamefont
  {Larkoski}(2011)}]{Jankowiak:2011qa}%
  \BibitemOpen
  \bibfield  {author} {\bibinfo {author} {\bibfnamefont {M.}~\bibnamefont
  {Jankowiak}}\ and\ \bibinfo {author} {\bibfnamefont {A.~J.}\ \bibnamefont
  {Larkoski}},\ }\bibfield  {title} {\bibinfo {title} {{Jet Substructure
  Without Trees}},\ }\href {https://doi.org/10.1007/JHEP06(2011)057} {\bibfield
   {journal} {\bibinfo  {journal} {JHEP}\ }\textbf {\bibinfo {volume} {06}},\
  \bibinfo {pages} {057}},\ \Eprint {https://arxiv.org/abs/1104.1646}
  {arXiv:1104.1646 [hep-ph]} \BibitemShut {NoStop}%
\bibitem [{\citenamefont {Lim}\ and\ \citenamefont
  {Nojiri}(2018)}]{Lim:2018toa}%
  \BibitemOpen
  \bibfield  {author} {\bibinfo {author} {\bibfnamefont {S.~H.}\ \bibnamefont
  {Lim}}\ and\ \bibinfo {author} {\bibfnamefont {M.~M.}\ \bibnamefont
  {Nojiri}},\ }\bibfield  {title} {\bibinfo {title} {{Spectral Analysis of Jet
  Substructure with Neural Networks: Boosted Higgs Case}},\ }\href
  {https://doi.org/10.1007/JHEP10(2018)181} {\bibfield  {journal} {\bibinfo
  {journal} {JHEP}\ }\textbf {\bibinfo {volume} {10}},\ \bibinfo {pages}
  {181}},\ \Eprint {https://arxiv.org/abs/1807.03312} {arXiv:1807.03312
  [hep-ph]} \BibitemShut {NoStop}%
\bibitem [{\citenamefont {Chakraborty}\ \emph {et~al.}(2019)\citenamefont
  {Chakraborty}, \citenamefont {Lim},\ and\ \citenamefont
  {Nojiri}}]{Chakraborty:2019imr}%
  \BibitemOpen
  \bibfield  {author} {\bibinfo {author} {\bibfnamefont {A.}~\bibnamefont
  {Chakraborty}}, \bibinfo {author} {\bibfnamefont {S.~H.}\ \bibnamefont
  {Lim}},\ and\ \bibinfo {author} {\bibfnamefont {M.~M.}\ \bibnamefont
  {Nojiri}},\ }\bibfield  {title} {\bibinfo {title} {{Interpretable deep
  learning for two-prong jet classification with jet spectra}},\ }\href
  {https://doi.org/10.1007/JHEP07(2019)135} {\bibfield  {journal} {\bibinfo
  {journal} {JHEP}\ }\textbf {\bibinfo {volume} {07}},\ \bibinfo {pages}
  {135}},\ \Eprint {https://arxiv.org/abs/1904.02092} {arXiv:1904.02092
  [hep-ph]} \BibitemShut {NoStop}%
\bibitem [{\citenamefont {Garcia~Caffaro}\ \emph {et~al.}(2025)\citenamefont
  {Garcia~Caffaro}, \citenamefont {Moult},\ and\ \citenamefont
  {Shimmin}}]{GarciaCaffaro:2025gkm}%
  \BibitemOpen
  \bibfield  {author} {\bibinfo {author} {\bibfnamefont {A.}~\bibnamefont
  {Garcia~Caffaro}}, \bibinfo {author} {\bibfnamefont {I.}~\bibnamefont
  {Moult}},\ and\ \bibinfo {author} {\bibfnamefont {C.}~\bibnamefont
  {Shimmin}},\ }\bibfield  {title} {\bibinfo {title} {{Energy-Energy Flow
  Networks}},\ }\href@noop {} {\  (\bibinfo {year} {2025})},\ \Eprint
  {https://arxiv.org/abs/2510.06314} {arXiv:2510.06314 [hep-ph]} \BibitemShut
  {NoStop}%
\bibitem [{\citenamefont {Altakach}\ \emph {et~al.}(2026)\citenamefont
  {Altakach}, \citenamefont {Hassan}, \citenamefont {Kraml}, \citenamefont
  {Sakurai},\ and\ \citenamefont {Zaraket}}]{Altakach:2026gdg}%
  \BibitemOpen
  \bibfield  {author} {\bibinfo {author} {\bibfnamefont {M.~M.}\ \bibnamefont
  {Altakach}}, \bibinfo {author} {\bibfnamefont {H.}~\bibnamefont {Hassan}},
  \bibinfo {author} {\bibfnamefont {S.}~\bibnamefont {Kraml}}, \bibinfo
  {author} {\bibfnamefont {K.}~\bibnamefont {Sakurai}},\ and\ \bibinfo {author}
  {\bibfnamefont {H.}~\bibnamefont {Zaraket}},\ }\bibfield  {title} {\bibinfo
  {title} {{Machine learning fully hadronic events with spectral functions}},\
  }\href@noop {} {\  (\bibinfo {year} {2026})},\ \Eprint
  {https://arxiv.org/abs/2606.27420} {arXiv:2606.27420 [hep-ph]} \BibitemShut
  {NoStop}%
\bibitem [{\citenamefont {Boutin}\ and\ \citenamefont
  {Kemper}(2004)}]{boutin2004reconstructing}%
  \BibitemOpen
  \bibfield  {author} {\bibinfo {author} {\bibfnamefont {M.}~\bibnamefont
  {Boutin}}\ and\ \bibinfo {author} {\bibfnamefont {G.}~\bibnamefont
  {Kemper}},\ }\bibfield  {title} {\bibinfo {title} {On reconstructing n-point
  configurations from the distribution of distances or areas},\ }\href@noop {}
  {\bibfield  {journal} {\bibinfo  {journal} {Advances in Applied Mathematics}\
  }\textbf {\bibinfo {volume} {32}},\ \bibinfo {pages} {709} (\bibinfo {year}
  {2004})}\BibitemShut {NoStop}%
\bibitem [{\citenamefont {Cesarotti}\ and\ \citenamefont
  {Thaler}(2020)}]{Cesarotti:2020hwb}%
  \BibitemOpen
  \bibfield  {author} {\bibinfo {author} {\bibfnamefont {C.}~\bibnamefont
  {Cesarotti}}\ and\ \bibinfo {author} {\bibfnamefont {J.}~\bibnamefont
  {Thaler}},\ }\bibfield  {title} {\bibinfo {title} {{A Robust Measure of Event
  Isotropy at Colliders}},\ }\href {https://doi.org/10.1007/JHEP08(2020)084}
  {\bibfield  {journal} {\bibinfo  {journal} {JHEP}\ }\textbf {\bibinfo
  {volume} {08}},\ \bibinfo {pages} {084}},\ \Eprint
  {https://arxiv.org/abs/2004.06125} {arXiv:2004.06125 [hep-ph]} \BibitemShut
  {NoStop}%
\bibitem [{\citenamefont {Atzori}\ \emph {et~al.}(2026)\citenamefont {Atzori},
  \citenamefont {Cacciari}, \citenamefont {Marzani},\ and\ \citenamefont
  {Soyez}}]{Atzori:2026blf}%
  \BibitemOpen
  \bibfield  {author} {\bibinfo {author} {\bibfnamefont {D.}~\bibnamefont
  {Atzori}}, \bibinfo {author} {\bibfnamefont {M.}~\bibnamefont {Cacciari}},
  \bibinfo {author} {\bibfnamefont {S.}~\bibnamefont {Marzani}},\ and\ \bibinfo
  {author} {\bibfnamefont {G.}~\bibnamefont {Soyez}},\ }\bibfield  {title}
  {\bibinfo {title} {{Event isotropy in perturbative QCD}},\ }\href@noop {} {\
  (\bibinfo {year} {2026})},\ \Eprint {https://arxiv.org/abs/2606.23924}
  {arXiv:2606.23924 [hep-ph]} \BibitemShut {NoStop}%
\bibitem [{\citenamefont {Doherty}\ \emph {et~al.}(2026)\citenamefont
  {Doherty}, \citenamefont {Hay}, \citenamefont {Jain}, \citenamefont
  {LeBlanc}, \citenamefont {Marrinan}, \citenamefont {Mauceri},\ and\
  \citenamefont {Roloff}}]{Doherty:2026fli}%
  \BibitemOpen
  \bibfield  {author} {\bibinfo {author} {\bibfnamefont {R.}~\bibnamefont
  {Doherty}}, \bibinfo {author} {\bibfnamefont {L.}~\bibnamefont {Hay}},
  \bibinfo {author} {\bibfnamefont {R.}~\bibnamefont {Jain}}, \bibinfo {author}
  {\bibfnamefont {M.}~\bibnamefont {LeBlanc}}, \bibinfo {author} {\bibfnamefont
  {J.}~\bibnamefont {Marrinan}}, \bibinfo {author} {\bibfnamefont
  {C.}~\bibnamefont {Mauceri}},\ and\ \bibinfo {author} {\bibfnamefont
  {J.}~\bibnamefont {Roloff}},\ }\bibfield  {title} {\bibinfo {title}
  {{Optimal-Transport-Based Cell Resampling for Negative and Pathological Event
  Weights}},\ }\href@noop {} {\  (\bibinfo {year} {2026})},\ \Eprint
  {https://arxiv.org/abs/2607.08723} {arXiv:2607.08723 [hep-ph]} \BibitemShut
  {NoStop}%
\bibitem [{\citenamefont {Anderson}(1972)}]{Anderson:1972pca}%
  \BibitemOpen
  \bibfield  {author} {\bibinfo {author} {\bibfnamefont {P.~W.}\ \bibnamefont
  {Anderson}},\ }\bibfield  {title} {\bibinfo {title} {{More Is Different}},\
  }\href {https://doi.org/10.1126/science.177.4047.393} {\bibfield  {journal}
  {\bibinfo  {journal} {Science}\ }\textbf {\bibinfo {volume} {177}},\ \bibinfo
  {pages} {393} (\bibinfo {year} {1972})}\BibitemShut {NoStop}%
\bibitem [{\citenamefont {Feickert}\ and\ \citenamefont
  {Nachman}(2021)}]{Feickert:2021ajf}%
  \BibitemOpen
  \bibfield  {author} {\bibinfo {author} {\bibfnamefont {M.}~\bibnamefont
  {Feickert}}\ and\ \bibinfo {author} {\bibfnamefont {B.}~\bibnamefont
  {Nachman}},\ }\bibfield  {title} {\bibinfo {title} {{A Living Review of
  Machine Learning for Particle Physics}},\ }\href@noop {} {\  (\bibinfo {year}
  {2021})},\ \Eprint {https://arxiv.org/abs/2102.02770} {arXiv:2102.02770
  [hep-ph]} \BibitemShut {NoStop}%
\bibitem [{\citenamefont {Aad}\ \emph {et~al.}(2012)\citenamefont {Aad} \emph
  {et~al.}}]{ATLAS:2012yve}%
  \BibitemOpen
  \bibfield  {author} {\bibinfo {author} {\bibfnamefont {G.}~\bibnamefont
  {Aad}} \emph {et~al.} (\bibinfo {collaboration} {ATLAS}),\ }\bibfield
  {title} {\bibinfo {title} {{Observation of a new particle in the search for
  the Standard Model Higgs boson with the ATLAS detector at the LHC}},\ }\href
  {https://doi.org/10.1016/j.physletb.2012.08.020} {\bibfield  {journal}
  {\bibinfo  {journal} {Phys. Lett. B}\ }\textbf {\bibinfo {volume} {716}},\
  \bibinfo {pages} {1} (\bibinfo {year} {2012})},\ \Eprint
  {https://arxiv.org/abs/1207.7214} {arXiv:1207.7214 [hep-ex]} \BibitemShut
  {NoStop}%
\bibitem [{\citenamefont {Chatrchyan}\ \emph {et~al.}(2012)\citenamefont
  {Chatrchyan} \emph {et~al.}}]{CMS:2012qbp}%
  \BibitemOpen
  \bibfield  {author} {\bibinfo {author} {\bibfnamefont {S.}~\bibnamefont
  {Chatrchyan}} \emph {et~al.} (\bibinfo {collaboration} {CMS}),\ }\bibfield
  {title} {\bibinfo {title} {{Observation of a New Boson at a Mass of 125 GeV
  with the CMS Experiment at the LHC}},\ }\href
  {https://doi.org/10.1016/j.physletb.2012.08.021} {\bibfield  {journal}
  {\bibinfo  {journal} {Phys. Lett. B}\ }\textbf {\bibinfo {volume} {716}},\
  \bibinfo {pages} {30} (\bibinfo {year} {2012})},\ \Eprint
  {https://arxiv.org/abs/1207.7235} {arXiv:1207.7235 [hep-ex]} \BibitemShut
  {NoStop}%
\bibitem [{\citenamefont {Park}\ \emph {et~al.}(2023)\citenamefont {Park},
  \citenamefont {Harris},\ and\ \citenamefont {Ostdiek}}]{Park:2022zov}%
  \BibitemOpen
  \bibfield  {author} {\bibinfo {author} {\bibfnamefont {S.~E.}\ \bibnamefont
  {Park}}, \bibinfo {author} {\bibfnamefont {P.}~\bibnamefont {Harris}},\ and\
  \bibinfo {author} {\bibfnamefont {B.}~\bibnamefont {Ostdiek}},\ }\bibfield
  {title} {\bibinfo {title} {{Neural embedding: learning the embedding of the
  manifold of physics data}},\ }\href {https://doi.org/10.1007/JHEP07(2023)108}
  {\bibfield  {journal} {\bibinfo  {journal} {JHEP}\ }\textbf {\bibinfo
  {volume} {07}},\ \bibinfo {pages} {108}},\ \Eprint
  {https://arxiv.org/abs/2208.05484} {arXiv:2208.05484 [hep-ph]} \BibitemShut
  {NoStop}%
\bibitem [{\citenamefont {Cai}\ \emph {et~al.}(2026)\citenamefont {Cai},
  \citenamefont {Bhargava},\ and\ \citenamefont {Nachman}}]{Cai:2025vxl}%
  \BibitemOpen
  \bibfield  {author} {\bibinfo {author} {\bibfnamefont {T.}~\bibnamefont
  {Cai}}, \bibinfo {author} {\bibfnamefont {A.}~\bibnamefont {Bhargava}},\ and\
  \bibinfo {author} {\bibfnamefont {B.}~\bibnamefont {Nachman}},\ }\bibfield
  {title} {\bibinfo {title} {{Optimal transport event representation for
  anomaly detection}},\ }\href {https://doi.org/10.1103/wwzk-25sy} {\bibfield
  {journal} {\bibinfo  {journal} {Phys. Rev. D}\ }\textbf {\bibinfo {volume}
  {114}},\ \bibinfo {pages} {016011} (\bibinfo {year} {2026})},\ \Eprint
  {https://arxiv.org/abs/2512.04839} {arXiv:2512.04839 [hep-ph]} \BibitemShut
  {NoStop}%
\bibitem [{\citenamefont {Komiske}\ \emph
  {et~al.}(2020{\natexlab{b}})\citenamefont {Komiske}, \citenamefont
  {Mastandrea}, \citenamefont {Metodiev}, \citenamefont {Naik},\ and\
  \citenamefont {Thaler}}]{Komiske:2019jim}%
  \BibitemOpen
  \bibfield  {author} {\bibinfo {author} {\bibfnamefont {P.~T.}\ \bibnamefont
  {Komiske}}, \bibinfo {author} {\bibfnamefont {R.}~\bibnamefont {Mastandrea}},
  \bibinfo {author} {\bibfnamefont {E.~M.}\ \bibnamefont {Metodiev}}, \bibinfo
  {author} {\bibfnamefont {P.}~\bibnamefont {Naik}},\ and\ \bibinfo {author}
  {\bibfnamefont {J.}~\bibnamefont {Thaler}},\ }\bibfield  {title} {\bibinfo
  {title} {{Exploring the Space of Jets with CMS Open Data}},\ }\href
  {https://doi.org/10.1103/PhysRevD.101.034009} {\bibfield  {journal} {\bibinfo
   {journal} {Phys. Rev. D}\ }\textbf {\bibinfo {volume} {101}},\ \bibinfo
  {pages} {034009} (\bibinfo {year} {2020}{\natexlab{b}})},\ \Eprint
  {https://arxiv.org/abs/1908.08542} {arXiv:1908.08542 [hep-ph]} \BibitemShut
  {NoStop}%
\bibitem [{\citenamefont {Dokshitzer}\ \emph {et~al.}(1991)\citenamefont
  {Dokshitzer}, \citenamefont {Khoze}, \citenamefont {Mueller},\ and\
  \citenamefont {Troyan}}]{dokshitzer1991basics}%
  \BibitemOpen
  \bibfield  {author} {\bibinfo {author} {\bibfnamefont {Y.~L.}\ \bibnamefont
  {Dokshitzer}}, \bibinfo {author} {\bibfnamefont {V.}~\bibnamefont {Khoze}},
  \bibinfo {author} {\bibfnamefont {A.}~\bibnamefont {Mueller}},\ and\ \bibinfo
  {author} {\bibfnamefont {S.}~\bibnamefont {Troyan}},\ }\bibfield  {title}
  {\bibinfo {title} {Basics of perturbative qcd, ed},\ }\href@noop {}
  {\bibfield  {journal} {\bibinfo  {journal} {Frontieres, Gif-sur-Yvette
  France}\ } (\bibinfo {year} {1991})}\BibitemShut {NoStop}%
\bibitem [{\citenamefont {Seymour}(1998)}]{Seymour:1997kj}%
  \BibitemOpen
  \bibfield  {author} {\bibinfo {author} {\bibfnamefont {M.~H.}\ \bibnamefont
  {Seymour}},\ }\bibfield  {title} {\bibinfo {title} {{Jet shapes in hadron
  collisions: Higher orders, resummation and hadronization}},\ }\href
  {https://doi.org/10.1016/S0550-3213(97)00711-6} {\bibfield  {journal}
  {\bibinfo  {journal} {Nucl. Phys. B}\ }\textbf {\bibinfo {volume} {513}},\
  \bibinfo {pages} {269} (\bibinfo {year} {1998})},\ \Eprint
  {https://arxiv.org/abs/hep-ph/9707338} {arXiv:hep-ph/9707338} \BibitemShut
  {NoStop}%
\bibitem [{\citenamefont {Amoroso}\ \emph {et~al.}(2021)\citenamefont {Amoroso}
  \emph {et~al.}}]{HSFPhysicsEventGeneratorWG:2020gxw}%
  \BibitemOpen
  \bibfield  {author} {\bibinfo {author} {\bibfnamefont {S.}~\bibnamefont
  {Amoroso}} \emph {et~al.} (\bibinfo {collaboration} {HSF Physics Event
  Generator WG}),\ }\bibfield  {title} {\bibinfo {title} {{Challenges in Monte
  Carlo Event Generator Software for High-Luminosity LHC}},\ }\href
  {https://doi.org/10.1007/s41781-021-00055-1} {\bibfield  {journal} {\bibinfo
  {journal} {Comput. Softw. Big Sci.}\ }\textbf {\bibinfo {volume} {5}},\
  \bibinfo {pages} {12} (\bibinfo {year} {2021})},\ \Eprint
  {https://arxiv.org/abs/2004.13687} {arXiv:2004.13687 [hep-ph]} \BibitemShut
  {NoStop}%
\bibitem [{\citenamefont {Campbell}\ \emph {et~al.}(2024)\citenamefont
  {Campbell} \emph {et~al.}}]{Campbell:2022qmc}%
  \BibitemOpen
  \bibfield  {author} {\bibinfo {author} {\bibfnamefont {J.~M.}\ \bibnamefont
  {Campbell}} \emph {et~al.},\ }\bibfield  {title} {\bibinfo {title} {{Event
  generators for high-energy physics experiments}},\ }\href
  {https://doi.org/10.21468/SciPostPhys.16.5.130} {\bibfield  {journal}
  {\bibinfo  {journal} {SciPost Phys.}\ }\textbf {\bibinfo {volume} {16}},\
  \bibinfo {pages} {130} (\bibinfo {year} {2024})},\ \Eprint
  {https://arxiv.org/abs/2203.11110} {arXiv:2203.11110 [hep-ph]} \BibitemShut
  {NoStop}%
\bibitem [{\citenamefont {Aad}\ \emph {et~al.}(2022)\citenamefont {Aad} \emph
  {et~al.}}]{ATLAS:2021yza}%
  \BibitemOpen
  \bibfield  {author} {\bibinfo {author} {\bibfnamefont {G.}~\bibnamefont
  {Aad}} \emph {et~al.} (\bibinfo {collaboration} {ATLAS}),\ }\bibfield
  {title} {\bibinfo {title} {{Modelling and computational improvements to the
  simulation of single vector-boson plus jet processes for the ATLAS
  experiment}},\ }\href {https://doi.org/10.1007/JHEP08(2022)089} {\bibfield
  {journal} {\bibinfo  {journal} {JHEP}\ }\textbf {\bibinfo {volume} {08}},\
  \bibinfo {pages} {089}},\ \Eprint {https://arxiv.org/abs/2112.09588}
  {arXiv:2112.09588 [hep-ex]} \BibitemShut {NoStop}%
\bibitem [{\citenamefont {Tumasyan}\ \emph {et~al.}(2023)\citenamefont
  {Tumasyan} \emph {et~al.}}]{CMS:2022psv}%
  \BibitemOpen
  \bibfield  {author} {\bibinfo {author} {\bibfnamefont {A.}~\bibnamefont
  {Tumasyan}} \emph {et~al.} (\bibinfo {collaboration} {CMS}),\ }\bibfield
  {title} {\bibinfo {title} {{Search for Higgs Boson Decay to a Charm
  Quark-Antiquark Pair in Proton-Proton Collisions at s=13{\,}{\,}TeV}},\
  }\href {https://doi.org/10.1103/PhysRevLett.131.061801} {\bibfield  {journal}
  {\bibinfo  {journal} {Phys. Rev. Lett.}\ }\textbf {\bibinfo {volume} {131}},\
  \bibinfo {pages} {061801} (\bibinfo {year} {2023})},\ \Eprint
  {https://arxiv.org/abs/2205.05550} {arXiv:2205.05550 [hep-ex]} \BibitemShut
  {NoStop}%
\bibitem [{\citenamefont {Nason}(2007)}]{Nason:2007vt}%
  \BibitemOpen
  \bibfield  {author} {\bibinfo {author} {\bibfnamefont {P.}~\bibnamefont
  {Nason}},\ }\bibfield  {title} {\bibinfo {title} {{MINT: A Computer program
  for adaptive Monte Carlo integration and generation of unweighted
  distributions}},\ }\href@noop {} {\  (\bibinfo {year} {2007})},\ \Eprint
  {https://arxiv.org/abs/0709.2085} {arXiv:0709.2085 [hep-ph]} \BibitemShut
  {NoStop}%
\bibitem [{\citenamefont {Borisyak}\ and\ \citenamefont
  {Kazeev}(2019)}]{Borisyak:2019vbz}%
  \BibitemOpen
  \bibfield  {author} {\bibinfo {author} {\bibfnamefont {M.}~\bibnamefont
  {Borisyak}}\ and\ \bibinfo {author} {\bibfnamefont {N.}~\bibnamefont
  {Kazeev}},\ }\bibfield  {title} {\bibinfo {title} {{Machine Learning on data
  with sPlot background subtraction}},\ }\href
  {https://doi.org/10.1088/1748-0221/14/08/P08020} {\bibfield  {journal}
  {\bibinfo  {journal} {JINST}\ }\textbf {\bibinfo {volume} {14}}\bibfield
  {number} {\bibinfo  {number} { (08)},\ \bibinfo {pages} {P08020}},\ }\Eprint
  {https://arxiv.org/abs/1905.11719} {arXiv:1905.11719 [cs.LG]} \BibitemShut
  {NoStop}%
\bibitem [{\citenamefont {Nachman}\ and\ \citenamefont
  {Thaler}(2020)}]{Nachman:2020fff}%
  \BibitemOpen
  \bibfield  {author} {\bibinfo {author} {\bibfnamefont {B.}~\bibnamefont
  {Nachman}}\ and\ \bibinfo {author} {\bibfnamefont {J.}~\bibnamefont
  {Thaler}},\ }\bibfield  {title} {\bibinfo {title} {{Neural resampler for
  Monte Carlo reweighting with preserved uncertainties}},\ }\href
  {https://doi.org/10.1103/PhysRevD.102.076004} {\bibfield  {journal} {\bibinfo
   {journal} {Phys. Rev. D}\ }\textbf {\bibinfo {volume} {102}},\ \bibinfo
  {pages} {076004} (\bibinfo {year} {2020})},\ \Eprint
  {https://arxiv.org/abs/2007.11586} {arXiv:2007.11586 [hep-ph]} \BibitemShut
  {NoStop}%
\bibitem [{\citenamefont {Frederix}\ \emph {et~al.}(2020)\citenamefont
  {Frederix}, \citenamefont {Frixione}, \citenamefont {Prestel},\ and\
  \citenamefont {Torrielli}}]{Frederix:2020trv}%
  \BibitemOpen
  \bibfield  {author} {\bibinfo {author} {\bibfnamefont {R.}~\bibnamefont
  {Frederix}}, \bibinfo {author} {\bibfnamefont {S.}~\bibnamefont {Frixione}},
  \bibinfo {author} {\bibfnamefont {S.}~\bibnamefont {Prestel}},\ and\ \bibinfo
  {author} {\bibfnamefont {P.}~\bibnamefont {Torrielli}},\ }\bibfield  {title}
  {\bibinfo {title} {{On the reduction of negative weights in MC@NLO-type
  matching procedures}},\ }\href {https://doi.org/10.1007/JHEP07(2020)238}
  {\bibfield  {journal} {\bibinfo  {journal} {JHEP}\ }\textbf {\bibinfo
  {volume} {07}},\ \bibinfo {pages} {238}},\ \Eprint
  {https://arxiv.org/abs/2002.12716} {arXiv:2002.12716 [hep-ph]} \BibitemShut
  {NoStop}%
\bibitem [{\citenamefont {Danziger}\ \emph {et~al.}(2021)\citenamefont
  {Danziger}, \citenamefont {H{\"o}che},\ and\ \citenamefont
  {Siegert}}]{Danziger:2021xvr}%
  \BibitemOpen
  \bibfield  {author} {\bibinfo {author} {\bibfnamefont {K.}~\bibnamefont
  {Danziger}}, \bibinfo {author} {\bibfnamefont {S.}~\bibnamefont
  {H{\"o}che}},\ and\ \bibinfo {author} {\bibfnamefont {F.}~\bibnamefont
  {Siegert}},\ }\bibfield  {title} {\bibinfo {title} {{Reducing negative
  weights in Monte Carlo event generation with Sherpa}},\ }\href@noop {} {\
  (\bibinfo {year} {2021})},\ \Eprint {https://arxiv.org/abs/2110.15211}
  {arXiv:2110.15211 [hep-ph]} \BibitemShut {NoStop}%
\bibitem [{\citenamefont {Frederix}\ and\ \citenamefont
  {Torrielli}(2023)}]{Frederix:2023hom}%
  \BibitemOpen
  \bibfield  {author} {\bibinfo {author} {\bibfnamefont {R.}~\bibnamefont
  {Frederix}}\ and\ \bibinfo {author} {\bibfnamefont {P.}~\bibnamefont
  {Torrielli}},\ }\bibfield  {title} {\bibinfo {title} {{A new way of reducing
  negative weights in MC@NLO}},\ }\href
  {https://doi.org/10.1140/epjc/s10052-023-12243-x} {\bibfield  {journal}
  {\bibinfo  {journal} {Eur. Phys. J. C}\ }\textbf {\bibinfo {volume} {83}},\
  \bibinfo {pages} {1051} (\bibinfo {year} {2023})},\ \Eprint
  {https://arxiv.org/abs/2310.04160} {arXiv:2310.04160 [hep-ph]} \BibitemShut
  {NoStop}%
\bibitem [{\citenamefont {Andersen}\ \emph {et~al.}(2023)\citenamefont
  {Andersen}, \citenamefont {Maier},\ and\ \citenamefont
  {Ma{\^\i}tre}}]{Andersen:2023cku}%
  \BibitemOpen
  \bibfield  {author} {\bibinfo {author} {\bibfnamefont {J.~R.}\ \bibnamefont
  {Andersen}}, \bibinfo {author} {\bibfnamefont {A.}~\bibnamefont {Maier}},\
  and\ \bibinfo {author} {\bibfnamefont {D.}~\bibnamefont {Ma{\^\i}tre}},\
  }\bibfield  {title} {\bibinfo {title} {{Efficient negative-weight elimination
  in large high-multiplicity Monte Carlo event samples}},\ }\href
  {https://doi.org/10.1140/epjc/s10052-023-11905-0} {\bibfield  {journal}
  {\bibinfo  {journal} {Eur. Phys. J. C}\ }\textbf {\bibinfo {volume} {83}},\
  \bibinfo {pages} {835} (\bibinfo {year} {2023})},\ \Eprint
  {https://arxiv.org/abs/2303.15246} {arXiv:2303.15246 [hep-ph]} \BibitemShut
  {NoStop}%
\bibitem [{\citenamefont {Andersen}\ \emph {et~al.}(2024)\citenamefont
  {Andersen}, \citenamefont {Cueto}, \citenamefont {Jones},\ and\ \citenamefont
  {Maier}}]{Andersen:2024mqh}%
  \BibitemOpen
  \bibfield  {author} {\bibinfo {author} {\bibfnamefont {J.~R.}\ \bibnamefont
  {Andersen}}, \bibinfo {author} {\bibfnamefont {A.}~\bibnamefont {Cueto}},
  \bibinfo {author} {\bibfnamefont {S.~P.}\ \bibnamefont {Jones}},\ and\
  \bibinfo {author} {\bibfnamefont {A.}~\bibnamefont {Maier}},\ }\bibfield
  {title} {\bibinfo {title} {{A Cell Resampler study of Negative Weights in
  Multi-jet Merged Samples}},\ }\href@noop {} {\  (\bibinfo {year} {2024})},\
  \Eprint {https://arxiv.org/abs/2411.11651} {arXiv:2411.11651 [hep-ph]}
  \BibitemShut {NoStop}%
\bibitem [{\citenamefont {Glazier}\ and\ \citenamefont
  {Tyson}(2026)}]{Glazier:2024ogg}%
  \BibitemOpen
  \bibfield  {author} {\bibinfo {author} {\bibfnamefont {D.~I.}\ \bibnamefont
  {Glazier}}\ and\ \bibinfo {author} {\bibfnamefont {R.}~\bibnamefont
  {Tyson}},\ }\bibfield  {title} {\bibinfo {title} {{Converting sWeights to
  probabilities with density ratios}},\ }\href
  {https://doi.org/10.1016/j.cpc.2025.109890} {\bibfield  {journal} {\bibinfo
  {journal} {Comput. Phys. Commun.}\ }\textbf {\bibinfo {volume} {318}},\
  \bibinfo {pages} {109890} (\bibinfo {year} {2026})},\ \Eprint
  {https://arxiv.org/abs/2409.08183} {arXiv:2409.08183 [physics.data-an]}
  \BibitemShut {NoStop}%
\bibitem [{\citenamefont {Drnevich}\ \emph {et~al.}(2025)\citenamefont
  {Drnevich}, \citenamefont {Jiggins}, \citenamefont {Katzy},\ and\
  \citenamefont {Cranmer}}]{Drnevich:2024vfj}%
  \BibitemOpen
  \bibfield  {author} {\bibinfo {author} {\bibfnamefont {M.}~\bibnamefont
  {Drnevich}}, \bibinfo {author} {\bibfnamefont {S.}~\bibnamefont {Jiggins}},
  \bibinfo {author} {\bibfnamefont {J.}~\bibnamefont {Katzy}},\ and\ \bibinfo
  {author} {\bibfnamefont {K.}~\bibnamefont {Cranmer}},\ }\bibfield  {title}
  {\bibinfo {title} {{Neural quasiprobabilistic likelihood ratio estimation
  with negatively weighted data}},\ }\href
  {https://doi.org/10.1088/2632-2153/ae0def} {\bibfield  {journal} {\bibinfo
  {journal} {Mach. Learn. Sci. Tech.}\ }\textbf {\bibinfo {volume} {6}},\
  \bibinfo {pages} {045023} (\bibinfo {year} {2025})},\ \Eprint
  {https://arxiv.org/abs/2410.10216} {arXiv:2410.10216 [stat.ML]} \BibitemShut
  {NoStop}%
\bibitem [{\citenamefont {Hayrapetyan}\ \emph {et~al.}(2025)\citenamefont
  {Hayrapetyan} \emph {et~al.}}]{CMS:2024jdl}%
  \BibitemOpen
  \bibfield  {author} {\bibinfo {author} {\bibfnamefont {A.}~\bibnamefont
  {Hayrapetyan}} \emph {et~al.} (\bibinfo {collaboration} {CMS}),\ }\bibfield
  {title} {\bibinfo {title} {{Reweighting simulated events using
  machine-learning techniques in the CMS experiment}},\ }\href
  {https://doi.org/10.1140/epjc/s10052-025-14097-x} {\bibfield  {journal}
  {\bibinfo  {journal} {Eur. Phys. J. C}\ }\textbf {\bibinfo {volume} {85}},\
  \bibinfo {pages} {495} (\bibinfo {year} {2025})},\ \Eprint
  {https://arxiv.org/abs/2411.03023} {arXiv:2411.03023 [hep-ex]} \BibitemShut
  {NoStop}%
\bibitem [{\citenamefont {Jan{\ss}en}\ \emph {et~al.}(2025)\citenamefont
  {Jan{\ss}en}, \citenamefont {Poncelet},\ and\ \citenamefont
  {Schumann}}]{Janssen:2025zke}%
  \BibitemOpen
  \bibfield  {author} {\bibinfo {author} {\bibfnamefont {T.}~\bibnamefont
  {Jan{\ss}en}}, \bibinfo {author} {\bibfnamefont {R.}~\bibnamefont
  {Poncelet}},\ and\ \bibinfo {author} {\bibfnamefont {S.}~\bibnamefont
  {Schumann}},\ }\bibfield  {title} {\bibinfo {title} {{Sampling NNLO QCD phase
  space with normalizing flows}},\ }\href
  {https://doi.org/10.1007/JHEP09(2025)194} {\bibfield  {journal} {\bibinfo
  {journal} {JHEP}\ }\textbf {\bibinfo {volume} {09}},\ \bibinfo {pages}
  {194}},\ \Eprint {https://arxiv.org/abs/2505.13608} {arXiv:2505.13608
  [hep-ph]} \BibitemShut {NoStop}%
\bibitem [{\citenamefont
  {Shyamsundar}(2025{\natexlab{a}})}]{Shyamsundar:2025nzn}%
  \BibitemOpen
  \bibfield  {author} {\bibinfo {author} {\bibfnamefont {P.}~\bibnamefont
  {Shyamsundar}},\ }\bibfield  {title} {\bibinfo {title} {{ARCANE Reweighting:
  A Monte Carlo Technique to Tackle the Negative Weights Problem in Collider
  Event Generation}},\ }\href@noop {} {\  (\bibinfo {year}
  {2025}{\natexlab{a}})},\ \Eprint {https://arxiv.org/abs/2502.08052}
  {arXiv:2502.08052 [hep-ph]} \BibitemShut {NoStop}%
\bibitem [{\citenamefont
  {Shyamsundar}(2025{\natexlab{b}})}]{Shyamsundar:2025mfw}%
  \BibitemOpen
  \bibfield  {author} {\bibinfo {author} {\bibfnamefont {P.}~\bibnamefont
  {Shyamsundar}},\ }\bibfield  {title} {\bibinfo {title} {{A Demonstration of
  ARCANE Reweighting: Reducing the Sign Problem in the MC@NLO Generation of
  $e^+ e^- \rightarrow q \bar{q} + 1\, jet$ Events}},\ }\href@noop {} {\
  (\bibinfo {year} {2025}{\natexlab{b}})},\ \Eprint
  {https://arxiv.org/abs/2502.08053} {arXiv:2502.08053 [hep-ph]} \BibitemShut
  {NoStop}%
\bibitem [{\citenamefont {van Beekveld}\ \emph {et~al.}(2025)\citenamefont {van
  Beekveld}, \citenamefont {Ferrario~Ravasio}, \citenamefont {Helliwell},
  \citenamefont {Karlberg}, \citenamefont {Salam}, \citenamefont {Scyboz},
  \citenamefont {Soto-Ontoso}, \citenamefont {Soyez},\ and\ \citenamefont
  {Zanoli}}]{vanBeekveld:2025lpz}%
  \BibitemOpen
  \bibfield  {author} {\bibinfo {author} {\bibfnamefont {M.}~\bibnamefont {van
  Beekveld}}, \bibinfo {author} {\bibfnamefont {S.}~\bibnamefont
  {Ferrario~Ravasio}}, \bibinfo {author} {\bibfnamefont {J.}~\bibnamefont
  {Helliwell}}, \bibinfo {author} {\bibfnamefont {A.}~\bibnamefont {Karlberg}},
  \bibinfo {author} {\bibfnamefont {G.~P.}\ \bibnamefont {Salam}}, \bibinfo
  {author} {\bibfnamefont {L.}~\bibnamefont {Scyboz}}, \bibinfo {author}
  {\bibfnamefont {A.}~\bibnamefont {Soto-Ontoso}}, \bibinfo {author}
  {\bibfnamefont {G.}~\bibnamefont {Soyez}},\ and\ \bibinfo {author}
  {\bibfnamefont {S.}~\bibnamefont {Zanoli}},\ }\bibfield  {title} {\bibinfo
  {title} {{Logarithmically-accurate and positive-definite NLO shower
  matching}},\ }\href {https://doi.org/10.1007/JHEP10(2025)038} {\bibfield
  {journal} {\bibinfo  {journal} {JHEP}\ }\textbf {\bibinfo {volume} {10}},\
  \bibinfo {pages} {038}},\ \Eprint {https://arxiv.org/abs/2504.05377}
  {arXiv:2504.05377 [hep-ph]} \BibitemShut {NoStop}%
\bibitem [{\citenamefont {Palmer}\ and\ \citenamefont
  {Kronheim}(2026)}]{Palmer:2025jmb}%
  \BibitemOpen
  \bibfield  {author} {\bibinfo {author} {\bibfnamefont {C.}~\bibnamefont
  {Palmer}}\ and\ \bibinfo {author} {\bibfnamefont {B.}~\bibnamefont
  {Kronheim}},\ }\bibfield  {title} {\bibinfo {title} {{Improving statistical
  precision in Monte~Carlo samples with negative weights via reweighting and
  uncertainty quantification}},\ }\href {https://doi.org/10.1103/k8w6-wn37}
  {\bibfield  {journal} {\bibinfo  {journal} {Phys. Rev. D}\ }\textbf {\bibinfo
  {volume} {113}},\ \bibinfo {pages} {012003} (\bibinfo {year} {2026})},\
  \Eprint {https://arxiv.org/abs/2510.16217} {arXiv:2510.16217 [hep-ex]}
  \BibitemShut {NoStop}%
\bibitem [{\citenamefont {Nachman}\ and\ \citenamefont
  {Noll}(2025)}]{Nachman:2025lid}%
  \BibitemOpen
  \bibfield  {author} {\bibinfo {author} {\bibfnamefont {B.}~\bibnamefont
  {Nachman}}\ and\ \bibinfo {author} {\bibfnamefont {D.}~\bibnamefont {Noll}},\
  }\bibfield  {title} {\bibinfo {title} {{Neural refinement of sample
  weights}},\ }\href {https://doi.org/10.1103/yx16-h7n9} {\bibfield  {journal}
  {\bibinfo  {journal} {Phys. Rev. D}\ }\textbf {\bibinfo {volume} {112}},\
  \bibinfo {pages} {096009} (\bibinfo {year} {2025})},\ \Eprint
  {https://arxiv.org/abs/2505.03724} {arXiv:2505.03724 [hep-ph]} \BibitemShut
  {NoStop}%
\bibitem [{\citenamefont {Gambhir}\ and\ \citenamefont
  {Mastandrea}(2026)}]{Gambhir:2025lka}%
  \BibitemOpen
  \bibfield  {author} {\bibinfo {author} {\bibfnamefont {R.}~\bibnamefont
  {Gambhir}}\ and\ \bibinfo {author} {\bibfnamefont {R.}~\bibnamefont
  {Mastandrea}},\ }\bibfield  {title} {\bibinfo {title} {{Resummed distribution
  functions: making perturbation theory positive and normalized}},\ }\href
  {https://doi.org/10.1007/JHEP06(2026)243} {\bibfield  {journal} {\bibinfo
  {journal} {JHEP}\ }\textbf {\bibinfo {volume} {06}},\ \bibinfo {pages}
  {243}},\ \Eprint {https://arxiv.org/abs/2512.04160} {arXiv:2512.04160
  [hep-ph]} \BibitemShut {NoStop}%
\bibitem [{\citenamefont {Heimel}\ \emph {et~al.}(2026)\citenamefont {Heimel},
  \citenamefont {Plehn}, \citenamefont {Revelli}, \citenamefont {Vent},\ and\
  \citenamefont {Winterhalder}}]{Heimel:2026cxh}%
  \BibitemOpen
  \bibfield  {author} {\bibinfo {author} {\bibfnamefont {T.}~\bibnamefont
  {Heimel}}, \bibinfo {author} {\bibfnamefont {T.}~\bibnamefont {Plehn}},
  \bibinfo {author} {\bibfnamefont {R.}~\bibnamefont {Revelli}}, \bibinfo
  {author} {\bibfnamefont {S.}~\bibnamefont {Vent}},\ and\ \bibinfo {author}
  {\bibfnamefont {R.}~\bibnamefont {Winterhalder}},\ }\bibfield  {title}
  {\bibinfo {title} {{Neural Control Variates at LO and NLO}},\ }\href@noop {}
  {\  (\bibinfo {year} {2026})},\ \Eprint {https://arxiv.org/abs/2607.23591}
  {arXiv:2607.23591 [hep-ph]} \BibitemShut {NoStop}%
\bibitem [{\citenamefont {Hausdorff}(1918)}]{hausdorff1918dimension}%
  \BibitemOpen
  \bibfield  {author} {\bibinfo {author} {\bibfnamefont {F.}~\bibnamefont
  {Hausdorff}},\ }\bibfield  {title} {\bibinfo {title} {Dimension und
  {\"a}u{\ss}eres ma{\ss}},\ }\href@noop {} {\bibfield  {journal} {\bibinfo
  {journal} {Mathematische Annalen}\ }\textbf {\bibinfo {volume} {79}},\
  \bibinfo {pages} {157} (\bibinfo {year} {1918})}\BibitemShut {NoStop}%
\bibitem [{\citenamefont {Mandelbrot}(1967)}]{mandelbrot1967long}%
  \BibitemOpen
  \bibfield  {author} {\bibinfo {author} {\bibfnamefont {B.}~\bibnamefont
  {Mandelbrot}},\ }\bibfield  {title} {\bibinfo {title} {How long is the coast
  of britain? statistical self-similarity and fractional dimension},\
  }\href@noop {} {\bibfield  {journal} {\bibinfo  {journal} {science}\ }\textbf
  {\bibinfo {volume} {156}},\ \bibinfo {pages} {636} (\bibinfo {year}
  {1967})}\BibitemShut {NoStop}%
\bibitem [{\citenamefont {Richardson}(1961)}]{richardson1961problem}%
  \BibitemOpen
  \bibfield  {author} {\bibinfo {author} {\bibfnamefont {L.~F.}\ \bibnamefont
  {Richardson}},\ }\bibfield  {title} {\bibinfo {title} {The problem of
  contiguity: an appendix of statistics of deadly quarrels},\ }\href@noop {}
  {\bibfield  {journal} {\bibinfo  {journal} {General system yearbook}\
  }\textbf {\bibinfo {volume} {6}},\ \bibinfo {pages} {139} (\bibinfo {year}
  {1961})}\BibitemShut {NoStop}%
\bibitem [{\citenamefont {Abbott}\ and\ \citenamefont
  {Wise}(1981)}]{Abbott:1979bh}%
  \BibitemOpen
  \bibfield  {author} {\bibinfo {author} {\bibfnamefont {L.~F.}\ \bibnamefont
  {Abbott}}\ and\ \bibinfo {author} {\bibfnamefont {M.~B.}\ \bibnamefont
  {Wise}},\ }\bibfield  {title} {\bibinfo {title} {{The Dimension of a Quantum
  Mechanical Path}},\ }\href {https://doi.org/10.1119/1.12657} {\bibfield
  {journal} {\bibinfo  {journal} {Am. J. Phys.}\ }\textbf {\bibinfo {volume}
  {49}},\ \bibinfo {pages} {37} (\bibinfo {year} {1981})}\BibitemShut {NoStop}%
\bibitem [{\citenamefont {Gustafson}\ and\ \citenamefont
  {Nilsson}(1991)}]{Gustafson:1991ru}%
  \BibitemOpen
  \bibfield  {author} {\bibinfo {author} {\bibfnamefont {G.}~\bibnamefont
  {Gustafson}}\ and\ \bibinfo {author} {\bibfnamefont {A.}~\bibnamefont
  {Nilsson}},\ }\bibfield  {title} {\bibinfo {title} {{Multifractal dimensions
  in QCD cascades}},\ }\href {https://doi.org/10.1007/BF01559451} {\bibfield
  {journal} {\bibinfo  {journal} {Z. Phys. C}\ }\textbf {\bibinfo {volume}
  {52}},\ \bibinfo {pages} {533} (\bibinfo {year} {1991})}\BibitemShut
  {NoStop}%
\bibitem [{\citenamefont {Bjorken}(1992)}]{Bjorken:1991ft}%
  \BibitemOpen
  \bibfield  {author} {\bibinfo {author} {\bibfnamefont {J.~D.}\ \bibnamefont
  {Bjorken}},\ }\bibfield  {title} {\bibinfo {title} {{A Plumber's view of
  perturbative QCD}},\ }\href {https://doi.org/10.1103/PhysRevD.45.4077}
  {\bibfield  {journal} {\bibinfo  {journal} {Phys. Rev. D}\ }\textbf {\bibinfo
  {volume} {45}},\ \bibinfo {pages} {4077} (\bibinfo {year}
  {1992})}\BibitemShut {NoStop}%
\bibitem [{\citenamefont {Ahmad}\ and\ \citenamefont
  {Tesauro}(1988)}]{ahmad1988scaling}%
  \BibitemOpen
  \bibfield  {author} {\bibinfo {author} {\bibfnamefont {S.}~\bibnamefont
  {Ahmad}}\ and\ \bibinfo {author} {\bibfnamefont {G.}~\bibnamefont
  {Tesauro}},\ }\bibfield  {title} {\bibinfo {title} {Scaling and
  generalization in neural networks: a case study},\ }\href@noop {} {\bibfield
  {journal} {\bibinfo  {journal} {Advances in neural information processing
  systems}\ }\textbf {\bibinfo {volume} {1}} (\bibinfo {year}
  {1988})}\BibitemShut {NoStop}%
\bibitem [{\citenamefont {Cohn}\ and\ \citenamefont
  {Tesauro}(1990)}]{cohn1990can}%
  \BibitemOpen
  \bibfield  {author} {\bibinfo {author} {\bibfnamefont {D.}~\bibnamefont
  {Cohn}}\ and\ \bibinfo {author} {\bibfnamefont {G.}~\bibnamefont {Tesauro}},\
  }\bibfield  {title} {\bibinfo {title} {Can neural networks do better than the
  vapnik-chervonenkis bounds?},\ }\href@noop {} {\bibfield  {journal} {\bibinfo
   {journal} {Advances in Neural Information Processing Systems}\ }\textbf
  {\bibinfo {volume} {3}} (\bibinfo {year} {1990})}\BibitemShut {NoStop}%
\bibitem [{\citenamefont {Hestness}\ \emph {et~al.}(2017)\citenamefont
  {Hestness}, \citenamefont {Narang}, \citenamefont {Ardalani}, \citenamefont
  {Diamos}, \citenamefont {Jun}, \citenamefont {Kianinejad}, \citenamefont
  {Patwary}, \citenamefont {Yang},\ and\ \citenamefont
  {Zhou}}]{hestness2017deep}%
  \BibitemOpen
  \bibfield  {author} {\bibinfo {author} {\bibfnamefont {J.}~\bibnamefont
  {Hestness}}, \bibinfo {author} {\bibfnamefont {S.}~\bibnamefont {Narang}},
  \bibinfo {author} {\bibfnamefont {N.}~\bibnamefont {Ardalani}}, \bibinfo
  {author} {\bibfnamefont {G.}~\bibnamefont {Diamos}}, \bibinfo {author}
  {\bibfnamefont {H.}~\bibnamefont {Jun}}, \bibinfo {author} {\bibfnamefont
  {H.}~\bibnamefont {Kianinejad}}, \bibinfo {author} {\bibfnamefont {M.~M.~A.}\
  \bibnamefont {Patwary}}, \bibinfo {author} {\bibfnamefont {Y.}~\bibnamefont
  {Yang}},\ and\ \bibinfo {author} {\bibfnamefont {Y.}~\bibnamefont {Zhou}},\
  }\bibfield  {title} {\bibinfo {title} {Deep learning scaling is predictable,
  empirically},\ }\href@noop {} {\bibfield  {journal} {\bibinfo  {journal}
  {arXiv preprint arXiv:1712.00409}\ } (\bibinfo {year} {2017})}\BibitemShut
  {NoStop}%
\bibitem [{\citenamefont {Rosenfeld}\ \emph {et~al.}(2019)\citenamefont
  {Rosenfeld}, \citenamefont {Rosenfeld}, \citenamefont {Belinkov},\ and\
  \citenamefont {Shavit}}]{rosenfeld2019constructive}%
  \BibitemOpen
  \bibfield  {author} {\bibinfo {author} {\bibfnamefont {J.~S.}\ \bibnamefont
  {Rosenfeld}}, \bibinfo {author} {\bibfnamefont {A.}~\bibnamefont
  {Rosenfeld}}, \bibinfo {author} {\bibfnamefont {Y.}~\bibnamefont
  {Belinkov}},\ and\ \bibinfo {author} {\bibfnamefont {N.}~\bibnamefont
  {Shavit}},\ }\bibfield  {title} {\bibinfo {title} {A constructive prediction
  of the generalization error across scales},\ }\href@noop {} {\bibfield
  {journal} {\bibinfo  {journal} {arXiv preprint arXiv:1909.12673}\ } (\bibinfo
  {year} {2019})}\BibitemShut {NoStop}%
\bibitem [{\citenamefont {Henighan}\ \emph {et~al.}(2020)\citenamefont
  {Henighan}, \citenamefont {Kaplan}, \citenamefont {Katz}, \citenamefont
  {Chen}, \citenamefont {Hesse}, \citenamefont {Jackson}, \citenamefont {Jun},
  \citenamefont {Brown}, \citenamefont {Dhariwal}, \citenamefont {Gray} \emph
  {et~al.}}]{henighan2020scaling}%
  \BibitemOpen
  \bibfield  {author} {\bibinfo {author} {\bibfnamefont {T.}~\bibnamefont
  {Henighan}}, \bibinfo {author} {\bibfnamefont {J.}~\bibnamefont {Kaplan}},
  \bibinfo {author} {\bibfnamefont {M.}~\bibnamefont {Katz}}, \bibinfo {author}
  {\bibfnamefont {M.}~\bibnamefont {Chen}}, \bibinfo {author} {\bibfnamefont
  {C.}~\bibnamefont {Hesse}}, \bibinfo {author} {\bibfnamefont
  {J.}~\bibnamefont {Jackson}}, \bibinfo {author} {\bibfnamefont
  {H.}~\bibnamefont {Jun}}, \bibinfo {author} {\bibfnamefont {T.~B.}\
  \bibnamefont {Brown}}, \bibinfo {author} {\bibfnamefont {P.}~\bibnamefont
  {Dhariwal}}, \bibinfo {author} {\bibfnamefont {S.}~\bibnamefont {Gray}},
  \emph {et~al.},\ }\bibfield  {title} {\bibinfo {title} {Scaling laws for
  autoregressive generative modeling},\ }\href@noop {} {\bibfield  {journal}
  {\bibinfo  {journal} {arXiv preprint arXiv:2010.14701}\ } (\bibinfo {year}
  {2020})}\BibitemShut {NoStop}%
\bibitem [{\citenamefont {Rosenfeld}\ \emph {et~al.}(2021)\citenamefont
  {Rosenfeld}, \citenamefont {Frankle}, \citenamefont {Carbin},\ and\
  \citenamefont {Shavit}}]{rosenfeld2021predictability}%
  \BibitemOpen
  \bibfield  {author} {\bibinfo {author} {\bibfnamefont {J.~S.}\ \bibnamefont
  {Rosenfeld}}, \bibinfo {author} {\bibfnamefont {J.}~\bibnamefont {Frankle}},
  \bibinfo {author} {\bibfnamefont {M.}~\bibnamefont {Carbin}},\ and\ \bibinfo
  {author} {\bibfnamefont {N.}~\bibnamefont {Shavit}},\ }\bibfield  {title}
  {\bibinfo {title} {On the predictability of pruning across scales},\ }in\
  \href@noop {} {\emph {\bibinfo {booktitle} {International conference on
  machine learning}}}\ (\bibinfo {organization} {PMLR},\ \bibinfo {year}
  {2021})\ pp.\ \bibinfo {pages} {9075--9083}\BibitemShut {NoStop}%
\bibitem [{\citenamefont {Batson}\ and\ \citenamefont
  {Kahn}(2025)}]{Batson:2023ohn}%
  \BibitemOpen
  \bibfield  {author} {\bibinfo {author} {\bibfnamefont {J.}~\bibnamefont
  {Batson}}\ and\ \bibinfo {author} {\bibfnamefont {Y.~F.}\ \bibnamefont
  {Kahn}},\ }\bibfield  {title} {\bibinfo {title} {{Scaling laws in jet
  classification}},\ }\href {https://doi.org/10.21468/SciPostPhysCore.8.1.034}
  {\bibfield  {journal} {\bibinfo  {journal} {SciPost Phys. Core}\ }\textbf
  {\bibinfo {volume} {8}},\ \bibinfo {pages} {034} (\bibinfo {year} {2025})},\
  \Eprint {https://arxiv.org/abs/2312.02264} {arXiv:2312.02264 [hep-ph]}
  \BibitemShut {NoStop}%
\bibitem [{ATL(2026)}]{ATLAS:2026zdb}%
  \BibitemOpen
  \bibfield  {title} {\bibinfo {title} {{Carpe Datum: Scaling behavior of
  transformers for heavy hadron flavor identification}},\ }\href@noop {} {\
  (\bibinfo {year} {2026})}\BibitemShut {NoStop}%
\bibitem [{\citenamefont {Bahl}\ \emph {et~al.}(2026)\citenamefont {Bahl},
  \citenamefont {Bres{\'o}-Pla}, \citenamefont {Butter},\ and\ \citenamefont
  {Ramirez}}]{Bahl:2026jvt}%
  \BibitemOpen
  \bibfield  {author} {\bibinfo {author} {\bibfnamefont {H.}~\bibnamefont
  {Bahl}}, \bibinfo {author} {\bibfnamefont {V.}~\bibnamefont {Bres{\'o}-Pla}},
  \bibinfo {author} {\bibfnamefont {A.}~\bibnamefont {Butter}},\ and\ \bibinfo
  {author} {\bibfnamefont {J.~I.}\ \bibnamefont {Ramirez}},\ }\bibfield
  {title} {\bibinfo {title} {{Scaling laws for amplitude surrogates}},\
  }\href@noop {} {\  (\bibinfo {year} {2026})},\ \Eprint
  {https://arxiv.org/abs/2601.13308} {arXiv:2601.13308 [hep-ph]} \BibitemShut
  {NoStop}%
\bibitem [{\citenamefont {Vigl}\ \emph {et~al.}(2026)\citenamefont {Vigl},
  \citenamefont {Hartman}, \citenamefont {Kagan},\ and\ \citenamefont
  {Heinrich}}]{Vigl:2026ppx}%
  \BibitemOpen
  \bibfield  {author} {\bibinfo {author} {\bibfnamefont {M.}~\bibnamefont
  {Vigl}}, \bibinfo {author} {\bibfnamefont {N.}~\bibnamefont {Hartman}},
  \bibinfo {author} {\bibfnamefont {M.}~\bibnamefont {Kagan}},\ and\ \bibinfo
  {author} {\bibfnamefont {L.}~\bibnamefont {Heinrich}},\ }\bibfield  {title}
  {\bibinfo {title} {{Neural Scaling Laws for Boosted Jet Tagging}},\
  }\href@noop {} {\  (\bibinfo {year} {2026})},\ \Eprint
  {https://arxiv.org/abs/2602.15781} {arXiv:2602.15781 [hep-ex]} \BibitemShut
  {NoStop}%
\bibitem [{\citenamefont {Amram}\ \emph {et~al.}(2026)\citenamefont {Amram},
  \citenamefont {Faroughy}, \citenamefont {Gerdes}, \citenamefont {Hallin},
  \citenamefont {Kasieczka}, \citenamefont {Kr{\"a}mer}, \citenamefont
  {Reyes-Gonzalez},\ and\ \citenamefont {Shih}}]{Amram:2026zzv}%
  \BibitemOpen
  \bibfield  {author} {\bibinfo {author} {\bibfnamefont {O.}~\bibnamefont
  {Amram}}, \bibinfo {author} {\bibfnamefont {D.~A.}\ \bibnamefont {Faroughy}},
  \bibinfo {author} {\bibfnamefont {T.}~\bibnamefont {Gerdes}}, \bibinfo
  {author} {\bibfnamefont {A.}~\bibnamefont {Hallin}}, \bibinfo {author}
  {\bibfnamefont {G.}~\bibnamefont {Kasieczka}}, \bibinfo {author}
  {\bibfnamefont {M.}~\bibnamefont {Kr{\"a}mer}}, \bibinfo {author}
  {\bibfnamefont {H.}~\bibnamefont {Reyes-Gonzalez}},\ and\ \bibinfo {author}
  {\bibfnamefont {D.}~\bibnamefont {Shih}},\ }\bibfield  {title} {\bibinfo
  {title} {{Neural Scaling Laws for Jet Generation}},\ }\href@noop {} {\
  (\bibinfo {year} {2026})},\ \Eprint {https://arxiv.org/abs/2605.28940}
  {arXiv:2605.28940 [hep-ph]} \BibitemShut {NoStop}%
\bibitem [{\citenamefont {Uslu}\ \emph
  {et~al.}(2026{\natexlab{a}})\citenamefont {Uslu}, \citenamefont {Greif},
  \citenamefont {Whiteson},\ and\ \citenamefont {Nachman}}]{Uslu:2026ywh}%
  \BibitemOpen
  \bibfield  {author} {\bibinfo {author} {\bibfnamefont {J.-L.}\ \bibnamefont
  {Uslu}}, \bibinfo {author} {\bibfnamefont {K.}~\bibnamefont {Greif}},
  \bibinfo {author} {\bibfnamefont {D.}~\bibnamefont {Whiteson}},\ and\
  \bibinfo {author} {\bibfnamefont {B.}~\bibnamefont {Nachman}},\ }\bibfield
  {title} {\bibinfo {title} {{Towards Engineering Scaling Laws with Pretraining
  Data Composition}},\ }\href@noop {} {\  (\bibinfo {year}
  {2026}{\natexlab{a}})},\ \Eprint {https://arxiv.org/abs/2606.19781}
  {arXiv:2606.19781 [hep-ex]} \BibitemShut {NoStop}%
\bibitem [{\citenamefont {Uslu}\ \emph
  {et~al.}(2026{\natexlab{b}})\citenamefont {Uslu}, \citenamefont {Nachman},\
  and\ \citenamefont {Re}}]{Uslu:2026nln}%
  \BibitemOpen
  \bibfield  {author} {\bibinfo {author} {\bibfnamefont {J.-L.}\ \bibnamefont
  {Uslu}}, \bibinfo {author} {\bibfnamefont {B.}~\bibnamefont {Nachman}},\ and\
  \bibinfo {author} {\bibfnamefont {C.}~\bibnamefont {Re}},\ }\bibfield
  {title} {\bibinfo {title} {{Predict before you train: Scaling Laws for
  particle physics foundation models}},\ }\href@noop {} {\  (\bibinfo {year}
  {2026}{\natexlab{b}})},\ \Eprint {https://arxiv.org/abs/2607.23377}
  {arXiv:2607.23377 [hep-ex]} \BibitemShut {NoStop}%
\bibitem [{\citenamefont {Larkoski}(2025{\natexlab{a}})}]{Larkoski:2025clo}%
  \BibitemOpen
  \bibfield  {author} {\bibinfo {author} {\bibfnamefont {A.~J.}\ \bibnamefont
  {Larkoski}},\ }\bibfield  {title} {\bibinfo {title} {{A step toward
  interpretability: smearing the likelihood}},\ }\href
  {https://doi.org/10.1007/JHEP03(2025)198} {\bibfield  {journal} {\bibinfo
  {journal} {JHEP}\ }\textbf {\bibinfo {volume} {03}},\ \bibinfo {pages}
  {198}},\ \Eprint {https://arxiv.org/abs/2501.07643} {arXiv:2501.07643
  [hep-ph]} \BibitemShut {NoStop}%
\bibitem [{\citenamefont {Grassberger}\ and\ \citenamefont
  {Procaccia}(1983)}]{Grassberger:1983zz}%
  \BibitemOpen
  \bibfield  {author} {\bibinfo {author} {\bibfnamefont {P.}~\bibnamefont
  {Grassberger}}\ and\ \bibinfo {author} {\bibfnamefont {I.}~\bibnamefont
  {Procaccia}},\ }\bibfield  {title} {\bibinfo {title} {{Characterization of
  Strange Attractors}},\ }\href {https://doi.org/10.1103/PhysRevLett.50.346}
  {\bibfield  {journal} {\bibinfo  {journal} {Phys. Rev. Lett.}\ }\textbf
  {\bibinfo {volume} {50}},\ \bibinfo {pages} {346} (\bibinfo {year}
  {1983})}\BibitemShut {NoStop}%
\bibitem [{\citenamefont {Theiler}(1990)}]{theiler1990estimating}%
  \BibitemOpen
  \bibfield  {author} {\bibinfo {author} {\bibfnamefont {J.}~\bibnamefont
  {Theiler}},\ }\bibfield  {title} {\bibinfo {title} {Estimating fractal
  dimension},\ }\href@noop {} {\bibfield  {journal} {\bibinfo  {journal}
  {Journal of the optical society of America A}\ }\textbf {\bibinfo {volume}
  {7}},\ \bibinfo {pages} {1055} (\bibinfo {year} {1990})}\BibitemShut
  {NoStop}%
\bibitem [{\citenamefont {Larkoski}(2025{\natexlab{b}})}]{Larkoski:2025idq}%
  \BibitemOpen
  \bibfield  {author} {\bibinfo {author} {\bibfnamefont {A.~J.}\ \bibnamefont
  {Larkoski}},\ }\bibfield  {title} {\bibinfo {title} {{Non-Gaussianities in
  Collider Metric Binning}},\ }\href@noop {} {\  (\bibinfo {year}
  {2025}{\natexlab{b}})},\ \Eprint {https://arxiv.org/abs/2503.03809}
  {arXiv:2503.03809 [hep-ph]} \BibitemShut {NoStop}%
\bibitem [{\citenamefont {Larkoski}(2025{\natexlab{c}})}]{Larkoski:2025pai}%
  \BibitemOpen
  \bibfield  {author} {\bibinfo {author} {\bibfnamefont {A.~J.}\ \bibnamefont
  {Larkoski}},\ }\bibfield  {title} {\bibinfo {title} {{Factorization for
  collider dataspace correlators}},\ }\href {https://doi.org/10.1103/6r43-xhtj}
  {\bibfield  {journal} {\bibinfo  {journal} {Phys. Rev. D}\ }\textbf {\bibinfo
  {volume} {112}},\ \bibinfo {pages} {054012} (\bibinfo {year}
  {2025}{\natexlab{c}})},\ \Eprint {https://arxiv.org/abs/2504.12380}
  {arXiv:2504.12380 [hep-ph]} \BibitemShut {NoStop}%
\bibitem [{\citenamefont {Komiske}\ \emph {et~al.}(2022)\citenamefont
  {Komiske}, \citenamefont {Kryhin},\ and\ \citenamefont
  {Thaler}}]{Komiske:2022vxg}%
  \BibitemOpen
  \bibfield  {author} {\bibinfo {author} {\bibfnamefont {P.~T.}\ \bibnamefont
  {Komiske}}, \bibinfo {author} {\bibfnamefont {S.}~\bibnamefont {Kryhin}},\
  and\ \bibinfo {author} {\bibfnamefont {J.}~\bibnamefont {Thaler}},\
  }\bibfield  {title} {\bibinfo {title} {{Disentangling quarks and gluons in
  CMS open data}},\ }\href {https://doi.org/10.1103/PhysRevD.106.094021}
  {\bibfield  {journal} {\bibinfo  {journal} {Phys. Rev. D}\ }\textbf {\bibinfo
  {volume} {106}},\ \bibinfo {pages} {094021} (\bibinfo {year} {2022})},\
  \Eprint {https://arxiv.org/abs/2205.04459} {arXiv:2205.04459 [hep-ph]}
  \BibitemShut {NoStop}%
\bibitem [{\citenamefont {Cox}\ and\ \citenamefont
  {Melia}(2018)}]{Cox:2018wce}%
  \BibitemOpen
  \bibfield  {author} {\bibinfo {author} {\bibfnamefont {P.}~\bibnamefont
  {Cox}}\ and\ \bibinfo {author} {\bibfnamefont {T.}~\bibnamefont {Melia}},\
  }\bibfield  {title} {\bibinfo {title} {{Independently Parameterised Momenta
  Variables and Monte Carlo IR Subtraction}},\ }\href
  {https://doi.org/10.1007/JHEP12(2018)038} {\bibfield  {journal} {\bibinfo
  {journal} {JHEP}\ }\textbf {\bibinfo {volume} {12}},\ \bibinfo {pages}
  {038}},\ \Eprint {https://arxiv.org/abs/1809.09325} {arXiv:1809.09325
  [hep-ph]} \BibitemShut {NoStop}%
\bibitem [{\citenamefont {Henning}\ and\ \citenamefont
  {Melia}(2019{\natexlab{a}})}]{Henning:2019mcv}%
  \BibitemOpen
  \bibfield  {author} {\bibinfo {author} {\bibfnamefont {B.}~\bibnamefont
  {Henning}}\ and\ \bibinfo {author} {\bibfnamefont {T.}~\bibnamefont
  {Melia}},\ }\bibfield  {title} {\bibinfo {title} {{Conformal-helicity duality
  \textbackslash{}\& the Hilbert space of free CFTs}},\ }\href@noop {} {\
  (\bibinfo {year} {2019}{\natexlab{a}})},\ \Eprint
  {https://arxiv.org/abs/1902.06747} {arXiv:1902.06747 [hep-th]} \BibitemShut
  {NoStop}%
\bibitem [{\citenamefont {Henning}\ and\ \citenamefont
  {Melia}(2019{\natexlab{b}})}]{Henning:2019enq}%
  \BibitemOpen
  \bibfield  {author} {\bibinfo {author} {\bibfnamefont {B.}~\bibnamefont
  {Henning}}\ and\ \bibinfo {author} {\bibfnamefont {T.}~\bibnamefont
  {Melia}},\ }\bibfield  {title} {\bibinfo {title} {{Constructing effective
  field theories via their harmonics}},\ }\href
  {https://doi.org/10.1103/PhysRevD.100.016015} {\bibfield  {journal} {\bibinfo
   {journal} {Phys. Rev. D}\ }\textbf {\bibinfo {volume} {100}},\ \bibinfo
  {pages} {016015} (\bibinfo {year} {2019}{\natexlab{b}})},\ \Eprint
  {https://arxiv.org/abs/1902.06754} {arXiv:1902.06754 [hep-ph]} \BibitemShut
  {NoStop}%
\bibitem [{\citenamefont {Batson}\ \emph {et~al.}(2021)\citenamefont {Batson},
  \citenamefont {Haaf}, \citenamefont {Kahn},\ and\ \citenamefont
  {Roberts}}]{Batson:2021agz}%
  \BibitemOpen
  \bibfield  {author} {\bibinfo {author} {\bibfnamefont {J.}~\bibnamefont
  {Batson}}, \bibinfo {author} {\bibfnamefont {C.~G.}\ \bibnamefont {Haaf}},
  \bibinfo {author} {\bibfnamefont {Y.}~\bibnamefont {Kahn}},\ and\ \bibinfo
  {author} {\bibfnamefont {D.~A.}\ \bibnamefont {Roberts}},\ }\bibfield
  {title} {\bibinfo {title} {{Topological Obstructions to Autoencoding}},\
  }\href {https://doi.org/10.1007/JHEP04(2021)280} {\bibfield  {journal}
  {\bibinfo  {journal} {JHEP}\ }\textbf {\bibinfo {volume} {04}},\ \bibinfo
  {pages} {280}},\ \Eprint {https://arxiv.org/abs/2102.08380} {arXiv:2102.08380
  [hep-ph]} \BibitemShut {NoStop}%
\bibitem [{\citenamefont {Abdesselam}\ \emph {et~al.}(2011)\citenamefont
  {Abdesselam} \emph {et~al.}}]{Abdesselam:2010pt}%
  \BibitemOpen
  \bibfield  {author} {\bibinfo {author} {\bibfnamefont {A.}~\bibnamefont
  {Abdesselam}} \emph {et~al.},\ }\bibfield  {title} {\bibinfo {title}
  {{Boosted Objects: A Probe of Beyond the Standard Model Physics}},\ }\href
  {https://doi.org/10.1140/epjc/s10052-011-1661-y} {\bibfield  {journal}
  {\bibinfo  {journal} {Eur. Phys. J. C}\ }\textbf {\bibinfo {volume} {71}},\
  \bibinfo {pages} {1661} (\bibinfo {year} {2011})},\ \Eprint
  {https://arxiv.org/abs/1012.5412} {arXiv:1012.5412 [hep-ph]} \BibitemShut
  {NoStop}%
\bibitem [{\citenamefont {Takahashi}\ \emph {et~al.}(2026)\citenamefont
  {Takahashi} \emph {et~al.}}]{ParticleDataGroup:2026aaa}%
  \BibitemOpen
  \bibfield  {author} {\bibinfo {author} {\bibfnamefont {F.}~\bibnamefont
  {Takahashi}} \emph {et~al.} (\bibinfo {collaboration} {Particle Data
  Group}),\ }\bibfield  {title} {\bibinfo {title} {{Review of Particle
  Physics}},\ }\href {https://doi.org/10.1142/S0217751X26300115} {\bibfield
  {journal} {\bibinfo  {journal} {Int. J. Mod. Phys. A}\ }\textbf {\bibinfo
  {volume} {41}},\ \bibinfo {pages} {2630011} (\bibinfo {year}
  {2026})}\BibitemShut {NoStop}%
\bibitem [{\citenamefont {Salam}(2010)}]{Salam:2010nqg}%
  \BibitemOpen
  \bibfield  {author} {\bibinfo {author} {\bibfnamefont {G.~P.}\ \bibnamefont
  {Salam}},\ }\bibfield  {title} {\bibinfo {title} {{Towards Jetography}},\
  }\href {https://doi.org/10.1140/epjc/s10052-010-1314-6} {\bibfield  {journal}
  {\bibinfo  {journal} {Eur. Phys. J. C}\ }\textbf {\bibinfo {volume} {67}},\
  \bibinfo {pages} {637} (\bibinfo {year} {2010})},\ \Eprint
  {https://arxiv.org/abs/0906.1833} {arXiv:0906.1833 [hep-ph]} \BibitemShut
  {NoStop}%
\bibitem [{\citenamefont {Ellis}\ \emph {et~al.}(2008)\citenamefont {Ellis},
  \citenamefont {Huston}, \citenamefont {Hatakeyama}, \citenamefont {Loch},\
  and\ \citenamefont {Tonnesmann}}]{Ellis:2007ib}%
  \BibitemOpen
  \bibfield  {author} {\bibinfo {author} {\bibfnamefont {S.~D.}\ \bibnamefont
  {Ellis}}, \bibinfo {author} {\bibfnamefont {J.}~\bibnamefont {Huston}},
  \bibinfo {author} {\bibfnamefont {K.}~\bibnamefont {Hatakeyama}}, \bibinfo
  {author} {\bibfnamefont {P.}~\bibnamefont {Loch}},\ and\ \bibinfo {author}
  {\bibfnamefont {M.}~\bibnamefont {Tonnesmann}},\ }\bibfield  {title}
  {\bibinfo {title} {{Jets in hadron-hadron collisions}},\ }\href
  {https://doi.org/10.1016/j.ppnp.2007.12.002} {\bibfield  {journal} {\bibinfo
  {journal} {Prog. Part. Nucl. Phys.}\ }\textbf {\bibinfo {volume} {60}},\
  \bibinfo {pages} {484} (\bibinfo {year} {2008})},\ \Eprint
  {https://arxiv.org/abs/0712.2447} {arXiv:0712.2447 [hep-ph]} \BibitemShut
  {NoStop}%
\bibitem [{\citenamefont {Gerbush}\ \emph {et~al.}(2008)\citenamefont
  {Gerbush}, \citenamefont {Khoo}, \citenamefont {Phalen}, \citenamefont
  {Pierce},\ and\ \citenamefont {Tucker-Smith}}]{Gerbush:2007fe}%
  \BibitemOpen
  \bibfield  {author} {\bibinfo {author} {\bibfnamefont {M.}~\bibnamefont
  {Gerbush}}, \bibinfo {author} {\bibfnamefont {T.~J.}\ \bibnamefont {Khoo}},
  \bibinfo {author} {\bibfnamefont {D.~J.}\ \bibnamefont {Phalen}}, \bibinfo
  {author} {\bibfnamefont {A.}~\bibnamefont {Pierce}},\ and\ \bibinfo {author}
  {\bibfnamefont {D.}~\bibnamefont {Tucker-Smith}},\ }\bibfield  {title}
  {\bibinfo {title} {{Color-octet scalars at the CERN LHC}},\ }\href
  {https://doi.org/10.1103/PhysRevD.77.095003} {\bibfield  {journal} {\bibinfo
  {journal} {Phys. Rev. D}\ }\textbf {\bibinfo {volume} {77}},\ \bibinfo
  {pages} {095003} (\bibinfo {year} {2008})},\ \Eprint
  {https://arxiv.org/abs/0710.3133} {arXiv:0710.3133 [hep-ph]} \BibitemShut
  {NoStop}%
\bibitem [{\citenamefont {Brooijmans}(2008)}]{Brooijmans:2008zza}%
  \BibitemOpen
  \bibfield  {author} {\bibinfo {author} {\bibfnamefont {G.}~\bibnamefont
  {Brooijmans}},\ }\bibfield  {title} {\bibinfo {title} {{High pT hadronic top
  quark identification. Part I: Jet mass and YSplitter}}\ }(\bibinfo {year}
  {2008})\BibitemShut {NoStop}%
\bibitem [{\citenamefont {Kaplan}\ \emph {et~al.}(2008)\citenamefont {Kaplan},
  \citenamefont {Rehermann}, \citenamefont {Schwartz},\ and\ \citenamefont
  {Tweedie}}]{Kaplan:2008ie}%
  \BibitemOpen
  \bibfield  {author} {\bibinfo {author} {\bibfnamefont {D.~E.}\ \bibnamefont
  {Kaplan}}, \bibinfo {author} {\bibfnamefont {K.}~\bibnamefont {Rehermann}},
  \bibinfo {author} {\bibfnamefont {M.~D.}\ \bibnamefont {Schwartz}},\ and\
  \bibinfo {author} {\bibfnamefont {B.}~\bibnamefont {Tweedie}},\ }\bibfield
  {title} {\bibinfo {title} {{Top Tagging: A Method for Identifying Boosted
  Hadronically Decaying Top Quarks}},\ }\href
  {https://doi.org/10.1103/PhysRevLett.101.142001} {\bibfield  {journal}
  {\bibinfo  {journal} {Phys. Rev. Lett.}\ }\textbf {\bibinfo {volume} {101}},\
  \bibinfo {pages} {142001} (\bibinfo {year} {2008})},\ \Eprint
  {https://arxiv.org/abs/0806.0848} {arXiv:0806.0848 [hep-ph]} \BibitemShut
  {NoStop}%
\bibitem [{\citenamefont {Thaler}\ and\ \citenamefont
  {Wang}(2008)}]{Thaler:2008ju}%
  \BibitemOpen
  \bibfield  {author} {\bibinfo {author} {\bibfnamefont {J.}~\bibnamefont
  {Thaler}}\ and\ \bibinfo {author} {\bibfnamefont {L.-T.}\ \bibnamefont
  {Wang}},\ }\bibfield  {title} {\bibinfo {title} {{Strategies to Identify
  Boosted Tops}},\ }\href {https://doi.org/10.1088/1126-6708/2008/07/092}
  {\bibfield  {journal} {\bibinfo  {journal} {JHEP}\ }\textbf {\bibinfo
  {volume} {07}},\ \bibinfo {pages} {092}},\ \Eprint
  {https://arxiv.org/abs/0806.0023} {arXiv:0806.0023 [hep-ph]} \BibitemShut
  {NoStop}%
\bibitem [{\citenamefont {Almeida}\ \emph
  {et~al.}(2009{\natexlab{a}})\citenamefont {Almeida}, \citenamefont {Lee},
  \citenamefont {Perez}, \citenamefont {Sterman}, \citenamefont {Sung},\ and\
  \citenamefont {Virzi}}]{Almeida:2008yp}%
  \BibitemOpen
  \bibfield  {author} {\bibinfo {author} {\bibfnamefont {L.~G.}\ \bibnamefont
  {Almeida}}, \bibinfo {author} {\bibfnamefont {S.~J.}\ \bibnamefont {Lee}},
  \bibinfo {author} {\bibfnamefont {G.}~\bibnamefont {Perez}}, \bibinfo
  {author} {\bibfnamefont {G.~F.}\ \bibnamefont {Sterman}}, \bibinfo {author}
  {\bibfnamefont {I.}~\bibnamefont {Sung}},\ and\ \bibinfo {author}
  {\bibfnamefont {J.}~\bibnamefont {Virzi}},\ }\bibfield  {title} {\bibinfo
  {title} {{Substructure of high-$p_T$ Jets at the LHC}},\ }\href
  {https://doi.org/10.1103/PhysRevD.79.074017} {\bibfield  {journal} {\bibinfo
  {journal} {Phys. Rev. D}\ }\textbf {\bibinfo {volume} {79}},\ \bibinfo
  {pages} {074017} (\bibinfo {year} {2009}{\natexlab{a}})},\ \Eprint
  {https://arxiv.org/abs/0807.0234} {arXiv:0807.0234 [hep-ph]} \BibitemShut
  {NoStop}%
\bibitem [{\citenamefont {Plehn}\ \emph
  {et~al.}(2010{\natexlab{a}})\citenamefont {Plehn}, \citenamefont {Salam},\
  and\ \citenamefont {Spannowsky}}]{Plehn:2009rk}%
  \BibitemOpen
  \bibfield  {author} {\bibinfo {author} {\bibfnamefont {T.}~\bibnamefont
  {Plehn}}, \bibinfo {author} {\bibfnamefont {G.~P.}\ \bibnamefont {Salam}},\
  and\ \bibinfo {author} {\bibfnamefont {M.}~\bibnamefont {Spannowsky}},\
  }\bibfield  {title} {\bibinfo {title} {{Fat Jets for a Light Higgs}},\ }\href
  {https://doi.org/10.1103/PhysRevLett.104.111801} {\bibfield  {journal}
  {\bibinfo  {journal} {Phys. Rev. Lett.}\ }\textbf {\bibinfo {volume} {104}},\
  \bibinfo {pages} {111801} (\bibinfo {year} {2010}{\natexlab{a}})},\ \Eprint
  {https://arxiv.org/abs/0910.5472} {arXiv:0910.5472 [hep-ph]} \BibitemShut
  {NoStop}%
\bibitem [{\citenamefont {Almeida}\ \emph
  {et~al.}(2009{\natexlab{b}})\citenamefont {Almeida}, \citenamefont {Lee},
  \citenamefont {Perez}, \citenamefont {Sung},\ and\ \citenamefont
  {Virzi}}]{Almeida:2008tp}%
  \BibitemOpen
  \bibfield  {author} {\bibinfo {author} {\bibfnamefont {L.~G.}\ \bibnamefont
  {Almeida}}, \bibinfo {author} {\bibfnamefont {S.~J.}\ \bibnamefont {Lee}},
  \bibinfo {author} {\bibfnamefont {G.}~\bibnamefont {Perez}}, \bibinfo
  {author} {\bibfnamefont {I.}~\bibnamefont {Sung}},\ and\ \bibinfo {author}
  {\bibfnamefont {J.}~\bibnamefont {Virzi}},\ }\bibfield  {title} {\bibinfo
  {title} {{Top Jets at the LHC}},\ }\href
  {https://doi.org/10.1103/PhysRevD.79.074012} {\bibfield  {journal} {\bibinfo
  {journal} {Phys. Rev. D}\ }\textbf {\bibinfo {volume} {79}},\ \bibinfo
  {pages} {074012} (\bibinfo {year} {2009}{\natexlab{b}})},\ \Eprint
  {https://arxiv.org/abs/0810.0934} {arXiv:0810.0934 [hep-ph]} \BibitemShut
  {NoStop}%
\bibitem [{CMS(2009)}]{CMS:2009lxa}%
  \BibitemOpen
  \bibfield  {title} {\bibinfo {title} {{A Cambridge-Aachen (C-A) based Jet
  Algorithm for boosted top-jet tagging}},\ }\href@noop {} {\  (\bibinfo {year}
  {2009})}\BibitemShut {NoStop}%
\bibitem [{ATL(2010)}]{ATLAS:2010rhr}%
  \BibitemOpen
  \bibfield  {title} {\bibinfo {title} {{Prospects for early $t\overline{t}$
  resonance searches in ATLAS}},\ }\href@noop {} {\  (\bibinfo {year}
  {2010})}\BibitemShut {NoStop}%
\bibitem [{\citenamefont {Plehn}\ \emph
  {et~al.}(2010{\natexlab{b}})\citenamefont {Plehn}, \citenamefont
  {Spannowsky}, \citenamefont {Takeuchi},\ and\ \citenamefont
  {Zerwas}}]{Plehn:2010st}%
  \BibitemOpen
  \bibfield  {author} {\bibinfo {author} {\bibfnamefont {T.}~\bibnamefont
  {Plehn}}, \bibinfo {author} {\bibfnamefont {M.}~\bibnamefont {Spannowsky}},
  \bibinfo {author} {\bibfnamefont {M.}~\bibnamefont {Takeuchi}},\ and\
  \bibinfo {author} {\bibfnamefont {D.}~\bibnamefont {Zerwas}},\ }\bibfield
  {title} {\bibinfo {title} {{Stop Reconstruction with Tagged Tops}},\ }\href
  {https://doi.org/10.1007/JHEP10(2010)078} {\bibfield  {journal} {\bibinfo
  {journal} {JHEP}\ }\textbf {\bibinfo {volume} {10}},\ \bibinfo {pages}
  {078}},\ \Eprint {https://arxiv.org/abs/1006.2833} {arXiv:1006.2833 [hep-ph]}
  \BibitemShut {NoStop}%
\bibitem [{\citenamefont {Thaler}\ and\ \citenamefont
  {Van~Tilburg}(2011)}]{Thaler:2010tr}%
  \BibitemOpen
  \bibfield  {author} {\bibinfo {author} {\bibfnamefont {J.}~\bibnamefont
  {Thaler}}\ and\ \bibinfo {author} {\bibfnamefont {K.}~\bibnamefont
  {Van~Tilburg}},\ }\bibfield  {title} {\bibinfo {title} {{Identifying Boosted
  Objects with N-subjettiness}},\ }\href
  {https://doi.org/10.1007/JHEP03(2011)015} {\bibfield  {journal} {\bibinfo
  {journal} {JHEP}\ }\textbf {\bibinfo {volume} {03}},\ \bibinfo {pages}
  {015}},\ \Eprint {https://arxiv.org/abs/1011.2268} {arXiv:1011.2268 [hep-ph]}
  \BibitemShut {NoStop}%
\bibitem [{\citenamefont {Almeida}\ \emph {et~al.}(2010)\citenamefont
  {Almeida}, \citenamefont {Lee}, \citenamefont {Perez}, \citenamefont
  {Sterman},\ and\ \citenamefont {Sung}}]{Almeida:2010pa}%
  \BibitemOpen
  \bibfield  {author} {\bibinfo {author} {\bibfnamefont {L.~G.}\ \bibnamefont
  {Almeida}}, \bibinfo {author} {\bibfnamefont {S.~J.}\ \bibnamefont {Lee}},
  \bibinfo {author} {\bibfnamefont {G.}~\bibnamefont {Perez}}, \bibinfo
  {author} {\bibfnamefont {G.}~\bibnamefont {Sterman}},\ and\ \bibinfo {author}
  {\bibfnamefont {I.}~\bibnamefont {Sung}},\ }\bibfield  {title} {\bibinfo
  {title} {{Template Overlap Method for Massive Jets}},\ }\href
  {https://doi.org/10.1103/PhysRevD.82.054034} {\bibfield  {journal} {\bibinfo
  {journal} {Phys. Rev. D}\ }\textbf {\bibinfo {volume} {82}},\ \bibinfo
  {pages} {054034} (\bibinfo {year} {2010})},\ \Eprint
  {https://arxiv.org/abs/1006.2035} {arXiv:1006.2035 [hep-ph]} \BibitemShut
  {NoStop}%
\bibitem [{\citenamefont {Thaler}\ and\ \citenamefont
  {Van~Tilburg}(2012)}]{Thaler:2011gf}%
  \BibitemOpen
  \bibfield  {author} {\bibinfo {author} {\bibfnamefont {J.}~\bibnamefont
  {Thaler}}\ and\ \bibinfo {author} {\bibfnamefont {K.}~\bibnamefont
  {Van~Tilburg}},\ }\bibfield  {title} {\bibinfo {title} {{Maximizing Boosted
  Top Identification by Minimizing N-subjettiness}},\ }\href
  {https://doi.org/10.1007/JHEP02(2012)093} {\bibfield  {journal} {\bibinfo
  {journal} {JHEP}\ }\textbf {\bibinfo {volume} {02}},\ \bibinfo {pages}
  {093}},\ \Eprint {https://arxiv.org/abs/1108.2701} {arXiv:1108.2701 [hep-ph]}
  \BibitemShut {NoStop}%
\bibitem [{\citenamefont {Soper}\ and\ \citenamefont
  {Spannowsky}(2013)}]{Soper:2012pb}%
  \BibitemOpen
  \bibfield  {author} {\bibinfo {author} {\bibfnamefont {D.~E.}\ \bibnamefont
  {Soper}}\ and\ \bibinfo {author} {\bibfnamefont {M.}~\bibnamefont
  {Spannowsky}},\ }\bibfield  {title} {\bibinfo {title} {{Finding top quarks
  with shower deconstruction}},\ }\href
  {https://doi.org/10.1103/PhysRevD.87.054012} {\bibfield  {journal} {\bibinfo
  {journal} {Phys. Rev. D}\ }\textbf {\bibinfo {volume} {87}},\ \bibinfo
  {pages} {054012} (\bibinfo {year} {2013})},\ \Eprint
  {https://arxiv.org/abs/1211.3140} {arXiv:1211.3140 [hep-ph]} \BibitemShut
  {NoStop}%
\bibitem [{\citenamefont {Larkoski}\ \emph {et~al.}(2013)\citenamefont
  {Larkoski}, \citenamefont {Salam},\ and\ \citenamefont
  {Thaler}}]{Larkoski:2013eya}%
  \BibitemOpen
  \bibfield  {author} {\bibinfo {author} {\bibfnamefont {A.~J.}\ \bibnamefont
  {Larkoski}}, \bibinfo {author} {\bibfnamefont {G.~P.}\ \bibnamefont
  {Salam}},\ and\ \bibinfo {author} {\bibfnamefont {J.}~\bibnamefont
  {Thaler}},\ }\bibfield  {title} {\bibinfo {title} {{Energy Correlation
  Functions for Jet Substructure}},\ }\href
  {https://doi.org/10.1007/JHEP06(2013)108} {\bibfield  {journal} {\bibinfo
  {journal} {JHEP}\ }\textbf {\bibinfo {volume} {06}},\ \bibinfo {pages}
  {108}},\ \Eprint {https://arxiv.org/abs/1305.0007} {arXiv:1305.0007 [hep-ph]}
  \BibitemShut {NoStop}%
\bibitem [{\citenamefont {Larkoski}\ \emph {et~al.}(2015)\citenamefont
  {Larkoski}, \citenamefont {Moult},\ and\ \citenamefont
  {Neill}}]{Larkoski:2014zma}%
  \BibitemOpen
  \bibfield  {author} {\bibinfo {author} {\bibfnamefont {A.~J.}\ \bibnamefont
  {Larkoski}}, \bibinfo {author} {\bibfnamefont {I.}~\bibnamefont {Moult}},\
  and\ \bibinfo {author} {\bibfnamefont {D.}~\bibnamefont {Neill}},\ }\bibfield
   {title} {\bibinfo {title} {{Building a Better Boosted Top Tagger}},\ }\href
  {https://doi.org/10.1103/PhysRevD.91.034035} {\bibfield  {journal} {\bibinfo
  {journal} {Phys. Rev. D}\ }\textbf {\bibinfo {volume} {91}},\ \bibinfo
  {pages} {034035} (\bibinfo {year} {2015})},\ \Eprint
  {https://arxiv.org/abs/1411.0665} {arXiv:1411.0665 [hep-ph]} \BibitemShut
  {NoStop}%
\bibitem [{\citenamefont {Dasgupta}\ \emph {et~al.}(2018)\citenamefont
  {Dasgupta}, \citenamefont {Guzzi}, \citenamefont {Rawling},\ and\
  \citenamefont {Soyez}}]{Dasgupta:2018emf}%
  \BibitemOpen
  \bibfield  {author} {\bibinfo {author} {\bibfnamefont {M.}~\bibnamefont
  {Dasgupta}}, \bibinfo {author} {\bibfnamefont {M.}~\bibnamefont {Guzzi}},
  \bibinfo {author} {\bibfnamefont {J.}~\bibnamefont {Rawling}},\ and\ \bibinfo
  {author} {\bibfnamefont {G.}~\bibnamefont {Soyez}},\ }\bibfield  {title}
  {\bibinfo {title} {{Top tagging : an analytical perspective}},\ }\href
  {https://doi.org/10.1007/JHEP09(2018)170} {\bibfield  {journal} {\bibinfo
  {journal} {JHEP}\ }\textbf {\bibinfo {volume} {09}},\ \bibinfo {pages}
  {170}},\ \Eprint {https://arxiv.org/abs/1807.04767} {arXiv:1807.04767
  [hep-ph]} \BibitemShut {NoStop}%
\bibitem [{\citenamefont {Gallicchio}\ and\ \citenamefont
  {Schwartz}(2010)}]{Gallicchio:2010sw}%
  \BibitemOpen
  \bibfield  {author} {\bibinfo {author} {\bibfnamefont {J.}~\bibnamefont
  {Gallicchio}}\ and\ \bibinfo {author} {\bibfnamefont {M.~D.}\ \bibnamefont
  {Schwartz}},\ }\bibfield  {title} {\bibinfo {title} {{Seeing in Color: Jet
  Superstructure}},\ }\href {https://doi.org/10.1103/PhysRevLett.105.022001}
  {\bibfield  {journal} {\bibinfo  {journal} {Phys. Rev. Lett.}\ }\textbf
  {\bibinfo {volume} {105}},\ \bibinfo {pages} {022001} (\bibinfo {year}
  {2010})},\ \Eprint {https://arxiv.org/abs/1001.5027} {arXiv:1001.5027
  [hep-ph]} \BibitemShut {NoStop}%
\bibitem [{\citenamefont {Hook}\ \emph {et~al.}(2012)\citenamefont {Hook},
  \citenamefont {Jankowiak},\ and\ \citenamefont {Wacker}}]{Hook:2011cq}%
  \BibitemOpen
  \bibfield  {author} {\bibinfo {author} {\bibfnamefont {A.}~\bibnamefont
  {Hook}}, \bibinfo {author} {\bibfnamefont {M.}~\bibnamefont {Jankowiak}},\
  and\ \bibinfo {author} {\bibfnamefont {J.~G.}\ \bibnamefont {Wacker}},\
  }\bibfield  {title} {\bibinfo {title} {{Jet Dipolarity: Top Tagging with
  Color Flow}},\ }\href {https://doi.org/10.1007/JHEP04(2012)007} {\bibfield
  {journal} {\bibinfo  {journal} {JHEP}\ }\textbf {\bibinfo {volume} {04}},\
  \bibinfo {pages} {007}},\ \Eprint {https://arxiv.org/abs/1102.1012}
  {arXiv:1102.1012 [hep-ph]} \BibitemShut {NoStop}%
\bibitem [{\citenamefont {Butter}\ \emph {et~al.}(2019)\citenamefont {Butter}
  \emph {et~al.}}]{Kasieczka:2019dbj}%
  \BibitemOpen
  \bibfield  {author} {\bibinfo {author} {\bibfnamefont {A.}~\bibnamefont
  {Butter}} \emph {et~al.},\ }\bibfield  {title} {\bibinfo {title} {{The
  Machine Learning landscape of top taggers}},\ }\href
  {https://doi.org/10.21468/SciPostPhys.7.1.014} {\bibfield  {journal}
  {\bibinfo  {journal} {SciPost Phys.}\ }\textbf {\bibinfo {volume} {7}},\
  \bibinfo {pages} {014} (\bibinfo {year} {2019})},\ \Eprint
  {https://arxiv.org/abs/1902.09914} {arXiv:1902.09914 [hep-ph]} \BibitemShut
  {NoStop}%
\bibitem [{\citenamefont {Qu}\ \emph {et~al.}(2022{\natexlab{a}})\citenamefont
  {Qu}, \citenamefont {Li},\ and\ \citenamefont {Qian}}]{Qu:2022mxj}%
  \BibitemOpen
  \bibfield  {author} {\bibinfo {author} {\bibfnamefont {H.}~\bibnamefont
  {Qu}}, \bibinfo {author} {\bibfnamefont {C.}~\bibnamefont {Li}},\ and\
  \bibinfo {author} {\bibfnamefont {S.}~\bibnamefont {Qian}},\ }\bibfield
  {title} {\bibinfo {title} {{Particle Transformer for Jet Tagging}},\
  }\href@noop {} {\  (\bibinfo {year} {2022}{\natexlab{a}})},\ \Eprint
  {https://arxiv.org/abs/2202.03772} {arXiv:2202.03772 [hep-ph]} \BibitemShut
  {NoStop}%
\bibitem [{\citenamefont {Qu}\ \emph {et~al.}(2022{\natexlab{b}})\citenamefont
  {Qu}, \citenamefont {Li},\ and\ \citenamefont {Qian}}]{qu_2022_6619768}%
  \BibitemOpen
  \bibfield  {author} {\bibinfo {author} {\bibfnamefont {H.}~\bibnamefont
  {Qu}}, \bibinfo {author} {\bibfnamefont {C.}~\bibnamefont {Li}},\ and\
  \bibinfo {author} {\bibfnamefont {S.}~\bibnamefont {Qian}},\ }\bibfield
  {title} {\bibinfo {title} {Jetclass: A large-scale dataset for deep learning
  in jet physics},\ }\href {https://doi.org/10.5281/zenodo.6619768}
  {10.5281/zenodo.6619768} (\bibinfo {year} {2022}{\natexlab{b}})\BibitemShut
  {NoStop}%
\bibitem [{\citenamefont {Larkoski}(2024{\natexlab{b}})}]{Larkoski:2024hfe}%
  \BibitemOpen
  \bibfield  {author} {\bibinfo {author} {\bibfnamefont {A.~J.}\ \bibnamefont
  {Larkoski}},\ }\bibfield  {title} {\bibinfo {title} {{Systematic
  Interpretability and the Likelihood for Boosted Top Quark Identification}},\
  }\href@noop {} {\  (\bibinfo {year} {2024}{\natexlab{b}})},\ \Eprint
  {https://arxiv.org/abs/2411.00104} {arXiv:2411.00104 [hep-ph]} \BibitemShut
  {NoStop}%
\bibitem [{\citenamefont {Aad}\ \emph {et~al.}(2026)\citenamefont {Aad} \emph
  {et~al.}}]{ATLAS:2025dkv}%
  \BibitemOpen
  \bibfield  {author} {\bibinfo {author} {\bibfnamefont {G.}~\bibnamefont
  {Aad}} \emph {et~al.} (\bibinfo {collaboration} {ATLAS}),\ }\bibfield
  {title} {\bibinfo {title} {{Transforming jet flavour tagging at ATLAS}},\
  }\href {https://doi.org/10.1038/s41467-025-65059-6} {\bibfield  {journal}
  {\bibinfo  {journal} {Nature Commun.}\ }\textbf {\bibinfo {volume} {17}},\
  \bibinfo {pages} {541} (\bibinfo {year} {2026})},\ \Eprint
  {https://arxiv.org/abs/2505.19689} {arXiv:2505.19689 [hep-ex]} \BibitemShut
  {NoStop}%
\bibitem [{CMS(2026)}]{CMS:2026mee}%
  \BibitemOpen
  \bibfield  {title} {\bibinfo {title} {{Performance of heavy-flavour jet
  identification in the CMS high-level trigger during LHC Run 3}},\ }\href@noop
  {} {\  (\bibinfo {year} {2026})}\BibitemShut {NoStop}%
\bibitem [{\citenamefont {Banfi}\ \emph {et~al.}(2006)\citenamefont {Banfi},
  \citenamefont {Salam},\ and\ \citenamefont {Zanderighi}}]{Banfi:2006hf}%
  \BibitemOpen
  \bibfield  {author} {\bibinfo {author} {\bibfnamefont {A.}~\bibnamefont
  {Banfi}}, \bibinfo {author} {\bibfnamefont {G.~P.}\ \bibnamefont {Salam}},\
  and\ \bibinfo {author} {\bibfnamefont {G.}~\bibnamefont {Zanderighi}},\
  }\bibfield  {title} {\bibinfo {title} {{Infrared safe definition of jet
  flavor}},\ }\href {https://doi.org/10.1140/epjc/s2006-02552-4} {\bibfield
  {journal} {\bibinfo  {journal} {Eur. Phys. J. C}\ }\textbf {\bibinfo {volume}
  {47}},\ \bibinfo {pages} {113} (\bibinfo {year} {2006})},\ \Eprint
  {https://arxiv.org/abs/hep-ph/0601139} {arXiv:hep-ph/0601139} \BibitemShut
  {NoStop}%
\bibitem [{\citenamefont {Caletti}\ \emph
  {et~al.}(2022{\natexlab{a}})\citenamefont {Caletti}, \citenamefont
  {Larkoski}, \citenamefont {Marzani},\ and\ \citenamefont
  {Reichelt}}]{Caletti:2022hnc}%
  \BibitemOpen
  \bibfield  {author} {\bibinfo {author} {\bibfnamefont {S.}~\bibnamefont
  {Caletti}}, \bibinfo {author} {\bibfnamefont {A.~J.}\ \bibnamefont
  {Larkoski}}, \bibinfo {author} {\bibfnamefont {S.}~\bibnamefont {Marzani}},\
  and\ \bibinfo {author} {\bibfnamefont {D.}~\bibnamefont {Reichelt}},\
  }\bibfield  {title} {\bibinfo {title} {{Practical jet flavour through
  NNLO}},\ }\href {https://doi.org/10.1140/epjc/s10052-022-10568-7} {\bibfield
  {journal} {\bibinfo  {journal} {Eur. Phys. J. C}\ }\textbf {\bibinfo {volume}
  {82}},\ \bibinfo {pages} {632} (\bibinfo {year} {2022}{\natexlab{a}})},\
  \Eprint {https://arxiv.org/abs/2205.01109} {arXiv:2205.01109 [hep-ph]}
  \BibitemShut {NoStop}%
\bibitem [{\citenamefont {Caletti}\ \emph
  {et~al.}(2022{\natexlab{b}})\citenamefont {Caletti}, \citenamefont
  {Larkoski}, \citenamefont {Marzani},\ and\ \citenamefont
  {Reichelt}}]{Caletti:2022glq}%
  \BibitemOpen
  \bibfield  {author} {\bibinfo {author} {\bibfnamefont {S.}~\bibnamefont
  {Caletti}}, \bibinfo {author} {\bibfnamefont {A.~J.}\ \bibnamefont
  {Larkoski}}, \bibinfo {author} {\bibfnamefont {S.}~\bibnamefont {Marzani}},\
  and\ \bibinfo {author} {\bibfnamefont {D.}~\bibnamefont {Reichelt}},\
  }\bibfield  {title} {\bibinfo {title} {{A fragmentation approach to jet
  flavor}},\ }\href {https://doi.org/10.1007/JHEP10(2022)158} {\bibfield
  {journal} {\bibinfo  {journal} {JHEP}\ }\textbf {\bibinfo {volume} {10}},\
  \bibinfo {pages} {158}},\ \Eprint {https://arxiv.org/abs/2205.01117}
  {arXiv:2205.01117 [hep-ph]} \BibitemShut {NoStop}%
\bibitem [{\citenamefont {Czakon}\ \emph {et~al.}(2023)\citenamefont {Czakon},
  \citenamefont {Mitov},\ and\ \citenamefont {Poncelet}}]{Czakon:2022wam}%
  \BibitemOpen
  \bibfield  {author} {\bibinfo {author} {\bibfnamefont {M.}~\bibnamefont
  {Czakon}}, \bibinfo {author} {\bibfnamefont {A.}~\bibnamefont {Mitov}},\ and\
  \bibinfo {author} {\bibfnamefont {R.}~\bibnamefont {Poncelet}},\ }\bibfield
  {title} {\bibinfo {title} {{Infrared-safe flavoured anti-k$_{T}$ jets}},\
  }\href {https://doi.org/10.1007/JHEP04(2023)138} {\bibfield  {journal}
  {\bibinfo  {journal} {JHEP}\ }\textbf {\bibinfo {volume} {04}},\ \bibinfo
  {pages} {138}},\ \Eprint {https://arxiv.org/abs/2205.11879} {arXiv:2205.11879
  [hep-ph]} \BibitemShut {NoStop}%
\bibitem [{\citenamefont {Gauld}\ \emph {et~al.}(2023)\citenamefont {Gauld},
  \citenamefont {Huss},\ and\ \citenamefont {Stagnitto}}]{Gauld:2022lem}%
  \BibitemOpen
  \bibfield  {author} {\bibinfo {author} {\bibfnamefont {R.}~\bibnamefont
  {Gauld}}, \bibinfo {author} {\bibfnamefont {A.}~\bibnamefont {Huss}},\ and\
  \bibinfo {author} {\bibfnamefont {G.}~\bibnamefont {Stagnitto}},\ }\bibfield
  {title} {\bibinfo {title} {{Flavor Identification of Reconstructed Hadronic
  Jets}},\ }\href {https://doi.org/10.1103/PhysRevLett.130.161901} {\bibfield
  {journal} {\bibinfo  {journal} {Phys. Rev. Lett.}\ }\textbf {\bibinfo
  {volume} {130}},\ \bibinfo {pages} {161901} (\bibinfo {year} {2023})},\
  \Eprint {https://arxiv.org/abs/2208.11138} {arXiv:2208.11138 [hep-ph]}
  \BibitemShut {NoStop}%
\bibitem [{\citenamefont {Caola}\ \emph {et~al.}(2023)\citenamefont {Caola},
  \citenamefont {Grabarczyk}, \citenamefont {Hutt}, \citenamefont {Salam},
  \citenamefont {Scyboz},\ and\ \citenamefont {Thaler}}]{Caola:2023wpj}%
  \BibitemOpen
  \bibfield  {author} {\bibinfo {author} {\bibfnamefont {F.}~\bibnamefont
  {Caola}}, \bibinfo {author} {\bibfnamefont {R.}~\bibnamefont {Grabarczyk}},
  \bibinfo {author} {\bibfnamefont {M.~L.}\ \bibnamefont {Hutt}}, \bibinfo
  {author} {\bibfnamefont {G.~P.}\ \bibnamefont {Salam}}, \bibinfo {author}
  {\bibfnamefont {L.}~\bibnamefont {Scyboz}},\ and\ \bibinfo {author}
  {\bibfnamefont {J.}~\bibnamefont {Thaler}},\ }\bibfield  {title} {\bibinfo
  {title} {{Flavored jets with exact anti-kt kinematics and tests of infrared
  and collinear safety}},\ }\href {https://doi.org/10.1103/PhysRevD.108.094010}
  {\bibfield  {journal} {\bibinfo  {journal} {Phys. Rev. D}\ }\textbf {\bibinfo
  {volume} {108}},\ \bibinfo {pages} {094010} (\bibinfo {year} {2023})},\
  \Eprint {https://arxiv.org/abs/2306.07314} {arXiv:2306.07314 [hep-ph]}
  \BibitemShut {NoStop}%
\bibitem [{\citenamefont {Larkoski}(2026)}]{Larkoski:2025afg}%
  \BibitemOpen
  \bibfield  {author} {\bibinfo {author} {\bibfnamefont {A.~J.}\ \bibnamefont
  {Larkoski}},\ }\bibfield  {title} {\bibinfo {title} {{Flavor-changing
  nonglobal logarithms}},\ }\href {https://doi.org/10.1103/4jvb-24hk}
  {\bibfield  {journal} {\bibinfo  {journal} {Phys. Rev. D}\ }\textbf {\bibinfo
  {volume} {113}},\ \bibinfo {pages} {034013} (\bibinfo {year} {2026})},\
  \Eprint {https://arxiv.org/abs/2510.06031} {arXiv:2510.06031 [hep-ph]}
  \BibitemShut {NoStop}%
\bibitem [{\citenamefont {Generet}(2026)}]{Generet:2025gdy}%
  \BibitemOpen
  \bibfield  {author} {\bibinfo {author} {\bibfnamefont {T.}~\bibnamefont
  {Generet}},\ }\bibfield  {title} {\bibinfo {title} {{IRC-safe jet flavour at
  leading power}},\ }\href {https://doi.org/10.1007/JHEP06(2026)065} {\bibfield
   {journal} {\bibinfo  {journal} {JHEP}\ }\textbf {\bibinfo {volume} {06}},\
  \bibinfo {pages} {065}},\ \Eprint {https://arxiv.org/abs/2511.23423}
  {arXiv:2511.23423 [hep-ph]} \BibitemShut {NoStop}%
\bibitem [{\citenamefont {Behring}\ \emph {et~al.}(2025)\citenamefont {Behring}
  \emph {et~al.}}]{Behring:2025ilo}%
  \BibitemOpen
  \bibfield  {author} {\bibinfo {author} {\bibfnamefont {A.}~\bibnamefont
  {Behring}} \emph {et~al.},\ }\bibfield  {title} {\bibinfo {title} {{Flavoured
  jet algorithms: a comparative study}},\ }\href
  {https://doi.org/10.1007/JHEP09(2025)149} {\bibfield  {journal} {\bibinfo
  {journal} {JHEP}\ }\textbf {\bibinfo {volume} {09}},\ \bibinfo {pages}
  {149}},\ \Eprint {https://arxiv.org/abs/2506.13449} {arXiv:2506.13449
  [hep-ph]} \BibitemShut {NoStop}%
\bibitem [{\citenamefont {Cybenko}(1989)}]{cybenko1989approximation}%
  \BibitemOpen
  \bibfield  {author} {\bibinfo {author} {\bibfnamefont {G.}~\bibnamefont
  {Cybenko}},\ }\bibfield  {title} {\bibinfo {title} {Approximation by
  superpositions of a sigmoidal function},\ }\href@noop {} {\bibfield
  {journal} {\bibinfo  {journal} {Mathematics of control, signals and systems}\
  }\textbf {\bibinfo {volume} {2}},\ \bibinfo {pages} {303} (\bibinfo {year}
  {1989})}\BibitemShut {NoStop}%
\bibitem [{\citenamefont {Hornik}(1991)}]{hornik1991approximation}%
  \BibitemOpen
  \bibfield  {author} {\bibinfo {author} {\bibfnamefont {K.}~\bibnamefont
  {Hornik}},\ }\bibfield  {title} {\bibinfo {title} {Approximation capabilities
  of multilayer feedforward networks},\ }\href@noop {} {\bibfield  {journal}
  {\bibinfo  {journal} {Neural networks}\ }\textbf {\bibinfo {volume} {4}},\
  \bibinfo {pages} {251} (\bibinfo {year} {1991})}\BibitemShut {NoStop}%
\bibitem [{\citenamefont {Leshno}\ \emph {et~al.}(1993)\citenamefont {Leshno},
  \citenamefont {Lin}, \citenamefont {Pinkus},\ and\ \citenamefont
  {Schocken}}]{leshno1993multilayer}%
  \BibitemOpen
  \bibfield  {author} {\bibinfo {author} {\bibfnamefont {M.}~\bibnamefont
  {Leshno}}, \bibinfo {author} {\bibfnamefont {V.~Y.}\ \bibnamefont {Lin}},
  \bibinfo {author} {\bibfnamefont {A.}~\bibnamefont {Pinkus}},\ and\ \bibinfo
  {author} {\bibfnamefont {S.}~\bibnamefont {Schocken}},\ }\bibfield  {title}
  {\bibinfo {title} {Multilayer feedforward networks with a nonpolynomial
  activation function can approximate any function},\ }\href@noop {} {\bibfield
   {journal} {\bibinfo  {journal} {Neural networks}\ }\textbf {\bibinfo
  {volume} {6}},\ \bibinfo {pages} {861} (\bibinfo {year} {1993})}\BibitemShut
  {NoStop}%
\bibitem [{\citenamefont {Bogatskiy}\ \emph
  {et~al.}(2022{\natexlab{a}})\citenamefont {Bogatskiy}, \citenamefont
  {Hoffman}, \citenamefont {Miller},\ and\ \citenamefont
  {Offermann}}]{Bogatskiy:2022czk}%
  \BibitemOpen
  \bibfield  {author} {\bibinfo {author} {\bibfnamefont {A.}~\bibnamefont
  {Bogatskiy}}, \bibinfo {author} {\bibfnamefont {T.}~\bibnamefont {Hoffman}},
  \bibinfo {author} {\bibfnamefont {D.~W.}\ \bibnamefont {Miller}},\ and\
  \bibinfo {author} {\bibfnamefont {J.~T.}\ \bibnamefont {Offermann}},\
  }\bibfield  {title} {\bibinfo {title} {{PELICAN: Permutation Equivariant and
  Lorentz Invariant or Covariant Aggregator Network for Particle Physics}},\
  }\href@noop {} {\  (\bibinfo {year} {2022}{\natexlab{a}})},\ \Eprint
  {https://arxiv.org/abs/2211.00454} {arXiv:2211.00454 [hep-ph]} \BibitemShut
  {NoStop}%
\bibitem [{\citenamefont {Dolan}\ and\ \citenamefont
  {Ore}(2021)}]{Dolan:2020qkr}%
  \BibitemOpen
  \bibfield  {author} {\bibinfo {author} {\bibfnamefont {M.~J.}\ \bibnamefont
  {Dolan}}\ and\ \bibinfo {author} {\bibfnamefont {A.}~\bibnamefont {Ore}},\
  }\bibfield  {title} {\bibinfo {title} {{Equivariant Energy Flow Networks for
  Jet Tagging}},\ }\href {https://doi.org/10.1103/PhysRevD.103.074022}
  {\bibfield  {journal} {\bibinfo  {journal} {Phys. Rev. D}\ }\textbf {\bibinfo
  {volume} {103}},\ \bibinfo {pages} {074022} (\bibinfo {year} {2021})},\
  \Eprint {https://arxiv.org/abs/2012.00964} {arXiv:2012.00964 [hep-ph]}
  \BibitemShut {NoStop}%
\bibitem [{\citenamefont {Craven}\ \emph {et~al.}(2022)\citenamefont {Craven},
  \citenamefont {Croon}, \citenamefont {Cutting},\ and\ \citenamefont
  {Houtz}}]{Craven:2021ems}%
  \BibitemOpen
  \bibfield  {author} {\bibinfo {author} {\bibfnamefont {S.}~\bibnamefont
  {Craven}}, \bibinfo {author} {\bibfnamefont {D.}~\bibnamefont {Croon}},
  \bibinfo {author} {\bibfnamefont {D.}~\bibnamefont {Cutting}},\ and\ \bibinfo
  {author} {\bibfnamefont {R.}~\bibnamefont {Houtz}},\ }\bibfield  {title}
  {\bibinfo {title} {{Machine learning a manifold}},\ }\href
  {https://doi.org/10.1103/PhysRevD.105.096030} {\bibfield  {journal} {\bibinfo
   {journal} {Phys. Rev. D}\ }\textbf {\bibinfo {volume} {105}},\ \bibinfo
  {pages} {096030} (\bibinfo {year} {2022})},\ \Eprint
  {https://arxiv.org/abs/2112.07673} {arXiv:2112.07673 [hep-ph]} \BibitemShut
  {NoStop}%
\bibitem [{\citenamefont {Gong}\ \emph {et~al.}(2022)\citenamefont {Gong},
  \citenamefont {Meng}, \citenamefont {Zhang}, \citenamefont {Qu},
  \citenamefont {Li}, \citenamefont {Qian}, \citenamefont {Du}, \citenamefont
  {Ma},\ and\ \citenamefont {Liu}}]{Gong:2022lye}%
  \BibitemOpen
  \bibfield  {author} {\bibinfo {author} {\bibfnamefont {S.}~\bibnamefont
  {Gong}}, \bibinfo {author} {\bibfnamefont {Q.}~\bibnamefont {Meng}}, \bibinfo
  {author} {\bibfnamefont {J.}~\bibnamefont {Zhang}}, \bibinfo {author}
  {\bibfnamefont {H.}~\bibnamefont {Qu}}, \bibinfo {author} {\bibfnamefont
  {C.}~\bibnamefont {Li}}, \bibinfo {author} {\bibfnamefont {S.}~\bibnamefont
  {Qian}}, \bibinfo {author} {\bibfnamefont {W.}~\bibnamefont {Du}}, \bibinfo
  {author} {\bibfnamefont {Z.-M.}\ \bibnamefont {Ma}},\ and\ \bibinfo {author}
  {\bibfnamefont {T.-Y.}\ \bibnamefont {Liu}},\ }\bibfield  {title} {\bibinfo
  {title} {{An efficient Lorentz equivariant graph neural network for jet
  tagging}},\ }\href {https://doi.org/10.1007/JHEP07(2022)030} {\bibfield
  {journal} {\bibinfo  {journal} {JHEP}\ }\textbf {\bibinfo {volume} {07}},\
  \bibinfo {pages} {030}},\ \Eprint {https://arxiv.org/abs/2201.08187}
  {arXiv:2201.08187 [hep-ph]} \BibitemShut {NoStop}%
\bibitem [{\citenamefont {Bogatskiy}\ \emph
  {et~al.}(2022{\natexlab{b}})\citenamefont {Bogatskiy} \emph
  {et~al.}}]{Bogatskiy:2022hub}%
  \BibitemOpen
  \bibfield  {author} {\bibinfo {author} {\bibfnamefont {A.}~\bibnamefont
  {Bogatskiy}} \emph {et~al.},\ }\bibfield  {title} {\bibinfo {title}
  {{Symmetry Group Equivariant Architectures for Physics}},\ }in\ \href@noop {}
  {\emph {\bibinfo {booktitle} {{Snowmass 2021}}}}\ (\bibinfo {year} {2022})\
  \Eprint {https://arxiv.org/abs/2203.06153} {arXiv:2203.06153 [cs.LG]}
  \BibitemShut {NoStop}%
\bibitem [{\citenamefont {Hao}\ \emph {et~al.}(2023)\citenamefont {Hao},
  \citenamefont {Kansal}, \citenamefont {Duarte},\ and\ \citenamefont
  {Chernyavskaya}}]{Hao:2022zns}%
  \BibitemOpen
  \bibfield  {author} {\bibinfo {author} {\bibfnamefont {Z.}~\bibnamefont
  {Hao}}, \bibinfo {author} {\bibfnamefont {R.}~\bibnamefont {Kansal}},
  \bibinfo {author} {\bibfnamefont {J.}~\bibnamefont {Duarte}},\ and\ \bibinfo
  {author} {\bibfnamefont {N.}~\bibnamefont {Chernyavskaya}},\ }\bibfield
  {title} {\bibinfo {title} {{Lorentz group equivariant autoencoders}},\ }\href
  {https://doi.org/10.1140/epjc/s10052-023-11633-5} {\bibfield  {journal}
  {\bibinfo  {journal} {Eur. Phys. J. C}\ }\textbf {\bibinfo {volume} {83}},\
  \bibinfo {pages} {485} (\bibinfo {year} {2023})},\ \Eprint
  {https://arxiv.org/abs/2212.07347} {arXiv:2212.07347 [hep-ex]} \BibitemShut
  {NoStop}%
\bibitem [{\citenamefont {Forestano}\ \emph
  {et~al.}(2023{\natexlab{a}})\citenamefont {Forestano}, \citenamefont
  {Matchev}, \citenamefont {Matcheva}, \citenamefont {Roman}, \citenamefont
  {Unlu},\ and\ \citenamefont {Verner}}]{Forestano:2023fpj}%
  \BibitemOpen
  \bibfield  {author} {\bibinfo {author} {\bibfnamefont {R.~T.}\ \bibnamefont
  {Forestano}}, \bibinfo {author} {\bibfnamefont {K.~T.}\ \bibnamefont
  {Matchev}}, \bibinfo {author} {\bibfnamefont {K.}~\bibnamefont {Matcheva}},
  \bibinfo {author} {\bibfnamefont {A.}~\bibnamefont {Roman}}, \bibinfo
  {author} {\bibfnamefont {E.~B.}\ \bibnamefont {Unlu}},\ and\ \bibinfo
  {author} {\bibfnamefont {S.}~\bibnamefont {Verner}},\ }\bibfield  {title}
  {\bibinfo {title} {{Deep learning symmetries and their Lie groups, algebras,
  and subalgebras from first principles}},\ }\href
  {https://doi.org/10.1088/2632-2153/acd989} {\bibfield  {journal} {\bibinfo
  {journal} {Mach. Learn. Sci. Tech.}\ }\textbf {\bibinfo {volume} {4}},\
  \bibinfo {pages} {025027} (\bibinfo {year} {2023}{\natexlab{a}})},\ \Eprint
  {https://arxiv.org/abs/2301.05638} {arXiv:2301.05638 [hep-ph]} \BibitemShut
  {NoStop}%
\bibitem [{\citenamefont {Buhmann}\ \emph {et~al.}(2023)\citenamefont
  {Buhmann}, \citenamefont {Kasieczka},\ and\ \citenamefont
  {Thaler}}]{Buhmann:2023pmh}%
  \BibitemOpen
  \bibfield  {author} {\bibinfo {author} {\bibfnamefont {E.}~\bibnamefont
  {Buhmann}}, \bibinfo {author} {\bibfnamefont {G.}~\bibnamefont {Kasieczka}},\
  and\ \bibinfo {author} {\bibfnamefont {J.}~\bibnamefont {Thaler}},\
  }\bibfield  {title} {\bibinfo {title} {{EPiC-GAN: Equivariant point cloud
  generation for particle jets}},\ }\href
  {https://doi.org/10.21468/SciPostPhys.15.4.130} {\bibfield  {journal}
  {\bibinfo  {journal} {SciPost Phys.}\ }\textbf {\bibinfo {volume} {15}},\
  \bibinfo {pages} {130} (\bibinfo {year} {2023})},\ \Eprint
  {https://arxiv.org/abs/2301.08128} {arXiv:2301.08128 [hep-ph]} \BibitemShut
  {NoStop}%
\bibitem [{\citenamefont {Forestano}\ \emph
  {et~al.}(2023{\natexlab{b}})\citenamefont {Forestano}, \citenamefont
  {Matchev}, \citenamefont {Matcheva}, \citenamefont {Roman}, \citenamefont
  {Unlu},\ and\ \citenamefont {Verner}}]{Forestano:2023qcy}%
  \BibitemOpen
  \bibfield  {author} {\bibinfo {author} {\bibfnamefont {R.~T.}\ \bibnamefont
  {Forestano}}, \bibinfo {author} {\bibfnamefont {K.~T.}\ \bibnamefont
  {Matchev}}, \bibinfo {author} {\bibfnamefont {K.}~\bibnamefont {Matcheva}},
  \bibinfo {author} {\bibfnamefont {A.}~\bibnamefont {Roman}}, \bibinfo
  {author} {\bibfnamefont {E.~B.}\ \bibnamefont {Unlu}},\ and\ \bibinfo
  {author} {\bibfnamefont {S.}~\bibnamefont {Verner}},\ }\bibfield  {title}
  {\bibinfo {title} {{Discovering sparse representations of Lie groups with
  machine learning}},\ }\href {https://doi.org/10.1016/j.physletb.2023.138086}
  {\bibfield  {journal} {\bibinfo  {journal} {Phys. Lett. B}\ }\textbf
  {\bibinfo {volume} {844}},\ \bibinfo {pages} {138086} (\bibinfo {year}
  {2023}{\natexlab{b}})},\ \Eprint {https://arxiv.org/abs/2302.05383}
  {arXiv:2302.05383 [hep-ph]} \BibitemShut {NoStop}%
\bibitem [{\citenamefont {Bogatskiy}\ \emph {et~al.}(2024)\citenamefont
  {Bogatskiy}, \citenamefont {Hoffman}, \citenamefont {Miller}, \citenamefont
  {Offermann},\ and\ \citenamefont {Liu}}]{Bogatskiy:2023nnw}%
  \BibitemOpen
  \bibfield  {author} {\bibinfo {author} {\bibfnamefont {A.}~\bibnamefont
  {Bogatskiy}}, \bibinfo {author} {\bibfnamefont {T.}~\bibnamefont {Hoffman}},
  \bibinfo {author} {\bibfnamefont {D.~W.}\ \bibnamefont {Miller}}, \bibinfo
  {author} {\bibfnamefont {J.~T.}\ \bibnamefont {Offermann}},\ and\ \bibinfo
  {author} {\bibfnamefont {X.}~\bibnamefont {Liu}},\ }\bibfield  {title}
  {\bibinfo {title} {{Explainable equivariant neural networks for particle
  physics: PELICAN}},\ }\href {https://doi.org/10.1007/JHEP03(2024)113}
  {\bibfield  {journal} {\bibinfo  {journal} {JHEP}\ }\textbf {\bibinfo
  {volume} {03}},\ \bibinfo {pages} {113}},\ \Eprint
  {https://arxiv.org/abs/2307.16506} {arXiv:2307.16506 [hep-ph]} \BibitemShut
  {NoStop}%
\bibitem [{\citenamefont {Bright-Thonney}\ \emph {et~al.}(2024)\citenamefont
  {Bright-Thonney}, \citenamefont {Nachman},\ and\ \citenamefont
  {Thaler}}]{Bright-Thonney:2023gdl}%
  \BibitemOpen
  \bibfield  {author} {\bibinfo {author} {\bibfnamefont {S.}~\bibnamefont
  {Bright-Thonney}}, \bibinfo {author} {\bibfnamefont {B.}~\bibnamefont
  {Nachman}},\ and\ \bibinfo {author} {\bibfnamefont {J.}~\bibnamefont
  {Thaler}},\ }\bibfield  {title} {\bibinfo {title} {{Infrared-safe energy
  weighting does not guarantee small nonperturbative effects}},\ }\href
  {https://doi.org/10.1103/PhysRevD.110.014029} {\bibfield  {journal} {\bibinfo
   {journal} {Phys. Rev. D}\ }\textbf {\bibinfo {volume} {110}},\ \bibinfo
  {pages} {014029} (\bibinfo {year} {2024})},\ \Eprint
  {https://arxiv.org/abs/2311.07652} {arXiv:2311.07652 [hep-ph]} \BibitemShut
  {NoStop}%
\bibitem [{\citenamefont {Bressler}\ \emph {et~al.}(2024)\citenamefont
  {Bressler}, \citenamefont {Savoray},\ and\ \citenamefont
  {Zurgil}}]{Bressler:2024wzc}%
  \BibitemOpen
  \bibfield  {author} {\bibinfo {author} {\bibfnamefont {S.}~\bibnamefont
  {Bressler}}, \bibinfo {author} {\bibfnamefont {I.}~\bibnamefont {Savoray}},\
  and\ \bibinfo {author} {\bibfnamefont {Y.}~\bibnamefont {Zurgil}},\
  }\bibfield  {title} {\bibinfo {title} {{Learning new physics from data: A
  symmetrized approach}},\ }\href {https://doi.org/10.1103/PhysRevD.110.095004}
  {\bibfield  {journal} {\bibinfo  {journal} {Phys. Rev. D}\ }\textbf {\bibinfo
  {volume} {110}},\ \bibinfo {pages} {095004} (\bibinfo {year} {2024})},\
  \Eprint {https://arxiv.org/abs/2401.09530} {arXiv:2401.09530 [hep-ex]}
  \BibitemShut {NoStop}%
\bibitem [{\citenamefont {Chatterjee}\ \emph {et~al.}(2024)\citenamefont
  {Chatterjee}, \citenamefont {Cruz}, \citenamefont {Sch{\"o}fbeck},\ and\
  \citenamefont {Schwarz}}]{Chatterjee:2024pbp}%
  \BibitemOpen
  \bibfield  {author} {\bibinfo {author} {\bibfnamefont {S.}~\bibnamefont
  {Chatterjee}}, \bibinfo {author} {\bibfnamefont {S.~S.}\ \bibnamefont
  {Cruz}}, \bibinfo {author} {\bibfnamefont {R.}~\bibnamefont
  {Sch{\"o}fbeck}},\ and\ \bibinfo {author} {\bibfnamefont {D.}~\bibnamefont
  {Schwarz}},\ }\bibfield  {title} {\bibinfo {title} {{Rotation-equivariant
  graph neural network for learning hadronic SMEFT effects}},\ }\href
  {https://doi.org/10.1103/PhysRevD.109.076012} {\bibfield  {journal} {\bibinfo
   {journal} {Phys. Rev. D}\ }\textbf {\bibinfo {volume} {109}},\ \bibinfo
  {pages} {076012} (\bibinfo {year} {2024})},\ \Eprint
  {https://arxiv.org/abs/2401.10323} {arXiv:2401.10323 [hep-ph]} \BibitemShut
  {NoStop}%
\bibitem [{\citenamefont {Bhardwaj}\ \emph {et~al.}(2024)\citenamefont
  {Bhardwaj}, \citenamefont {Englert}, \citenamefont {Naskar}, \citenamefont
  {Ngairangbam},\ and\ \citenamefont {Spannowsky}}]{Bhardwaj:2024djv}%
  \BibitemOpen
  \bibfield  {author} {\bibinfo {author} {\bibfnamefont {A.}~\bibnamefont
  {Bhardwaj}}, \bibinfo {author} {\bibfnamefont {C.}~\bibnamefont {Englert}},
  \bibinfo {author} {\bibfnamefont {W.}~\bibnamefont {Naskar}}, \bibinfo
  {author} {\bibfnamefont {V.~S.}\ \bibnamefont {Ngairangbam}},\ and\ \bibinfo
  {author} {\bibfnamefont {M.}~\bibnamefont {Spannowsky}},\ }\bibfield  {title}
  {\bibinfo {title} {{Equivariant, safe and sensitive {\textemdash} graph
  networks for new physics}},\ }\href {https://doi.org/10.1007/JHEP07(2024)245}
  {\bibfield  {journal} {\bibinfo  {journal} {JHEP}\ }\textbf {\bibinfo
  {volume} {07}},\ \bibinfo {pages} {245}},\ \Eprint
  {https://arxiv.org/abs/2402.12449} {arXiv:2402.12449 [hep-ph]} \BibitemShut
  {NoStop}%
\bibitem [{\citenamefont {Sahu}(2025)}]{Sahu:2024sts}%
  \BibitemOpen
  \bibfield  {author} {\bibinfo {author} {\bibfnamefont {R.}~\bibnamefont
  {Sahu}},\ }\bibfield  {title} {\bibinfo {title} {{Integrating physics
  inspired features with graph convolution}},\ }\href
  {https://doi.org/10.1103/PhysRevD.111.036037} {\bibfield  {journal} {\bibinfo
   {journal} {Phys. Rev. D}\ }\textbf {\bibinfo {volume} {111}},\ \bibinfo
  {pages} {036037} (\bibinfo {year} {2025})},\ \Eprint
  {https://arxiv.org/abs/2403.11826} {arXiv:2403.11826 [hep-ph]} \BibitemShut
  {NoStop}%
\bibitem [{\citenamefont {Spinner}\ \emph {et~al.}(2024)\citenamefont
  {Spinner}, \citenamefont {Bres{\'o}}, \citenamefont {de~Haan}, \citenamefont
  {Plehn}, \citenamefont {Thaler},\ and\ \citenamefont
  {Brehmer}}]{Spinner:2024hjm}%
  \BibitemOpen
  \bibfield  {author} {\bibinfo {author} {\bibfnamefont {J.}~\bibnamefont
  {Spinner}}, \bibinfo {author} {\bibfnamefont {V.}~\bibnamefont {Bres{\'o}}},
  \bibinfo {author} {\bibfnamefont {P.}~\bibnamefont {de~Haan}}, \bibinfo
  {author} {\bibfnamefont {T.}~\bibnamefont {Plehn}}, \bibinfo {author}
  {\bibfnamefont {J.}~\bibnamefont {Thaler}},\ and\ \bibinfo {author}
  {\bibfnamefont {J.}~\bibnamefont {Brehmer}},\ }\bibfield  {title} {\bibinfo
  {title} {{Lorentz-Equivariant Geometric Algebra Transformers for High-Energy
  Physics}},\ }in\ \href@noop {} {\emph {\bibinfo {booktitle} {{38th conference
  on Neural Information Processing Systems}}}}\ (\bibinfo {year} {2024})\
  \Eprint {https://arxiv.org/abs/2405.14806} {arXiv:2405.14806
  [physics.data-an]} \BibitemShut {NoStop}%
\bibitem [{\citenamefont {Ma{\^\i}tre}\ \emph {et~al.}(2025)\citenamefont
  {Ma{\^\i}tre}, \citenamefont {Ngairangbam},\ and\ \citenamefont
  {Spannowsky}}]{Maitre:2024hzp}%
  \BibitemOpen
  \bibfield  {author} {\bibinfo {author} {\bibfnamefont {D.}~\bibnamefont
  {Ma{\^\i}tre}}, \bibinfo {author} {\bibfnamefont {V.~S.}\ \bibnamefont
  {Ngairangbam}},\ and\ \bibinfo {author} {\bibfnamefont {M.}~\bibnamefont
  {Spannowsky}},\ }\bibfield  {title} {\bibinfo {title} {{Optimal equivariant
  architectures from the symmetries of matrix-element likelihoods}},\ }\href
  {https://doi.org/10.1088/2632-2153/adbab1} {\bibfield  {journal} {\bibinfo
  {journal} {Mach. Learn. Sci. Tech.}\ }\textbf {\bibinfo {volume} {6}},\
  \bibinfo {pages} {015059} (\bibinfo {year} {2025})},\ \Eprint
  {https://arxiv.org/abs/2410.18553} {arXiv:2410.18553 [hep-ph]} \BibitemShut
  {NoStop}%
\bibitem [{\citenamefont {Brehmer}\ \emph {et~al.}(2025)\citenamefont
  {Brehmer}, \citenamefont {Bres{\'o}}, \citenamefont {de~Haan}, \citenamefont
  {Plehn}, \citenamefont {Qu}, \citenamefont {Spinner},\ and\ \citenamefont
  {Thaler}}]{Brehmer:2024yqw}%
  \BibitemOpen
  \bibfield  {author} {\bibinfo {author} {\bibfnamefont {J.}~\bibnamefont
  {Brehmer}}, \bibinfo {author} {\bibfnamefont {V.}~\bibnamefont {Bres{\'o}}},
  \bibinfo {author} {\bibfnamefont {P.}~\bibnamefont {de~Haan}}, \bibinfo
  {author} {\bibfnamefont {T.}~\bibnamefont {Plehn}}, \bibinfo {author}
  {\bibfnamefont {H.}~\bibnamefont {Qu}}, \bibinfo {author} {\bibfnamefont
  {J.}~\bibnamefont {Spinner}},\ and\ \bibinfo {author} {\bibfnamefont
  {J.}~\bibnamefont {Thaler}},\ }\bibfield  {title} {\bibinfo {title} {{A
  Lorentz-equivariant transformer for all of the LHC}},\ }\href
  {https://doi.org/10.21468/SciPostPhys.19.4.108} {\bibfield  {journal}
  {\bibinfo  {journal} {SciPost Phys.}\ }\textbf {\bibinfo {volume} {19}},\
  \bibinfo {pages} {108} (\bibinfo {year} {2025})},\ \Eprint
  {https://arxiv.org/abs/2411.00446} {arXiv:2411.00446 [hep-ph]} \BibitemShut
  {NoStop}%
\bibitem [{\citenamefont {Woodward}\ \emph {et~al.}(2024)\citenamefont
  {Woodward}, \citenamefont {Park}, \citenamefont {Grosso}, \citenamefont
  {Krupa},\ and\ \citenamefont {Harris}}]{Woodward:2024dxb}%
  \BibitemOpen
  \bibfield  {author} {\bibinfo {author} {\bibfnamefont {N.~S.}\ \bibnamefont
  {Woodward}}, \bibinfo {author} {\bibfnamefont {S.~E.}\ \bibnamefont {Park}},
  \bibinfo {author} {\bibfnamefont {G.}~\bibnamefont {Grosso}}, \bibinfo
  {author} {\bibfnamefont {J.}~\bibnamefont {Krupa}},\ and\ \bibinfo {author}
  {\bibfnamefont {P.}~\bibnamefont {Harris}},\ }\bibfield  {title} {\bibinfo
  {title} {{Product Manifold Machine Learning for Physics}},\ }\href@noop {} {\
   (\bibinfo {year} {2024})},\ \Eprint {https://arxiv.org/abs/2412.07033}
  {arXiv:2412.07033 [hep-ph]} \BibitemShut {NoStop}%
\bibitem [{\citenamefont {Nabat}\ \emph {et~al.}(2025)\citenamefont {Nabat},
  \citenamefont {Ghosh}, \citenamefont {Witkowski}, \citenamefont {Kasieczka},\
  and\ \citenamefont {Whiteson}}]{Nabat:2024nce}%
  \BibitemOpen
  \bibfield  {author} {\bibinfo {author} {\bibfnamefont {S.}~\bibnamefont
  {Nabat}}, \bibinfo {author} {\bibfnamefont {A.}~\bibnamefont {Ghosh}},
  \bibinfo {author} {\bibfnamefont {E.}~\bibnamefont {Witkowski}}, \bibinfo
  {author} {\bibfnamefont {G.}~\bibnamefont {Kasieczka}},\ and\ \bibinfo
  {author} {\bibfnamefont {D.}~\bibnamefont {Whiteson}},\ }\bibfield  {title}
  {\bibinfo {title} {{Learning broken symmetries with approximate
  invariance}},\ }\href {https://doi.org/10.1103/PhysRevD.111.072002}
  {\bibfield  {journal} {\bibinfo  {journal} {Phys. Rev. D}\ }\textbf {\bibinfo
  {volume} {111}},\ \bibinfo {pages} {072002} (\bibinfo {year} {2025})},\
  \Eprint {https://arxiv.org/abs/2412.18773} {arXiv:2412.18773 [hep-ph]}
  \BibitemShut {NoStop}%
\bibitem [{\citenamefont {Sanz}(2026)}]{Sanz:2025sld}%
  \BibitemOpen
  \bibfield  {author} {\bibinfo {author} {\bibfnamefont {V.}~\bibnamefont
  {Sanz}},\ }\bibfield  {title} {\bibinfo {title} {{Learning Symmetries in
  Datasets}},\ }\href {https://doi.org/10.3390/app16041930} {\bibfield
  {journal} {\bibinfo  {journal} {Appl. Sciences}\ }\textbf {\bibinfo {volume}
  {16}},\ \bibinfo {pages} {1930} (\bibinfo {year} {2026})},\ \Eprint
  {https://arxiv.org/abs/2504.05174} {arXiv:2504.05174 [cs.LG]} \BibitemShut
  {NoStop}%
\bibitem [{\citenamefont {Spinner}\ \emph {et~al.}(2025)\citenamefont
  {Spinner}, \citenamefont {Favaro}, \citenamefont {Lippmann}, \citenamefont
  {Pitz}, \citenamefont {Gerhartz}, \citenamefont {Plehn},\ and\ \citenamefont
  {Hamprecht}}]{Spinner:2025prg}%
  \BibitemOpen
  \bibfield  {author} {\bibinfo {author} {\bibfnamefont {J.}~\bibnamefont
  {Spinner}}, \bibinfo {author} {\bibfnamefont {L.}~\bibnamefont {Favaro}},
  \bibinfo {author} {\bibfnamefont {P.}~\bibnamefont {Lippmann}}, \bibinfo
  {author} {\bibfnamefont {S.}~\bibnamefont {Pitz}}, \bibinfo {author}
  {\bibfnamefont {G.}~\bibnamefont {Gerhartz}}, \bibinfo {author}
  {\bibfnamefont {T.}~\bibnamefont {Plehn}},\ and\ \bibinfo {author}
  {\bibfnamefont {F.~A.}\ \bibnamefont {Hamprecht}},\ }\bibfield  {title}
  {\bibinfo {title} {{Lorentz Local Canonicalization: How to Make Any Network
  Lorentz-Equivariant}},\ }\href@noop {} {\  (\bibinfo {year} {2025})},\
  \Eprint {https://arxiv.org/abs/2505.20280} {arXiv:2505.20280 [stat.ML]}
  \BibitemShut {NoStop}%
\bibitem [{\citenamefont {Favaro}\ \emph {et~al.}(2025)\citenamefont {Favaro},
  \citenamefont {Gerhartz}, \citenamefont {Hamprecht}, \citenamefont
  {Lippmann}, \citenamefont {Pitz}, \citenamefont {Plehn}, \citenamefont {Qu},\
  and\ \citenamefont {Spinner}}]{Favaro:2025pgz}%
  \BibitemOpen
  \bibfield  {author} {\bibinfo {author} {\bibfnamefont {L.}~\bibnamefont
  {Favaro}}, \bibinfo {author} {\bibfnamefont {G.}~\bibnamefont {Gerhartz}},
  \bibinfo {author} {\bibfnamefont {F.~A.}\ \bibnamefont {Hamprecht}}, \bibinfo
  {author} {\bibfnamefont {P.}~\bibnamefont {Lippmann}}, \bibinfo {author}
  {\bibfnamefont {S.}~\bibnamefont {Pitz}}, \bibinfo {author} {\bibfnamefont
  {T.}~\bibnamefont {Plehn}}, \bibinfo {author} {\bibfnamefont
  {H.}~\bibnamefont {Qu}},\ and\ \bibinfo {author} {\bibfnamefont
  {J.}~\bibnamefont {Spinner}},\ }\bibfield  {title} {\bibinfo {title}
  {{Lorentz-Equivariance without Limitations}},\ }\href@noop {} {\  (\bibinfo
  {year} {2025})},\ \Eprint {https://arxiv.org/abs/2508.14898}
  {arXiv:2508.14898 [hep-ph]} \BibitemShut {NoStop}%
\bibitem [{\citenamefont {Hebbar}\ \emph {et~al.}(2025)\citenamefont {Hebbar},
  \citenamefont {Madula}, \citenamefont {Mikuni}, \citenamefont {Nachman},
  \citenamefont {Outmezguine},\ and\ \citenamefont {Savoray}}]{Hebbar:2025adf}%
  \BibitemOpen
  \bibfield  {author} {\bibinfo {author} {\bibfnamefont {P.}~\bibnamefont
  {Hebbar}}, \bibinfo {author} {\bibfnamefont {T.}~\bibnamefont {Madula}},
  \bibinfo {author} {\bibfnamefont {V.}~\bibnamefont {Mikuni}}, \bibinfo
  {author} {\bibfnamefont {B.}~\bibnamefont {Nachman}}, \bibinfo {author}
  {\bibfnamefont {N.}~\bibnamefont {Outmezguine}},\ and\ \bibinfo {author}
  {\bibfnamefont {I.}~\bibnamefont {Savoray}},\ }\bibfield  {title} {\bibinfo
  {title} {{SEAL - A Symmetry EncourAging Loss for High Energy Physics}},\
  }\href@noop {} {\  (\bibinfo {year} {2025})},\ \Eprint
  {https://arxiv.org/abs/2511.01982} {arXiv:2511.01982 [hep-ph]} \BibitemShut
  {NoStop}%
\bibitem [{\citenamefont {Petitjean}\ \emph {et~al.}(2025)\citenamefont
  {Petitjean}, \citenamefont {Plehn}, \citenamefont {Spinner},\ and\
  \citenamefont {K{\"o}the}}]{Petitjean:2025zjf}%
  \BibitemOpen
  \bibfield  {author} {\bibinfo {author} {\bibfnamefont {A.}~\bibnamefont
  {Petitjean}}, \bibinfo {author} {\bibfnamefont {T.}~\bibnamefont {Plehn}},
  \bibinfo {author} {\bibfnamefont {J.}~\bibnamefont {Spinner}},\ and\ \bibinfo
  {author} {\bibfnamefont {U.}~\bibnamefont {K{\"o}the}},\ }\bibfield  {title}
  {\bibinfo {title} {{Economical Jet Taggers -- Equivariant, Slim, and
  Quantized}},\ }\href@noop {} {\  (\bibinfo {year} {2025})},\ \Eprint
  {https://arxiv.org/abs/2512.17011} {arXiv:2512.17011 [hep-ph]} \BibitemShut
  {NoStop}%
\bibitem [{\citenamefont {Breso-Pla}\ \emph {et~al.}(2026)\citenamefont
  {Breso-Pla}, \citenamefont {Greif}, \citenamefont {Mikuni}, \citenamefont
  {Nachman}, \citenamefont {Plehn}, \citenamefont {Wamorkar},\ and\
  \citenamefont {Whiteson}}]{Breso-Pla:2026tlz}%
  \BibitemOpen
  \bibfield  {author} {\bibinfo {author} {\bibfnamefont {V.}~\bibnamefont
  {Breso-Pla}}, \bibinfo {author} {\bibfnamefont {K.}~\bibnamefont {Greif}},
  \bibinfo {author} {\bibfnamefont {V.}~\bibnamefont {Mikuni}}, \bibinfo
  {author} {\bibfnamefont {B.}~\bibnamefont {Nachman}}, \bibinfo {author}
  {\bibfnamefont {T.}~\bibnamefont {Plehn}}, \bibinfo {author} {\bibfnamefont
  {T.}~\bibnamefont {Wamorkar}},\ and\ \bibinfo {author} {\bibfnamefont
  {D.}~\bibnamefont {Whiteson}},\ }\bibfield  {title} {\bibinfo {title}
  {{Explicit or Implicit? Encoding Physics at the Precision Frontier}},\
  }\href@noop {} {\  (\bibinfo {year} {2026})},\ \Eprint
  {https://arxiv.org/abs/2603.08802} {arXiv:2603.08802 [hep-ph]} \BibitemShut
  {NoStop}%
\bibitem [{\citenamefont {Abasov}\ \emph {et~al.}(2026)\citenamefont {Abasov},
  \citenamefont {Dudko}, \citenamefont {Grigoryev}, \citenamefont {Volkov},\
  and\ \citenamefont {Zaborenko}}]{Abasov:2026jed}%
  \BibitemOpen
  \bibfield  {author} {\bibinfo {author} {\bibfnamefont {E.}~\bibnamefont
  {Abasov}}, \bibinfo {author} {\bibfnamefont {L.}~\bibnamefont {Dudko}},
  \bibinfo {author} {\bibfnamefont {F.}~\bibnamefont {Grigoryev}}, \bibinfo
  {author} {\bibfnamefont {P.}~\bibnamefont {Volkov}},\ and\ \bibinfo {author}
  {\bibfnamefont {A.}~\bibnamefont {Zaborenko}},\ }\bibfield  {title} {\bibinfo
  {title} {{Geometric algebra as the input language of collider foundation
  models}},\ }\href@noop {} {\  (\bibinfo {year} {2026})},\ \Eprint
  {https://arxiv.org/abs/2605.15910} {arXiv:2605.15910 [hep-ph]} \BibitemShut
  {NoStop}%
\bibitem [{\citenamefont {Kato}\ \emph {et~al.}(2026)\citenamefont {Kato},
  \citenamefont {Urqu{\'\i}a-Calder{\'o}n}, \citenamefont {Timiryasov},\ and\
  \citenamefont {Ruchayskiy}}]{Kato:2026txd}%
  \BibitemOpen
  \bibfield  {author} {\bibinfo {author} {\bibfnamefont {M.}~\bibnamefont
  {Kato}}, \bibinfo {author} {\bibfnamefont {K.~A.}\ \bibnamefont
  {Urqu{\'\i}a-Calder{\'o}n}}, \bibinfo {author} {\bibfnamefont
  {I.}~\bibnamefont {Timiryasov}},\ and\ \bibinfo {author} {\bibfnamefont
  {O.}~\bibnamefont {Ruchayskiy}},\ }\bibfield  {title} {\bibinfo {title}
  {{Learning Standard Model structure from LHC data with Riemannian flow
  matching}},\ }\href@noop {} {\  (\bibinfo {year} {2026})},\ \Eprint
  {https://arxiv.org/abs/2607.16144} {arXiv:2607.16144 [hep-ph]} \BibitemShut
  {NoStop}%
\bibitem [{\citenamefont {Kleiss}\ \emph {et~al.}(1986)\citenamefont {Kleiss},
  \citenamefont {Stirling},\ and\ \citenamefont {Ellis}}]{Kleiss:1985gy}%
  \BibitemOpen
  \bibfield  {author} {\bibinfo {author} {\bibfnamefont {R.}~\bibnamefont
  {Kleiss}}, \bibinfo {author} {\bibfnamefont {W.~J.}\ \bibnamefont
  {Stirling}},\ and\ \bibinfo {author} {\bibfnamefont {S.~D.}\ \bibnamefont
  {Ellis}},\ }\bibfield  {title} {\bibinfo {title} {{A New Monte Carlo
  Treatment of Multiparticle Phase Space at High-energies}},\ }\href
  {https://doi.org/10.1016/0010-4655(86)90119-0} {\bibfield  {journal}
  {\bibinfo  {journal} {Comput. Phys. Commun.}\ }\textbf {\bibinfo {volume}
  {40}},\ \bibinfo {pages} {359} (\bibinfo {year} {1986})}\BibitemShut
  {NoStop}%
\bibitem [{\citenamefont {Bogorad}\ \emph {et~al.}(2026)\citenamefont
  {Bogorad}, \citenamefont {Elsharkawy}, \citenamefont {Kahn}, \citenamefont
  {Larkoski},\ and\ \citenamefont {Levi}}]{Bogorad:2026oxa}%
  \BibitemOpen
  \bibfield  {author} {\bibinfo {author} {\bibfnamefont {Z.}~\bibnamefont
  {Bogorad}}, \bibinfo {author} {\bibfnamefont {I.}~\bibnamefont {Elsharkawy}},
  \bibinfo {author} {\bibfnamefont {Y.}~\bibnamefont {Kahn}}, \bibinfo {author}
  {\bibfnamefont {A.~J.}\ \bibnamefont {Larkoski}},\ and\ \bibinfo {author}
  {\bibfnamefont {N.}~\bibnamefont {Levi}},\ }\bibfield  {title} {\bibinfo
  {title} {{Generative models on phase space}},\ }\href@noop {} {\  (\bibinfo
  {year} {2026})},\ \Eprint {https://arxiv.org/abs/2604.02415}
  {arXiv:2604.02415 [hep-ph]} \BibitemShut {NoStop}%
\bibitem [{\citenamefont {Zamolodchikov}(1986)}]{Zamolodchikov:1986gt}%
  \BibitemOpen
  \bibfield  {author} {\bibinfo {author} {\bibfnamefont {A.~B.}\ \bibnamefont
  {Zamolodchikov}},\ }\bibfield  {title} {\bibinfo {title} {{Irreversibility of
  the Flux of the Renormalization Group in a 2D Field Theory}},\ }\href@noop {}
  {\bibfield  {journal} {\bibinfo  {journal} {JETP Lett.}\ }\textbf {\bibinfo
  {volume} {43}},\ \bibinfo {pages} {730} (\bibinfo {year} {1986})}\BibitemShut
  {NoStop}%
\bibitem [{\citenamefont {Polchinski}(1988)}]{Polchinski:1987dy}%
  \BibitemOpen
  \bibfield  {author} {\bibinfo {author} {\bibfnamefont {J.}~\bibnamefont
  {Polchinski}},\ }\bibfield  {title} {\bibinfo {title} {{Scale and Conformal
  Invariance in Quantum Field Theory}},\ }\href
  {https://doi.org/10.1016/0550-3213(88)90179-4} {\bibfield  {journal}
  {\bibinfo  {journal} {Nucl. Phys. B}\ }\textbf {\bibinfo {volume} {303}},\
  \bibinfo {pages} {226} (\bibinfo {year} {1988})}\BibitemShut {NoStop}%
\bibitem [{\citenamefont {Nakayama}(2015)}]{Nakayama:2013is}%
  \BibitemOpen
  \bibfield  {author} {\bibinfo {author} {\bibfnamefont {Y.}~\bibnamefont
  {Nakayama}},\ }\bibfield  {title} {\bibinfo {title} {{Scale invariance vs
  conformal invariance}},\ }\href
  {https://doi.org/10.1016/j.physrep.2014.12.003} {\bibfield  {journal}
  {\bibinfo  {journal} {Phys. Rept.}\ }\textbf {\bibinfo {volume} {569}},\
  \bibinfo {pages} {1} (\bibinfo {year} {2015})},\ \Eprint
  {https://arxiv.org/abs/1302.0884} {arXiv:1302.0884 [hep-th]} \BibitemShut
  {NoStop}%
\bibitem [{\citenamefont {Dymarsky}\ \emph {et~al.}(2015)\citenamefont
  {Dymarsky}, \citenamefont {Komargodski}, \citenamefont {Schwimmer},\ and\
  \citenamefont {Theisen}}]{Dymarsky:2013pqa}%
  \BibitemOpen
  \bibfield  {author} {\bibinfo {author} {\bibfnamefont {A.}~\bibnamefont
  {Dymarsky}}, \bibinfo {author} {\bibfnamefont {Z.}~\bibnamefont
  {Komargodski}}, \bibinfo {author} {\bibfnamefont {A.}~\bibnamefont
  {Schwimmer}},\ and\ \bibinfo {author} {\bibfnamefont {S.}~\bibnamefont
  {Theisen}},\ }\bibfield  {title} {\bibinfo {title} {{On Scale and Conformal
  Invariance in Four Dimensions}},\ }\href
  {https://doi.org/10.1007/JHEP10(2015)171} {\bibfield  {journal} {\bibinfo
  {journal} {JHEP}\ }\textbf {\bibinfo {volume} {10}},\ \bibinfo {pages}
  {171}},\ \Eprint {https://arxiv.org/abs/1309.2921} {arXiv:1309.2921 [hep-th]}
  \BibitemShut {NoStop}%
\bibitem [{\citenamefont {Peskin}\ and\ \citenamefont
  {Schroeder}(1995)}]{Peskin:1995ev}%
  \BibitemOpen
  \bibfield  {author} {\bibinfo {author} {\bibfnamefont {M.~E.}\ \bibnamefont
  {Peskin}}\ and\ \bibinfo {author} {\bibfnamefont {D.~V.}\ \bibnamefont
  {Schroeder}},\ }\href {https://doi.org/10.1201/9780429503559} {\emph
  {\bibinfo {title} {{An Introduction to quantum field theory}}}}\ (\bibinfo
  {publisher} {Addison-Wesley},\ \bibinfo {address} {Reading, USA},\ \bibinfo
  {year} {1995})\BibitemShut {NoStop}%
\bibitem [{\citenamefont {Mangano}\ and\ \citenamefont
  {Parke}(1991)}]{Mangano:1990by}%
  \BibitemOpen
  \bibfield  {author} {\bibinfo {author} {\bibfnamefont {M.~L.}\ \bibnamefont
  {Mangano}}\ and\ \bibinfo {author} {\bibfnamefont {S.~J.}\ \bibnamefont
  {Parke}},\ }\bibfield  {title} {\bibinfo {title} {{Multiparton Amplitudes in
  Gauge Theories}},\ }\href {https://doi.org/10.1016/0370-1573(91)90091-Y}
  {\bibfield  {journal} {\bibinfo  {journal} {Phys. Rept.}\ }\textbf {\bibinfo
  {volume} {200}},\ \bibinfo {pages} {301} (\bibinfo {year} {1991})},\ \Eprint
  {https://arxiv.org/abs/hep-th/0509223} {arXiv:hep-th/0509223} \BibitemShut
  {NoStop}%
\bibitem [{\citenamefont {Dixon}(1996)}]{Dixon:1996wi}%
  \BibitemOpen
  \bibfield  {author} {\bibinfo {author} {\bibfnamefont {L.~J.}\ \bibnamefont
  {Dixon}},\ }\bibfield  {title} {\bibinfo {title} {{Calculating scattering
  amplitudes efficiently}},\ }in\ \href@noop {} {\emph {\bibinfo {booktitle}
  {{Theoretical Advanced Study Institute in Elementary Particle Physics (TASI
  95): QCD and Beyond}}}}\ (\bibinfo {year} {1996})\ pp.\ \bibinfo {pages}
  {539--584},\ \Eprint {https://arxiv.org/abs/hep-ph/9601359}
  {arXiv:hep-ph/9601359} \BibitemShut {NoStop}%
\bibitem [{\citenamefont {Banfi}\ \emph {et~al.}(2005)\citenamefont {Banfi},
  \citenamefont {Salam},\ and\ \citenamefont {Zanderighi}}]{Banfi:2004yd}%
  \BibitemOpen
  \bibfield  {author} {\bibinfo {author} {\bibfnamefont {A.}~\bibnamefont
  {Banfi}}, \bibinfo {author} {\bibfnamefont {G.~P.}\ \bibnamefont {Salam}},\
  and\ \bibinfo {author} {\bibfnamefont {G.}~\bibnamefont {Zanderighi}},\
  }\bibfield  {title} {\bibinfo {title} {{Principles of general final-state
  resummation and automated implementation}},\ }\href
  {https://doi.org/10.1088/1126-6708/2005/03/073} {\bibfield  {journal}
  {\bibinfo  {journal} {JHEP}\ }\textbf {\bibinfo {volume} {03}},\ \bibinfo
  {pages} {073}},\ \Eprint {https://arxiv.org/abs/hep-ph/0407286}
  {arXiv:hep-ph/0407286} \BibitemShut {NoStop}%
\bibitem [{\citenamefont {Komiske}\ \emph {et~al.}(2018)\citenamefont
  {Komiske}, \citenamefont {Metodiev},\ and\ \citenamefont
  {Thaler}}]{Komiske:2017aww}%
  \BibitemOpen
  \bibfield  {author} {\bibinfo {author} {\bibfnamefont {P.~T.}\ \bibnamefont
  {Komiske}}, \bibinfo {author} {\bibfnamefont {E.~M.}\ \bibnamefont
  {Metodiev}},\ and\ \bibinfo {author} {\bibfnamefont {J.}~\bibnamefont
  {Thaler}},\ }\bibfield  {title} {\bibinfo {title} {{Energy flow polynomials:
  A complete linear basis for jet substructure}},\ }\href
  {https://doi.org/10.1007/JHEP04(2018)013} {\bibfield  {journal} {\bibinfo
  {journal} {JHEP}\ }\textbf {\bibinfo {volume} {04}},\ \bibinfo {pages}
  {013}},\ \Eprint {https://arxiv.org/abs/1712.07124} {arXiv:1712.07124
  [hep-ph]} \BibitemShut {NoStop}%
\end{thebibliography}%

\end{document}